\documentclass[preprint,12pt,number]{elsarticle}

\usepackage[utf8]{inputenc}
\usepackage{blindtext}
\usepackage{todonotes}
\usepackage{subcaption}
\usepackage{amsmath}
\usepackage{amsthm}
\usepackage{stmaryrd}
\usepackage{hyperref}
\usepackage{import}
\usepackage{amsfonts}
\usepackage{amssymb}
\usepackage{yhmath}
\usepackage{algorithm}
\usepackage{algpseudocode}
\usepackage{mathrsfs}
\usepackage{enumitem}
\usepackage{tabularx} %
\usepackage{array}    %
\usepackage{geometry}
\usetikzlibrary{arrows.meta}
\usetikzlibrary{decorations}
\usetikzlibrary{decorations.pathmorphing}
\usetikzlibrary[patterns]
\usetikzlibrary{shapes.misc, positioning}
\usepackage{cleveref}
\usepackage{xcolor}

\usepackage{appendix}
\newtheorem{theorem}{Theorem}[section]
\newtheorem{proposition}[theorem]{Proposition}
\newtheorem{lemma}[theorem]{Lemma}
\newtheorem{claim}[theorem]{Claim}
\newtheorem{remark}[theorem]{Remark}
\newtheorem{invariant}[theorem]{Invariant}
\newtheorem{definition}[theorem]{Definition}
\newtheorem{property}[theorem]{Property}
\newtheorem{example}[theorem]{Example}

\Crefname{property}{Property}{Properties} 
\crefname{property}{Property}{Properties}

\newcommand{\tsat}{3-SAT\xspace}

\newcommand{\MPTSAT}{Monotone Planar \tsat}

\newcommand{\LPTSAT}{Linear Planar \tsat}
\newcommand{\PLLPTSAT}{Linear Literal-Planar \tsat}
\newcommand{\PLLPTformula}{linear literal-planar 3-CNF\xspace}
\newcommand{\npc}{NP-complete\xspace}
\newcommand{\pspacec}{PSPACE-complete\xspace}

\newcommand{\NCL}{Nondeterministic Constraint Logic\xspace}

\newcommand{\ctoc}{Configuration To Configuration\xspace}

\newcommand{\inciGraph}{variable-clause graph\xspace}
\newcommand{\litClausGraph}{literal-clause graph\xspace}
\newcommand{\InciGraph}{Variable-clause graph\xspace}
\newcommand{\LitClausGraph}{Literal-clause graph\xspace}

\newcommand{\resp}{resp.\xspace}

\newcommand{\assignment}{\nu}

\def\vars{V}
\newcommand{\variableedge}[1]{\mathsf{v}_{#1}}

\newcommand{\litClauGraphSign}[1]{{G_{\textrm{\tt lit}}(#1)}}
\newcommand{\inciGraphSign}[1]{G_{\textrm{\tt var}}(#1)}

\newcommand\set[1]{\{#1\}}
\newcommand\literal{\ell}

\newcommand{\acycle}{\pi}

\newcommand{\faces}{\textrm{\tt dual}}
\newcommand{\face}{\textrm{\tt face}}
\newcommand{\multiset}[1]{\{\!\!\{#1\}\!\!\}}

\newcommand{\bijectionedgetodualG}{\faces_{G}} %
\newcommand{\bijectionedgetodual}{\faces_{\Gamma_G}} %
\newcommand{\superleftrightarrow}[2]{
	\begin{tikzpicture}[baseline=0mm]
		\node (v1) at (-2, 0) {};
		\node (v2) at (2,0) {};
		\draw[-latex, line width=0.5mm] (v1) edge[bend left=20] node[above] {#1} (v2);
		\draw[-latex, line width=0.5mm] (v2) edge[bend left=20] node[below] {#2} (v1);
	\end{tikzpicture}
}

\tikzstyle{primal} = [fill=white,inner sep=0.5mm]
\tikzstyle{dual} = [red, inner sep=0.5mm]
\newcommand{\facenode}[2]{
	\draw[draw=none,pattern=north west lines, pattern color=gray!30!white] (#1, #2) circle (6mm);
	\node[fill=gray!10!white, circle, inner sep=2.5mm] at (#1, #2) {};}

	\newcommand{\fleche}[1]{\tikz{\draw (0, 0) edge[line width=1mm, -latex] node[above] {\footnotesize #1} (2, 0);)}}

\newcommand{\radius}{\rho}
\newcommand{\setZ}{\mathbb Z}

\newcommand{\rightarrowcmapf}[1]{\rightsquigarrow_{#1}}
\newcommand{\rightarrowcmapfbounded}[2]{\rightsquigarrow_{#1}^{\leq #2}}

\newcommand{\bridge}[1]{\ensuremath{\textrm{\tt BRIDGE}_{#1}}}
\newcommand{\switch}[1]{\ensuremath{\textrm{\tt SWITCH}_{#1}}}
\newcommand{\splitgadget}[1]{\ensuremath{\textrm{\tt SPLIT}_{#1}}}
\newcommand{\splitagent}[1]{s_{#1}}
\newcommand{\connectorgadget}[1]{\ensuremath{\textrm{\tt CONNECTOR}_{#1}}}

\def\face{\textrm{\tt face}}
\def\other{\textrm{\tt Other}}

\newcommand\singlerequester{
\begin{tikzpicture}
\node [style={green_agent}] at (0, 0) {};
\end{tikzpicture}
}

\newcommand{\wireproviders}{
\begin{tikzpicture}
\foreach \x in {0, 0.5, ..., 2} {
\node [style={blue_agent}] at (\x, 0) {};
}
\end{tikzpicture}}

\newcommand{\wirerequesters}{
\begin{tikzpicture}
\foreach \x in {0, 0.5, ..., 2} {
\node [style={green_agent}] at (\x, 0) {};
}
\end{tikzpicture} }

\newcommand\firstvariable{\alpha}
\newcommand\lastvariable{\omega}

\journal{Theoretical Computer Science}

\begin{document}

\begin{frontmatter}

\title{Linear Planar 3-SAT and Its Applications in Planning}

\author[irisa,numasoft]{Victorien Desbois}{}
\author[irisa,inria]{Ocan Sankur}{}
\author[enslyon]{Fran\c cois Schwarzentruber}{}

\affiliation[enslyon]{organization={ENS de Lyon, CNRS, University Claude Bernard Lyon 1, Inria, LIP, UMR 5668},
city={Lyon},
postcode={69342},
country={France}}

\affiliation[irisa]{organization={Univ Rennes, CNRS, IRISA},%
city={Rennes},
postcode={35000},
country={France}}

\affiliation[numasoft]{organization={NumaSoft},city={Bruz},postcode={35170},country={France}}

\affiliation[inria]{organization={Inria},%
city={Rennes},
postcode={35000},
country={France}}

\begin{abstract}
	Several fragments of the satisfiability problem have been studied in the literature.
	Among these, Linear \tsat is a satisfaction problem in which each clause (viewed as a set of literals) intersects with at most one other clause;
	moreover, any pair of clauses have at most one literal in common. 
	Planar \tsat is a fragment which requires that the so-called variable-clause graph is planar.
	Both fragments are NP-complete and have applications in encoding NP-hard planning problems.
	In this paper, we investigate the complexity and applications of the fragment obtained combining both features. We define Linear Planar \tsat and prove its NP-completeness. 
	We also study the reconfiguration problem of Linear Planar \tsat and show that it is PSPACE-complete. 
	As an application, we use these new results to prove 
	the NP-completeness of Bounded Connected Multi-Agent Pathfinding and the PSPACE-completeness of Connected Multi-Agent Pathfinding
	in two-dimensional grids.
\end{abstract}

\end{frontmatter}

\tikzstyle{none}=[]

\tikzstyle{dual_g_face}=[fill=white, draw=black, shape=circle, densely dotted, inner sep=0.5mm]
\tikzstyle{variable_node}=[fill=white, draw=black, minimum height=6mm, inner sep=1mm]
\tikzstyle{clause_node}=[fill=none, draw=black, shape=rectangle]

\tikzstyle{requester_agent}=
[fill=white, draw=black, shape=circle, dotted, execute at begin node={$\bullet$}, inner sep=0pt, minimum size=0.4cm]

\tikzstyle{provider_agent}=
[
fill=gray!20!white, draw=gray,
 shape=rectangle, densely dotted, 
execute at begin node={$\bullet$}, 
minimum size=4mm, inner sep=0pt]

\tikzstyle{blue_agent} = [provider_agent]
\tikzstyle{green_agent} = [requester_agent]

\tikzstyle{movement_node}=[very thick, fill=white, draw=black, shape=circle, minimum size=0.4cm, inner sep=0pt]
\tikzstyle{gadget}=[fill=gray!10!white, draw=gray, shape=rectangle]
\tikzstyle{target}=[very thick, fill=lightgray, draw=black, shape=circle, minimum size=0.4cm, inner sep=0pt]

\tikzstyle{positiveclause} = 
[clause, 
fill=gray!20!white,
minimum height=0.75cm]

\tikzstyle{negativeclause} = [clause, fill=white, minimum height=0.75cm]

\tikzstyle{litteral} = [draw, circle, minimum height=8mm, inner sep=0.2mm]
\tikzstyle{variable_node} = [draw, circle, minimum height=8mm, inner sep=0.2mm]

\tikzstyle{clause} = [draw, rectangle, minimum size=1cm, scale=0.7]
\tikzstyle{variablecycle} = %
 [draw=gray, 
 line width=0.2mm, 
 double=gray!20!white,
  double distance = 0.5mm ]
  
\tikzstyle{paired} = [decorate,decoration={snake,amplitude=.4mm,segment length=1mm
}]

\tikzstyle{treeTour} = [very thick, densely dotted]

\tikzstyle{redface} = [
draw=none,
fill=gray!20!white
]

\tikzstyle{blueface} = [
draw=none,
fill=white
]

\tikzstyle{blueedge} = [
thin
]
\tikzstyle{rededge} = [
line width=1mm,
]

\tikzstyle{edge}=[draw=black, fill=none, ->, densely dotted, >={Stealth[length=1mm, width=1.5mm]}]
\tikzstyle{directed_edge}=[->, dashed]
\tikzstyle{dotted_edge}=[-, densely dotted]
\tikzstyle{tree_edge}=[-, very thick, densely dashed]
\tikzstyle{paired_planar}=[-, dashed]

\tikzstyle{connectivity_requester_wire}=[-, dashed]
\tikzstyle{connectivity_provider_wire}= [
 line width=0.2mm ]

\tikzstyle{movement_wire}=[
 line width=0.4mm, double=gray!50!white, double distance = 0.4mm ]

\tikzstyle{explanation} = [gray, font=\tiny]

\pgfdeclarelayer{nodelayer}
\pgfdeclarelayer{edgelayer}

\tikzstyle{inVprime} = [draw, fill=white,shape=circle, inner sep=0.7mm]
\tikzstyle{kite} = [dashed]
\tikzstyle{node} = [draw,inner sep=0.5mm]
\tikzstyle{dual} = %
[draw,gray,inner sep=0.5mm, fill=gray!20!white]

\tikzstyle{face} =
[gray,draw=none,fill=gray!20!white]

\tikzstyle{two weig edge}=[
draw, 
-latex, double, ultra thick]
\tikzstyle{one weig edge}=[
-latex, thick]

\tikzstyle{nclnode} = [rounded corners=0.9mm, draw]

\tikzstyle{vfgedge} = [line width=1mm, orange]
\tikzstyle{dualvfgvertex} = [fill=green, opacity=0.4, circle, minimum height=8mm]
\tikzstyle{dualvfgvertexclause} = [fill=yellow, opacity=0.4, circle, minimum height=8mm]

\section{Introduction}
The boolean satisfiability problem (SAT) is a major tool to classify decision problems with respect to their feasibility.
For instance, many graph problems are proven to be NP-hard by reduction from SAT~\cite{garey2002computers}.
Despite this versatility of SAT, reductions from SAT are not always natural or immediate for some classes of problems. 
This is the case, for example, of hard problems on \emph{planar} graphs.
While some NP-hard problems such as node cover, and directed Hamiltonian path remain NP-hard when the input graph is assumed to be planar~\cite{lichtenstein1982planar},
existing polynomial reductions from SAT for general graphs do not produce planar graphs, and proving hardness for planar graphs
thus requires designing reductions adapted for planar graphs. 
This has motivated defining subclasses of the SAT problem. Accordingly, \cite{lichtenstein1982planar} first defines the  NP-complete \emph{planar SAT} problem 
which is the SAT problem restricted to instances that induce a planar \emph{incidence} graph\footnote{This is a bipartite graph connecting clauses to variables they contain,
defined formally in \Cref{sec:background}.}
Once the NP-completeness of planar SAT is established, then reductions to planar graph problems are done easily.
Thus planar SAT has become a natural tool for proving the hardness of planar graph problems \cite{dyer1986planar,mahajan2012planar,shi2005rectilinear} but also planning problems which are naturally cast in planar domains~\cite{yu2015intractability,adler2020tatamibari,allen2018sto}.

\emph{Linear 3-SAT} (LSAT) \cite{arkin2018selecting} is the SAT problem in which each clause intersects at most one other clause, and where this intersection is at most a singleton. 
This particular restriction was motivated by 
graph problems \cite{arkin2018selecting,simon2021algorithmic} and also applied to planning problems;
for instance, when modeling collision detection on two-dimensional grids the linearity restriction arises naturally \cite{gupta2024collision}. 
Other applications in planning include the complexity of multi-agent path execution with uncertainty~\cite{liu2024multi}. As for the planar case, resulting reductions from linear SAT are much more natural than encoding the general SAT problem in the graph or planning problems at hand.

Another type of problems related to planning is the \emph{SAT reconfiguration} problem.
In this problem, a Boolean formula $\phi$ is given along with two satisfying assignments~$\nu,\nu'$,
and one needs to determine whether one can define a sequence of assignments obtained from $\nu$
by flipping the truth value of a single variable at each step and obtain $\nu'$,
while satisfying $\phi$ at each step. 
The \emph{3-SAT reconfiguration} problem consists in solving this for 3-CNF formulas, and is PSPACE-complete~\cite{monotonNAEtSATRecon}.
Planning problems in which the arena is succinctly described (by a Boolean formula) can naturally be reduced to SAT reconfiguration problems. It was shown that the planar 3-SAT reconfiguration problem is PSPACE-complete \cite{reconfigComplexity}.
Note that reconfiguration problems are not restricted to SAT; one can lift any decision problem to a reconfiguration problem; \cite{reconfigComplexity} shows that many reconfiguration problems derived from NP-complete problems are PSPACE-complete.

In this paper, we investigate subclasses of the SAT and SAT reconfiguration problems restricted to both planar \emph{and} linear formulas. Our goal is to investigate restricted classes of these problems that are still NP-complete and PSPACE-complete respectively, while providing new tools for proving the hardness of planning problems;
and to present applications of these problems to connected multi-agent path finding (CMAPF)~\cite{cmapfIsse}.

For the SAT problem, we present two variants: \emph{linear planar 3-SAT} and \emph{linear literal-planar 3-SAT}.
Both require formulas to be linear but both make planarity assumptions on different graphs.
Linear planar 3-SAT is the 3-SAT problem restricted to formulas that are both linear and planar.
Therefore, it requires that the \emph{variable-clause graph} augmented with a \emph{variable cycle} is planar~\cite{lichtenstein1982planar}.
The variable-clause graph is a bipartite graph connecting clauses to variables they contain while a variable cycle is a cycle visiting each variable exactly once.
In linear literal-planar 3-SAT, 
we require that the \emph{literal-clause graph} augmented with a \emph{literal cycle} is planar.
The literal-clause graph is defined in analogy with the variable-clause graph by connecting clauses to literals they contain;
and a literal cycle is a cycle visitng each literal exactly once.
The literal-clause graph can be seen as a more refined version of the variable-clause graph and has not been considered
in the literature to our knowledge. 

We prove that both new variants are NP-complete. We obtain this result first for linear literal-planar 3-SAT by reduction 
from monotone planar 3-SAT \cite{de2010optimal}; then we reduce linear literal-planar 3-SAT to 
monotone linear planar 3-SAT which is a restriction of linear planar 3-SAT to monotone formulas.
The NP-hardness of linear planar 3-SAT follows.

We also study the complexity of the reconfiguration problems derived from both variants of 3-SAT
and prove them to be PSPACE-complete following a similar path. We first prove the PSPACE-hardness of
linear literal-planar 3-SAT reconfiguration by reduction from NCL~\cite{gamesNCLthesis}, and then reduce it to monotone linear planar 3-SAT reconfiguration,
which is a subset of the linear planar 3-SAT reconfiguration.

As an application, we consider the \emph{connected multi-agent path finding} (CMAPF) problem
over two-dimensional grids and with radius communication. In this problem, a 2D grid is given along with an
initial and a final configuration of $n$ agents, and a communication radius $\radius$. The goal is to find a sequence of configurations starting at the initial configuration and ending in the final configuration, where each consecutive configuration is obtained by moving each agent to a neighboring cell while avoiding collisions with obstacles (some cells are obstacles) and with other agents.
Moreover, at each configuration, any pair of agents $a$ and $b$ must remain connected in the following sense:
either $a$ and $b$ are within a distance of at most $\radius$, or $a$ is within a distance of at most $\radius$ to a third agent~$c$,
and $c$ and $b$ are connected.
When $\radius=\infty$, this problem is known simply as MAPF~\cite{LiPhD22}.
On general graphs, and when communication is defined by an additional arbitrary graph rather than by radius, this problem is PSPACE-complete~\cite{cmapfTateo}, while it remains PSPACE-complete on 3D grids with radius communication~\cite{cmapfIsse}.
The bounded version of these problems are NP-complete when the bound on the length of the execution is given in unary~\cite{cmapfTateo,cmapfIsse}.

In both \cite{cmapfTateo} and \cite{cmapfIsse} the PSPACE-hardness was obtained by reduction from the configuration to configuration problem for NCL. While the latter work manages to present a planar encoding, 
getting rid of the third dimension was difficult given the constraints of the reduction.
In fact, in \cite{cmapfIsse} it was possible to connect the gadgets arbitrarily thanks to the additional freedom brought by the third dimension
(an additional \emph{activation} gadget could be placed at a different height and interact with each gadget placed below).
In two dimensions, this trick no longer works: gadgets now cannot be designed independently, and their connectivity and the mentioned activation features must be a part of their design.
We achieve this instead thanks to the planar variable cycles.
We thus prove the PSPACE-hardness of the 2D case by reduction from the
linear literal-planar 3-SAT reconfiguration problem.
The NP-hardness of the bounded version was also only established in the 3D case;
and we prove the NP-hardness in 2D as well by reduction from linear literal-planar 3-SAT.

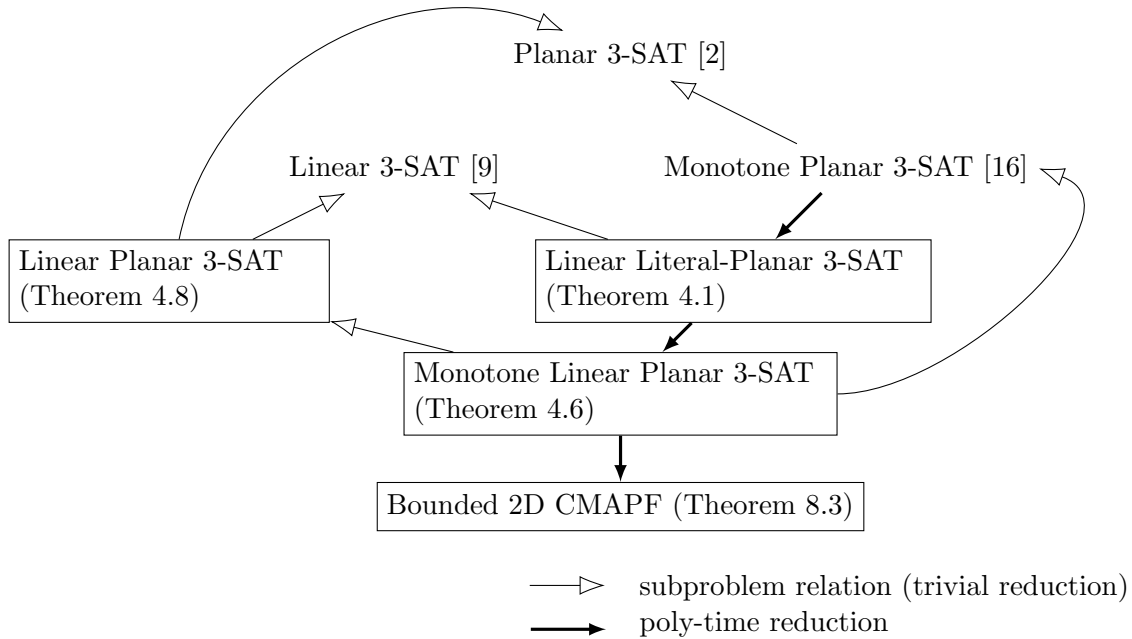
\begin{figure}
	\begin{center}
		\tikzstyle{inheritsfrom} = [-{Latex[open,scale=2]}]
		\tikzstyle{reducesto} = [-latex, very thick]
		\small
		\begin{tikzpicture}[scale=1.5]

			\node (planar3-SAT) at (0,0) {Planar 3-SAT \cite{lichtenstein1982planar}};
			
			\node (linear3-SAT) at (-2, -1) {Linear 3-SAT \cite{arkin2018selecting}};
			\node (planarmonotone3-SAT) at (2, -1) {Monotone Planar 3-SAT \cite{de2010optimal}};
			\node[draw, text width=40mm] (planarlinear3-SAT) at (-4, -2) {Linear Planar 3-SAT \\
			 (\Cref{th:planar_linear_tsat_npc})};
			\node[draw, text width=55mm] (constraintedmonotoneplanarlinear) at (0, -3) {Monotone Linear Planar 3-SAT (\Cref{th:constrainedmonotoneplanarlinearthreesatnpcomplete})};
			\node[draw, text width=50mm] (linearliteralplanar) at (1, -2) {Linear Literal-Planar 3-SAT (\Cref{th:paired_planar_linear_tsat_npc})};
			\node[draw] (boundedCMAPF) at (0, -4) {Bounded 2D CMAPF (\Cref{th:bounded2DCMAPFNPcomplete})};
			\draw[inheritsfrom] (planarlinear3-SAT) edge[bend left=50] (planar3-SAT);
			\draw[inheritsfrom] (planarlinear3-SAT) edge (linear3-SAT);
			\draw[inheritsfrom] (linearliteralplanar) edge (linear3-SAT);
			\draw[inheritsfrom] (planarmonotone3-SAT) edge (planar3-SAT);
			\draw[inheritsfrom] (constraintedmonotoneplanarlinear) edge (planarlinear3-SAT);
			\draw[inheritsfrom] (constraintedmonotoneplanarlinear) edge[in=340, out=0] (planarmonotone3-SAT.east);
			\draw[reducesto] (planarmonotone3-SAT) -- (linearliteralplanar);
			\draw (linearliteralplanar) edge[reducesto] (constraintedmonotoneplanarlinear);
			\draw[reducesto] (constraintedmonotoneplanarlinear) -- (boundedCMAPF);
		\end{tikzpicture}

		~

		\hfill
		\begin{tabular}{ll}
			\tikz{\draw (0, 0) edge[inheritsfrom] (1, 0)} & subproblem relation (trivial reduction) \\
			\tikz{\draw (0, 0) edge[reducesto] (1, 0)} & poly-time reduction
		\end{tabular}
	\end{center}
	\caption{NP-complete 3-SAT problems, and the NP-completeness of bounded grid CMAPF.
		Problems shown in a box are introduced in this paper, while others are from the literature.
		Thick edges show the reductions we build to prove our hardness results, while gray edges 
		are only given for information.
	\label{figure:3-SATproblems}}
\end{figure}

\begin{figure}
	\begin{center}
		\tikzstyle{inheritsfrom} = [-{Latex[open,scale=2]}]
		\tikzstyle{reducesto} = [-latex]
		\small
		\begin{tikzpicture}[scale=1.5]
			\node (NCL) at (1.35, -0.7) {C2C in NCL \cite{gamesNCLthesis}};
			\node (3-SAT) at (0,0) {(Cycle-free) Planar 3-SAT reconfiguration \cite{monotonNAEtSATRecon}};
			\node[draw, text width=4cm] (planarlinear3-SAT) at (-4, -1) {Linear Planar 3-SAT Reconfiguration};
			\node[draw, text width=6cm] (constraintedmonotoneplanarlinear) at (-5, -3) {Monotone Linear Planar 3-SAT Reconfiguration (\Cref{th:linear_planar_tsat_pspace_c})};
			\node[draw, text width=6cm] (linearliteralplanar) at (0, -2) {Linear Literal-Planar 3-SAT Reconfiguration (\Cref{th:proof_ltsat_reconfig_pspace})};
			\node[draw] (CMAPF) at (0, -3) {2D CMAPF (\Cref{th:2D CMAPFPSPACEcomplete})};
			\draw[inheritsfrom] (linearliteralplanar) edge (3-SAT);
			\draw[inheritsfrom] (planarlinear3-SAT) edge (3-SAT);
			\draw[reducesto, very thick] (NCL) -- (3-SAT);
			\draw[reducesto, very thick] (NCL) -- (linearliteralplanar);
			\draw[reducesto, very thick] (constraintedmonotoneplanarlinear) -- (CMAPF);
			\draw[reducesto, very thick] (linearliteralplanar) edge[reducesto] (constraintedmonotoneplanarlinear);
			\draw[inheritsfrom
			] (constraintedmonotoneplanarlinear) edge (planarlinear3-SAT);
			
		\end{tikzpicture}
	\end{center}
	\caption{PSPACE-complete of reconfiguration problems.\label{figure:reconfigurationproblems}}
\end{figure}
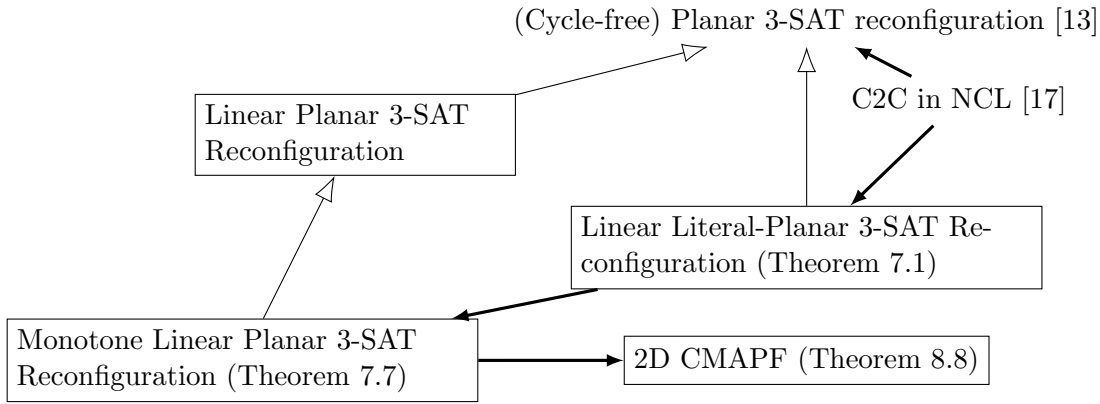

All reductions established in this paper are illustrated in Figures \ref{figure:3-SATproblems} and~\ref{figure:reconfigurationproblems}.

Apart from the fragments of SAT and reductions presented above, we also take a close look in this work at \emph{variable cycles} and \emph{literal cycles}
which augment and constrain the variable-clause and literal-clause graphs associated to SAT formulas.
Variable cycles appear in \cite{lichtenstein1982planar} and are used to constrain the Monotone Planar 3-SAT problem by requiring it to separate
positive clauses from negative clauses in considered embeddings \cite{de2010optimal}. 
In these proofs, the given gadgets ensure by construction that a variable cycle is formed.
In the reconfiguration problems studied in the literature, these cycles are not used in the reductions.
In our reconfiguration problems in this paper, we reduce from NCL and not from 3-SAT, therefore the variable cycle becomes nontrivial to build (in fact, the existence of
such a cycle is NP-complete; see \Cref{th:general_planar_cycle_construction_npc}).
Thus, we establish here a sufficient criterion, called \emph{kite graphs}, for variable-clause and literal-clause graphs to admit cycles that preserve their planarity. We furthermore
identify a case where such a cycle always exists and can be computed in polynomial time. These results are crucial steps in our reductions while they have interest on their own.

\paragraph{Outline}
We first present background on SAT problems and their known variants in \Cref{sec:background}.
We introduce our variants in \Cref{sec:new-variants}, and prove their NP-hardness in \Cref{sec:complexity-tsat}.
The cycle augmentation problem is then studied in \Cref{section:cycleaugmentation}.
Reconfiguration problems are presented in \Cref{sec:background-reconfiguration} and PSPACE-hardness proved in our case
in \Cref{sec:complexity-reconfiguration}.
Last, complexity results for the CMAPF problems in 2D are presented in \Cref{sec:complexity-cmapf}.

\section{Background on Satisfaction Problems}
\label{sec:background}

Let $\vars$ be a set of variables. We consider propositional formulas over variables in~$\vars$, in conjunctive normal form (CNF), that is, each formula has the form $\phi=\land_{i=1}^mc_i$. Moreover we see each clause as a set of literals.
Furthermore, we consider 3-CNF formulas whose clauses contain at most three literals.
The \emph{SAT problem} consists in deciding if a CNF formula is satisfiable, and the \emph{3-SAT} problem is the restriction of this problem to 3-CNF formulas.

In this section, we present some NP-complete classes of 3-SAT problems from the literature, namely, the \emph{linear 3-SAT}, \emph{planar 3-SAT}, and \emph{monotone planar 3-SAT}.
We will then define our own variants which combine features from these classes.

\subsection{Linear 3-SAT}
\label{subsection:linear}
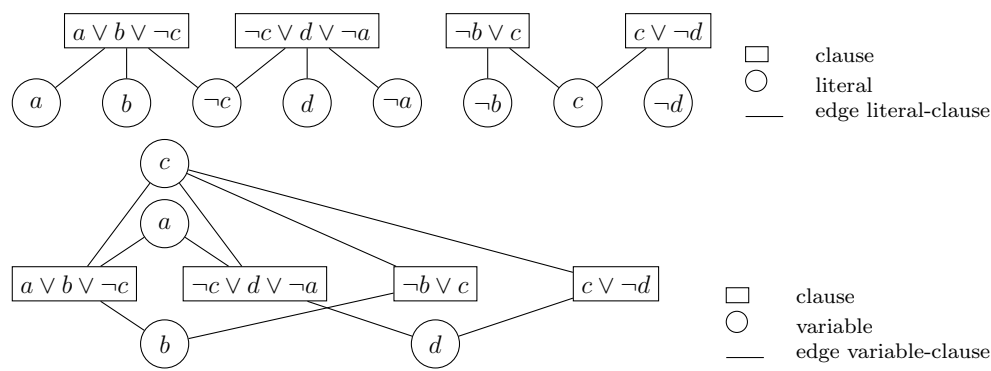
\begin{figure}[h]
	\centering
	
	\begin{tabular}{l}
	\begin{tikzpicture}[yscale=0.8,scale=0.8, every node/.style={scale=0.8}]

		\node (n1) at (0, 0) [litteral] {$a$};
		\node (n2) at (1.5, 0) [litteral] {$b$};
		\node (n3) at (3, 0) [litteral] {$\neg c$};
		\node (n4) at (4.5, 0) [litteral] {$d$};
		\node (n5) at (6, 0) [litteral] {$\neg a$};
		\node (n6) at (7.5, 0) [litteral] {$\neg b$};
		\node (n7) at (9, 0) [litteral] {$c$};
		\node (n8) at (10.5, 0) [litteral] {$\neg d$};

		\node (c1) at (1.5, 1.5) [draw, rectangle] {$a \lor b \lor \neg c$};
		\draw[-] (n1) -- (c1);
		\draw[-] (n2) -- (c1);
		\draw[-] (n3) -- (c1);
		
		\node (c2) at (4.5, 1.5) [draw, rectangle] {$\neg c \lor d \lor \neg a$};
		\draw[-] (n3) -- (c2);
		\draw[-] (n4) -- (c2);
		\draw[-] (n5) -- (c2);
		
		\node (c3) at (7.5, 1.5) [draw, rectangle] {$\neg b \lor c$};
		\draw[-] (n6) -- (c3);
		\draw[-] (n7) -- (c3);
		
		\node (c4) at (10.5, 1.5) [draw, rectangle] {$c \lor \neg d$};
		\draw[-] (n7) -- (c4);
		\draw[-] (n8) -- (c4);
		
	\end{tikzpicture} 	\hfill
\scriptsize
\raisebox{5mm}{
\begin{tabular}{ll}
\tikz{\node[draw]{~~};}&	clause \\
\tikz{\node[circle, draw]{};}&	literal \\
\tikz{\draw (0, 0) -- (0.5, 0);} & edge literal-clause \\
\end{tabular}} \\
	\begin{tikzpicture}[scale=0.8, every node/.style={scale=0.8}]

	\node (n1) at (3, 2.5) [litteral] {$a$};
	\node (n2) at (3, 0.5) [litteral] {$b$};
	\node (n3) at (3, 3.5) [litteral] {$c$};
	\node (n4) at (7.5, 0.5) [litteral] {$d$};

	\node (c1) at (1.5, 1.5) [draw, rectangle] {$a \lor b \lor \neg c$};
	\draw[-] (n1) -- (c1);
	\draw[-] (n2) -- (c1);
	\draw[-] (n3) -- (c1);
	
	\node (c2) at (4.5, 1.5) [draw, rectangle] {$\neg c \lor d \lor \neg a$};
	\draw[-] (n3) -- (c2);
	\draw[-] (n4) -- (c2);
	\draw[-] (n1) -- (c2);
	
	\node (c3) at (7.5, 1.5) [draw, rectangle] {$\neg b \lor c$};
	\draw[-] (n2) -- (c3);
	\draw[-] (n3) -- (c3);
	
	\node (c4) at (10.5, 1.5) [draw, rectangle] {$c \lor \neg d$};
	\draw[-] (n3) -- (c4);
	\draw[-] (n4) -- (c4);
	
\end{tikzpicture}
	\hfill
\scriptsize
\raisebox{5mm}{
\begin{tabular}{ll}
	\tikz{\node[draw]{~~};}&	clause \\
	\tikz{\node[circle, draw]{};}&	variable \\
	\tikz{\draw (0, 0) -- (0.5, 0);} & edge variable-clause \\
\end{tabular}}
\end{tabular}
	\caption{\LitClausGraph (up) and \inciGraph (down) of the CNF formula $(a \lor b \lor \lnot c) \land (\lnot c \lor d \lor \lnot a) \land (\lnot b \lor c) \land (c \lor \lnot d)$.}
	
	\label{fig:literalclausegraph}
\end{figure}

A CNF formula $\phi$ is \emph{linear} if 
1) each clause intersects with at most one other clause, and 
2) each such intersection contains at most one literal.

\begin{example}
	The formula $(a \lor b \lor \lnot c) \land (\lnot c \lor d \lor \lnot a) \land (\lnot b \lor c) \land (c \lor \lnot d)$ is linear. %
	 For instance $(a \lor b \lor \lnot c)$ intersects only with $(\lnot c \lor d \lor \lnot a)$ and the intersection is~$\{\lnot c\}$.
\end{example}

Linear \tsat is the restriction of 3-SAT to inputs that are linear 3-CNF. It is \npc  \cite{arkin2018selecting}.

\subsection{Literal-Clause Graph and Variable-Clause Graph}

\begin{definition}[Literal-Clause Graph] %
    Let $\phi=\wedge_{i=1}^m c_i$ be a  CNF formula. The \emph{\litClausGraph of $\phi$}, denoted $\litClauGraphSign{\phi}$ is the bipartite graph where:
    \begin{itemize}
    	\item the vertices are the literals occurring in $\phi$ and the clauses in $\phi$;
    	\item there is an edge between a literal $\ell$ and a clause $c_i$ iff $\ell$ appears in $c_i$.
    \end{itemize}
    \label{def:lit_claus_graph}
\end{definition}

\begin{definition}[Variable-Clause Graph]
    Let $\phi=\wedge_{i=1}^m c_i$ a  CNF formula. The \emph{variable-clause graph of $\phi$}, denoted $\inciGraphSign{\phi}$, is the graph obtained from the \litClausGraph of $\phi$ by merging each literal vertex $\ell$ with $\neg \ell$.
    \label{def:inci_graph}
\end{definition}

\begin{example}
	Figure~\ref{fig:literalclausegraph} shows both the literal-clause and variable-clause graphs of formula $(a \lor b \lor \lnot c) \land (\lnot c \lor d \lor \lnot a) \land (\lnot b \lor c) \land (c \lor \lnot d)$. A merged vertex is simply represented by the corresponding variable.
\end{example}

While variable-clause graphs appear in the literature (e.g \cite{lichtenstein1982planar})
as \emph{incidence graphs}, we introduce literal-clause graphs here as a more refined version of variable-clause graphs.
We will use the term variable-clause graphs instead of incidence graphs to emphasize the relation to literal-clause graphs.

\subsection{Planar 3-SAT}
\label{subsection:planar}

\begin{figure}[h]
    \centering
    
    \begin{tikzpicture}[
        scale=0.6, 
        var/.style={draw, circle, minimum size=1cm, scale=0.6},
        edge/.style={-, scale=0.8}
    ]

        \node[var] (x1) at (0, 0) {$a$};
        \node[var] (x2) at (2, 0) {$b$};
        \node[var] (x3) at (4, 0) {$c$};
        \node[var] (x4) at (6, 0) {$d$};

        \node[clause, minimum width=2.5cm] (C1) at (2, 1.25) {$\neg a \vee b \vee \neg c$};
        \node[clause, minimum width=6cm] (C2) at (3, 2.75) {$a \vee c \vee \neg d$};

        \node[clause, minimum width=6cm] (C3) at (3, -2.75) {$a \vee \neg b \vee \neg d$};
        \node[clause, minimum width=2.5cm] (C4) at (4, -1.25) {$b \vee c \vee \neg d$};

        \draw[edge] (x1) -- (C1);
        \draw[edge] (x2) -- (x2 |- C1.south);
        \draw[edge] (x3) -- (C1);

        \draw[edge] (x1) -- (x1 |- C2.south);
        \draw[edge] (x3) -- (x3 |- C2.south);
        \draw[edge] (x4) -- (x4 |- C2.south);

        \draw[edge] (x1) -- (x1 |- C3.north);
        \draw[edge] (x2) -- (x2 |- C3.north);
        \draw[edge] (x4) -- (x4 |- C3.north);

        \draw[edge] (x2) -- (C4);
        \draw[edge] (x3) -- (x3 |- C4.north);
        \draw[edge] (x4) -- (C4);

        \draw[variablecycle] (x1) -- (x2);
        \draw[variablecycle] (x2) -- (x3);
        \draw[variablecycle] (x3) -- (x4);
        \draw[variablecycle] (x4) -- (7, 0) -- (7, 3.75) -- (-1, 3.75) -- (-1,0) -- (x1);
        
    \end{tikzpicture}
    \hspace{1cm}
	\scriptsize
    \raisebox{2cm}{
    \begin{tabular}{ll}
        \tikz{\node[clause, minimum width=0.5cm, minimum height=0.5cm]{};}&	clause \\
        \tikz{\node[circle, draw]{};}&	variable \\
        \tikz{\draw (0, 0) -- (0.5, 0);} & edge variable-clause \\
        \tikz{\draw (0, 0) edge[variablecycle] (0.5, 0);} &  variable cycle 
    \end{tabular}
    }
    
    \caption{\InciGraph $\inciGraphSign{\phi, \pi}$ augmented with variable cycle $(a, b, c, d)$ for the planar 3-CNF formula $(\neg a \vee b \vee \neg c) \land (a \vee c \vee \neg d) \land (a \vee \neg b \vee \neg d) \land (b \vee c \vee \neg d) $}
    
    \label{fig:planar_sat_example}
\end{figure}
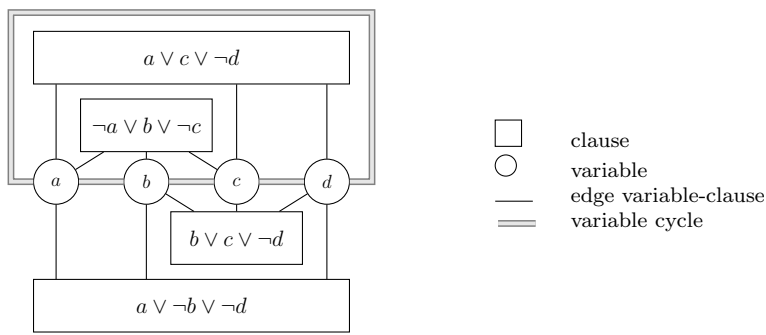

\begin{definition}[Variable Cycle]
    A \emph{variable cycle} $\pi$ is a permutation $x_{\pi_1}, \dots, x_{\pi_n}$ of a given sequence of variables $x_1, \dots, x_n$.
\end{definition}

Given $\phi$ a 3-CNF formula with variables $x_1, \dots, x_n$, the \emph{variable-clause graph $\inciGraphSign{\phi}$ augmented with $\pi$}, denoted $\inciGraphSign{\phi, \pi}$, is the graph obtained from $\inciGraphSign{\phi}$ by adding the edges $(x_{\pi_1}, x_{\pi_2}), (x_{\pi_2}, x_{\pi_3}), \dots, (x_{\pi_n}, x_{\pi_1})$. The cycle of edges $$(x_{\pi_1}, x_{\pi_2}), (x_{\pi_2}, x_{\pi_3}), \dots, (x_{\pi_n}, x_{\pi_1})$$ is also called the \emph{variable-cycle}. 

A CNF $\phi$ is \emph{planar} if there exists some variable cycle $\pi$ such that $\inciGraphSign{\phi, \pi}$ is planar.
\emph{Planar \tsat} \cite{lichtenstein1982planar} is the restriction of 3-SAT to inputs that are planar 3-CNF. More precisely, Planar \tsat takes a planar embedding of $\inciGraphSign{\phi, \pi}$ for some $\pi$ as input. That is, the input contains not only the formula $\phi$ but also a given variable cycle $\pi$, and the integer coordinates of the vertices of $\inciGraphSign{\phi, \pi}$ on the plane. Planar \tsat is NP-complete.

\begin{example}
	The formula $\phi = (\neg a \vee b \vee \neg c) \land (a \vee c \vee \neg d) \land (a \vee \neg b \vee \neg d) \land (b \vee c \vee \neg d)$ is a planar 3-CNF. Figure~\ref{fig:planar_sat_example} shows the \inciGraph of formula $\phi$ augmented with the variable cycle $\pi = (a,b,c,d)$. The augmented graph $\inciGraphSign{\phi, \pi}$ is planar.
\end{example}

\subsection{Monotone Planar 3-SAT}
\label{subsection:mplanar}
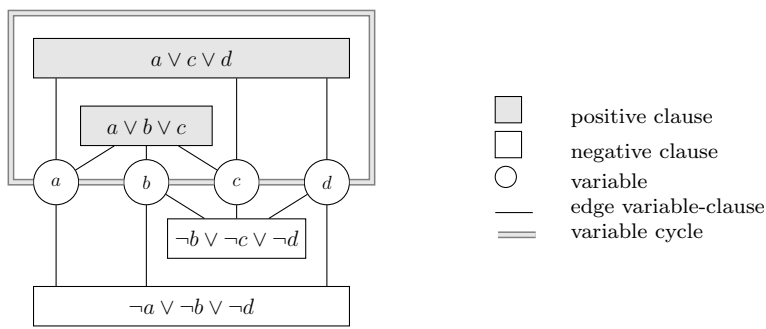
\begin{figure}[h]
    \centering
    
    \begin{tikzpicture}[
        scale=0.6, 
        var/.style={draw, circle, minimum size=1cm, scale=0.6},
        clause/.style={draw, rectangle, minimum size=1cm, scale=0.7},
        edge/.style={-, scale=0.8}
    ]

        \node[var] (x1) at (0, 0) {$a$};
        \node[var] (x2) at (2, 0) {$b$};
        \node[var] (x3) at (4, 0) {$c$};
        \node[var] (x4) at (6, 0) {$d$};

        \node[positiveclause, minimum width=2.5cm] (C1) at (2, 1.25) {$a \vee b \vee c$};
        \node[positiveclause, minimum width=6cm] (C2) at (3, 2.75) {$a \vee c \vee d$};

        \node[negativeclause, minimum width=6cm] (C3) at (3, -2.75) {$\neg a \vee \neg b \vee \neg d$};
        \node[negativeclause, minimum width=2.5cm] (C4) at (4, -1.25) {$\neg b \vee \neg c \vee \neg d$};

        \draw[edge] (x1) -- (C1);
        \draw[edge] (x2) -- (x2 |- C1.south);
        \draw[edge] (x3) -- (C1);

        \draw[edge] (x1) -- (x1 |- C2.south);
        \draw[edge] (x3) -- (x3 |- C2.south);
        \draw[edge] (x4) -- (x4 |- C2.south);

        \draw[edge] (x1) -- (x1 |- C3.north);
        \draw[edge] (x2) -- (x2 |- C3.north);
        \draw[edge] (x4) -- (x4 |- C3.north);

        \draw[edge] (x2) -- (C4);
        \draw[edge] (x3) -- (x3 |- C4.north);
        \draw[edge] (x4) -- (C4);

        \draw[variablecycle] (x1) -- (x2);
        \draw[variablecycle] (x2) -- (x3);
        \draw[variablecycle] (x3) -- (x4);
        \draw[variablecycle] (x4) -- (7, 0) -- (7, 3.75) -- (-1, 3.75) -- (-1,0) -- (x1);
        
    \end{tikzpicture}
    \hspace{1cm}
	\scriptsize
    \raisebox{2cm}{
    \begin{tabular}{ll}
        \tikz{\node[positiveclause, minimum width=0.5cm, minimum height=0.5cm]{};}& positive	clause \\
        \tikz{\node[negativeclause, minimum width=0.5cm, minimum height=0.5cm]{};}& negative	clause \\
        \tikz{\node[circle, draw]{};}&	variable \\
        \tikz{\draw (0, 0) -- (0.5, 0);} & edge variable-clause \\
        \tikz{\draw (0, 0) edge[variablecycle]  (0.5, 0);} & variable cycle 
    \end{tabular}
    }
    
    \caption{\InciGraph of the monotone planar 3-CNF formula $(a \vee b \vee c) \land (a \vee c \vee d) \land (\neg a \vee \neg b \vee \neg d) \land (\neg b \vee \neg c \vee \neg d) $ and variable cycle $(1, 2, 3, 4)$.}
    
    \label{fig:monotone_planar_sat_example}
\end{figure}

A CNF formula $\phi$ is \emph{monotone} if the literals of each clause $c_i$  are either all positive, or all negative.
The so-called \emph{negative clauses} contain only negative literals while \emph{positive clauses} contain only positive literals. 

A CNF formula $\phi$ is \emph{monotone planar} if there exists a variable-cycle $\pi$ such that the variable-clause graph $\inciGraphSign{\phi, \pi}$ 
has a planar embedding in which %
positive clause nodes are inside (\resp outside) the curve formed by the variable cycle, while negative clauses nodes are outside (\resp inside).

\MPTSAT asks: given a monotone planar 3-CNF formula $\phi$, a cycle $\pi$, a planar embedding of $\inciGraphSign{\phi, \pi}$, is $\phi$ satisfiable?
This problem is \npc~\cite{de2010optimal}.

\begin{example}
    The formula $\phi = (a \vee b \vee c) \land (a \vee c \vee d) \land (\neg a \vee \neg b \vee \neg d) \land (\neg b \vee \neg c \vee \neg d)$ is a monotone planar 3-CNF. Indeed, for the variable-cycle $(a,b,c,d)$, 
    Figure~\ref{fig:monotone_planar_sat_example} shows a planar embedding of \inciGraph of $\phi$ and $\pi$ where all the positive clauses are inside the curve formed by the variable cycle while all the negative ones are outside.
\end{example}

\section{New Variants of \tsat}
\label{sec:new-variants}
In this section, we introduce two new variants of 3-SAT combining some of the features of previously seen variants.
We first define \emph{linear planar 3-SAT}, and then \emph{linear literal-planar 3-SAT}.

These new variants correspond to types of formulas that appear naturally when encoding multi-agent planning problems to SAT.
In Section~\ref{section:npc}, we prove both problems to be NP-complete (Theorems~\ref{th:paired_planar_linear_tsat_npc} and~\ref{th:planar_linear_tsat_npc}).
In Section~\ref{subsection:reconfiguration}, we will study reconfiguration problems and prove that these are PSPACE-complete.

\subsection{Linear Planar \tsat}
\label{subsection:lplanar}
A 3-CNF $\phi$ is \emph{linear planar} if $\phi$ is linear and planar.
Let \emph{Linear Planar \tsat} denote the restriction of SAT to linear planar 3-CNF formulas.

\begin{figure}[h]
    \centering
    
    \begin{tikzpicture}[
        scale=0.6, 
        var/.style={draw, circle, minimum size=1cm, scale=0.6},
        edge/.style={-, scale=0.8}
    ]

        \node[var] (x1) at (0, 0) {$a$};
        \node[var] (x2) at (2, 0) {$b$};
        \node[var] (x3) at (4, 0) {$c$};
        \node[var] (x4) at (6, 0) {$d$};

        \node[clause, minimum width=1.5cm] (C1) at (2, 1.25) {$b \vee c$};
        \node[clause, minimum width=6cm] (C2) at (3, 2.75) {$a \vee c \vee \neg d$};

        \node[clause, minimum width=6cm] (C3) at (3, -2.75) {$\neg a \vee \neg b \vee d$};
        \node[clause, minimum width=2cm] (C4) at (4, -1.25) {$\neg c \vee d$};

        \draw[edge] (x2) -- (x2 |- C1.south);
        \draw[edge] (x3) -- (C1);

        \draw[edge] (x1) -- (x1 |- C2.south);
        \draw[edge] (x3) -- (x3 |- C2.south);
        \draw[edge] (x4) -- (x4 |- C2.south);

        \draw[edge] (x1) -- (x1 |- C3.north);
        \draw[edge] (x2) -- (x2 |- C3.north);
        \draw[edge] (x4) -- (x4 |- C3.north);

        \draw[edge] (x3) -- (x3 |- C4.north);
        \draw[edge] (x4) -- (C4);

        \draw[variablecycle] (x1) -- (x2);
        \draw[variablecycle] (x2) -- (x3);
        \draw[variablecycle] (x3) -- (x4);
        \draw[variablecycle] (x4) -- (7, 0) -- (7, 3.75) -- (-1, 3.75) -- (-1,0) -- (x1);
        
    \end{tikzpicture}
    \hspace{1cm}
	\scriptsize
    \raisebox{2cm}{
    \begin{tabular}{ll}
        \tikz{\node[clause, minimum width=0.5cm, minimum height=0.5cm]{};}&	clause \\
        \tikz{\node[circle, draw]{};}&	variable \\
        \tikz{\draw (0, 0) -- (0.5, 0);} & edge variable-clause \\
        \tikz{\draw (0, 0) edge[variablecycle] (0.5, 0);} & variable cycle 
    \end{tabular}
    }
    
    \caption{\InciGraph of the planar linear 3-CNF formula $(a \vee c \vee \neg d) \land (b \vee c) \land (\neg a \vee \neg b \vee d) \land (\neg c \vee d) $ and variable-cycle $(1, 2, 3, 4)$.}
    
    \label{fig:planar_linear_tsat_exmp}
\end{figure}
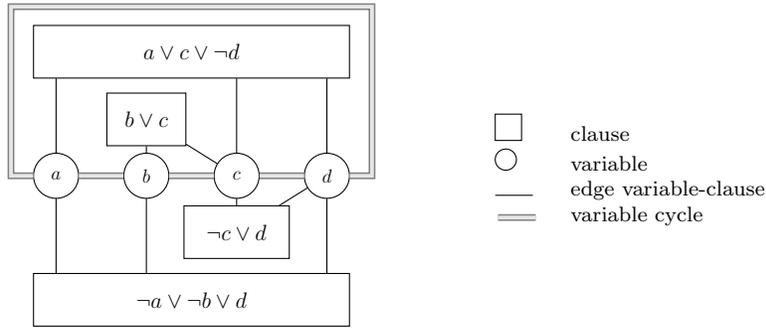 

\begin{example}
    The formula  $\phi = (a \vee c \vee \neg d) \land (b \vee c) \land (\neg a \vee \neg b \vee d) \land (\neg c \vee d)$ is a planar linear 3-CNF. Indeed $\phi$ is linear, and 
    Figure~\ref{fig:planar_linear_tsat_exmp} shows a planar embedding of the \inciGraph $\inciGraphSign{\phi, \pi}$ for $\pi =  (a,b,c,d)$.
\end{example}

\begin{figure}[h]
    \centering
    
    \begin{tikzpicture}[scale=0.8, every node/.style={scale=0.8}]
        
        \node (a) at (0, 0) [draw, circle] {$a$};
        \node (b) at (1.5, 0) [draw, circle] {$b$};
        \node (c) at (3, 0) [draw, circle]{$c$};
        \node (d) at (4.5, 0) [draw, circle] {$d$};
        \node (e) at (6, 0) [draw, circle] {$e$};
        \node (f) at (7.5, 0) [draw, circle] {$f$};
        \node (g) at (9, 0) [draw, circle] {$g$};
        \node (h) at (10.5, 0) [draw, circle] {$h$};
        \node (i) at (12, 0) [draw, circle] {$i$};
        
        \node (c1) at (0, 1) [draw, rectangle] {$a \lor b \lor c$};
        \node (c2) at (6, 1) [draw, rectangle] {$d \lor e \lor f$};
        \node (c3) at (12, 1) [draw, rectangle] {$g \lor h \lor j$};

        \node (c4) at (0, -3) [draw, rectangle] {$\neg a \lor \neg d \lor \neg g$};
        \node (c5) at (6, -3) [draw, rectangle] {$\neg b \lor \neg e \lor \neg h$};
        \node (c6) at (12, -3) [draw, rectangle] {$\neg c \lor \neg f \lor \neg i$};

        \draw[-] (c1) -- (a);
        \draw[-] (c1) -- (b);
        \draw[-] (c1) -- (c);
        
        \draw[-] (c2) -- (d);
        \draw[-] (c2) -- (e);
        \draw[-] (c2) -- (f);

        \draw[-] (c3) -- (g);
        \draw[-] (c3) -- (h);
        \draw[-] (c3) -- (i);
        
        \draw[-] (a) -- (c4);
        \draw[-] (b) -- (c5);
        \draw[-] (c) -- (c6);

        \draw[-] (d) -- (c4);
        \draw[-] (e) -- (c5);
        \draw[-] (f) -- (c6);

        \draw[-] (g) -- (c4);
        \draw[-] (h) -- (c5);
        \draw[-] (i) -- (c6);
    \end{tikzpicture}

    \caption{\InciGraph $\inciGraphSign{\phi}$ of the linear formula $\phi$, which is a subdivision of $K_{3,3}$.
    }
    
    \label{fig:lsat_nonplanar}
\end{figure}
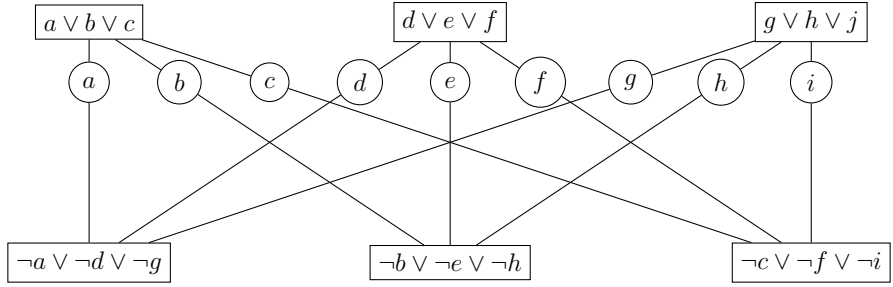

Let us first argue that this class is nontrivial, that is, not all linear 3-CNF formulas are planar.
Consider the following formula.

\begin{equation}
    \label{eqn:phi-nonplanar}
	\begin{split}
        \phi := & (a \lor b \lor c) \lor (d \lor e \lor f) \lor (g \lor h \lor j) \\
        & \lor (\neg a \lor \neg d \lor \neg g) \lor (\neg b \lor \neg e \lor \neg h) \lor (\neg c \lor \neg f \lor \neg i) 
	\end{split}
\end{equation}

\begin{proposition}
	Formula $\phi$ is linear and $\inciGraphSign{\phi}$ is not planar.
	\label{th:lsat_nonplanar}
\end{proposition}

\begin{proof}
Formula	$\phi$ is a linear since there are three unique literals per clause so there is no intersection between any pair clauses. 

	The graph $\inciGraphSign{\phi}$ is given in Figure~\ref{fig:lsat_nonplanar}. 
    This can be seen to be a \emph{subdivision} of the complete bipartite graph $K_{3,3}$. In other terms, $\inciGraphSign{\phi}$
     can be obtained from $K_{3,3}$ by adding `intermediate' vertices $a$, $b$, etc.%
    By Kuratowski's Theorem \cite{kuratowski1930probleme}, $\inciGraphSign{\phi}$ is nonplanar. 
\end{proof}

\Cref{th:lsat_nonplanar} shows that Planar Linear \tsat has strictly less possible inputs than Linear \tsat. 	   

\subsection{\PLLPTSAT}
\label{subsection:llplanar}
\begin{figure}[h]
    \centering
    
    \begin{tikzpicture}[scale=0.7, every node/.style={fill=white, scale=0.8}]
        \pgfdeclarelayer{background}
        \pgfdeclarelayer{foreground}
        \pgfsetlayers{background,main,foreground}

        \node (n1) at (0, 0) [litteral] {$a$};
        \node (n2) at (1.5, 0) [litteral] {$\neg b$};
        \node (n3) at (3, 0) [litteral] {$b$};
        \node (n4) at (4.5, 0) [litteral] {$\neg c$};
        \node (n5) at (6, 0) [litteral] {$d$};
        \node (n6) at (7.5, 0) [litteral] {$\neg d$};
        \node (n7) at (9, 0) [litteral] {$c$};
        \node (n8) at (10.5, 0) [litteral] {$\neg a$};

        \node (c1) at (3, 2.25) [draw, rectangle, minimum width=4cm] {$a \lor \neg b \lor d$};
        \node (c2) at (4.5, 1.5) [draw, rectangle] {$b \lor \neg c \lor d$};
        \node (c3) at (7.5, 1.5) [draw, rectangle] {$\neg d \lor c$};
        \node (c4) at (10.5, 1.5) [draw, rectangle] {$c \lor \neg a$};

        \begin{pgfonlayer}{background}
            \draw[variablecycle] (n1) -- (n2);
            \draw[variablecycle] (n2) -- (n3);
            \draw[variablecycle] (n3) -- (n4);
            \draw[variablecycle] (n4) -- (n5);
            \draw[variablecycle] (n5) -- (n6);
            \draw[variablecycle] (n6) -- (n7);
            \draw[variablecycle] (n7) -- (n8);
            \draw[variablecycle] (n1) -- (-1,0) -- (-1,-2) -- (11.5,-2) -- (11.5,0) -- (n8);

            \draw[] (n1) edge[paired, bend right=30] (n8);
            \draw[] (n2) edge[paired, bend right=20] (n3);
            \draw[] (n4) edge[paired, bend right=30] (n7);
            \draw[] (n5) edge[paired, bend right=20] (n6);
            
            \draw[-] (n1) -- (c1);
            \draw[-] (n2) -- (c1);
            \draw[-] (n5) |- (c1);
    
            \draw[-] (n3) -- (c2);
            \draw[-] (n4) -- (c2);
            \draw[-] (n5) -- (c2);
    
            \draw[-] (n6) -- (c3);
            \draw[-] (n7) -- (c3);
    
            \draw[-] (n7) -- (c4);
            \draw[-] (n8) -- (c4);

        \end{pgfonlayer}
    \end{tikzpicture}
    \scriptsize
    \raisebox{1cm}{
    \begin{tabular}{ll}
        \tikz{\node[draw]{~~};}&	clause \\
        \tikz{\node[circle, draw]{};}&	litteral \\
        \tikz{\draw (0, 0) -- (0.5, 0);} & edge litteral-clause \\
            \tikz{\draw (0, 0) edge[variablecycle]  (0.5, 0);} & variable cycle \\
            \tikz{\draw (0, 0) edge[paired] (0.5, 0);} & pair edge \\
    \end{tabular}}
    \caption{\LitClausGraph $\litClauGraphSign{\phi, \pi}$ augmented with the literal cycle $(a, \lnot b, b, \lnot c, d, \lnot d, c, \lnot a)$ and paired edges. As this graph is planar, and that paired edges are inside $\pi$ and the clauses are outside $\pi$ in the shown embedding, the corresponding formula $\phi$ is paired-literal planar linear.}
    
    \label{fig:planar_paired_lit_lsat_example}
\end{figure}
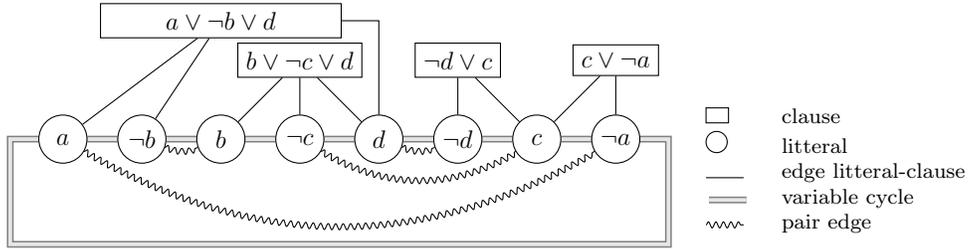 

Let us now introduce a variant of linear planar \tsat formulas.
Given a formula $\phi$ with variables $x_1, \dots, x_n$, a \emph{literal cycle} $\pi$ is a permutation of the literals $x_1, \lnot x_1, \dots, x_n, \lnot x_n$.
Given a formula $\phi$ and a literal cycle~$\pi$, 
the \emph{\litClausGraph of $\litClauGraphSign{\phi}$ augmented with $\pi$},
denoted by $\litClauGraphSign{\phi, \pi}$, is obtained from 
$\litClauGraphSign{\phi}$
by adding the literal cycle~$\pi$, edges between opposite literals $\ell$ and $\neg \ell$ called \emph{paired edges}.

The graph $\litClauGraphSign{\phi, \pi}$ is \emph{valid} if 
there exists a planar embedding of $\litClauGraphSign{\phi, \pi}$,
all paired edges are inside the cycle $\pi$, and all clauses are outside $\pi$.
A 3-CNF formula is said to be \emph{literal-planar} if there exists a literal cycle $\pi$ 
such that $\litClauGraphSign{\phi, \pi}$ is valid.

\emph{\PLLPTSAT} is the restriction of 3-SAT to \PLLPTformula. More precisely, the input contains a formula $\phi$, a cycle $\pi$, and a planar embedding of $\litClauGraphSign{\phi, \pi}$. W.l.o.g. the input could also be $\phi$ and a literal cycle $\pi$, as a planar embedding of $\litClauGraphSign{\phi, \pi}$ is computable from $\phi$ and $\pi$ in polynomial time in the size of $\phi$ and $\pi$ (see \cite{hopcroft1974efficient} for a poly-time algorithm to compute a planar embedding from an abstract graph).

\begin{example}
    The formula $\phi = (a \vee \neg b \vee d) \land (b \vee \neg c \vee d) \land (\neg d \vee c) \land (c \vee \neg a)$ is a \PLLPTformula. 
    Figure~\ref{fig:planar_paired_lit_lsat_example} shows the \litClausGraph $\litClauGraphSign{\phi, \pi}$ with $\pi = (a, \lnot b, b, \lnot c, d, \lnot d, c, \lnot a)$.
    In this embedding of $\litClauGraphSign{\phi, \pi}$ the paired edges are inside the curve formed by the literal cycle while all the clauses are outside.
\end{example}

\subsection{Summary of Variants}
Main features of all presented \tsat variants are summarized in \Cref{table:tsat-summary}.
\begin{table}[h]
    \caption{Summary of 3-SAT variants considered in this paper.}
    \begin{center}
    \begin{tabular}{|p{3.4cm}|p{9.6cm}|}
        \hline 3-CNF Fragments & Features \\
        \hline\hline \small Linear (Sect.~\ref{subsection:linear}) & Each clause $c$ intersects at most one other clause $c'$, such that $|c\cap c'| \leq 1$.\\
        \hline \small Planar (Sect.~\ref{subsection:planar}) & $\exists$ variable cycle $\pi$ such that $\inciGraphSign{\phi, \pi}$ is planar.\\
        \hline \small Monotone Planar (Sect.~\ref{subsection:mplanar}) & Planar, and each clause has only positive literals or only negative literals. The clauses that contains positive literals are separated from the clauses that contains negative literals by $\pi$.\\
        \hline \bf \small Linear Planar (Sect.~\ref{subsection:lplanar}) & Linear and planar.\\
        \hline \bf \small Linear Literal-Planar (Sect.~\ref{subsection:llplanar}) &
        $\exists$ literal cycle $\pi$ such that $\litClauGraphSign{\phi, \pi}$ is planar and valid.
        \\
        \hline \bf \small Monotone Linear Planar (Def.~\ref{def:cmpl})& Linear, and monotone planar. \\
        \hline
    \end{tabular}
    \end{center}
    \label{table:tsat-summary}
\end{table} 

\section{Complexity Results for Our \tsat Variants}
\label{sec:complexity-tsat}
\subsection{NP-completeness of \PLLPTSAT}
\label{section:npc}

In this section, we prove Theorem.~\ref{th:paired_planar_linear_tsat_npc}.
To prove the NP-hardness of \PLLPTSAT, we construct a polynomial-time reduction from the \MPTSAT proved to be NP-complete \cite{de2010optimal}.

\begin{theorem}
	\PLLPTSAT is \npc.
	\label{th:paired_planar_linear_tsat_npc}
\end{theorem}

    Let $\phi$ be a monotone planar 3-CNF with $n$ positive clauses and $m$ negative clauses, and a corresponding variable cycle $\pi$. 
    We consider an embedding of $\inciGraphSign{\phi,\pi}$ where positive clauses are inside and negative clauses are outside the variable cycle.
    
    We denote by $V$ the set of variables and by $C$ the set of clauses occurring in $\phi$. 
    For each variable $v$ in $V$,
    we introduce fresh variables $p^v_i$ and $a^v_i$ for $1 \leq i \leq n$, and variables $n^v_j$ and $b^v_j$ for $1 \leq j \leq m$.

    Intuitively, clauses of $\phi$ can have arbitrary intersections since $\phi$ is not assumed to be linear. Because we want to reduce the problem to 
    the satisfaction of a planar literal-linear formula, we will replace all occurrences of the variables by fresh variables we just introduced,
    in such a way that the resulting formula is literal-linear.

    \newcommand{\phiprimer}{\psi'}
    Formally, we define the 3-CNF formula  $\phiprimer$, of the same size as $\phi$, obtained as follows from $\phi$.
    For all variables $v \in V$, 
    we replace the $i$-th positive occurrence of $v$ by $\lnot p^{v}_i$ and the $j$-th negative occurrence of $v$ by $n^v_j$.
    Note that the obtained formula~$\phiprimer$ contains exclusively negative literals.

    For instance, consider the following monotone formula:
    \begin{equation}\label{eq:5}
        \phi = (x \lor y \lor z) \land (\neg x \lor \neg y \lor \neg w) \land (x \lor y \lor w)
    \end{equation}
    We get:
    \begin{equation}\label{eq:6}
        \phiprimer = (\neg p^{x}_1 \lor \neg p^{y}_1 \lor \neg p^{z}_1) \land (\neg n^{x}_1 \lor \neg n^{y}_1 \lor \neg n^{w}_1) \land (\neg p^{x}_2 \lor \neg p^{y}_2 \lor \neg p^{w}_1)
    \end{equation}  
   We further add clauses to $\phiprimer$ to link the fresh variables to $V$.
    For each variable $v \in V$, let us define a 3-CNF linear formula $\phi'_v$ given below.
    Here, all clauses have two literals and are shown in the form of an implication
    (recall that $x \rightarrow y$ means $\lnot x \lor y$).
    \begin{equation}\label{eq:phi_v}
        \begin{split}
            \phi'_v :=  & (\neg a_1^v \rightarrow v) \land (\neg b_1^v \rightarrow \neg v )\\ 
                    & \land~(\bigwedge^{n}_{i=1} (\lnot p_i^v \rightarrow \neg a_i^v) \land \bigwedge^{n-1}_{i=1} (\neg a_{i+1}^v \rightarrow \neg a_i^v ))\\
                    & \land~(\bigwedge^{m}_{j=1} (\lnot n_j^v \rightarrow \neg b_j^v) \land \bigwedge^{m-1}_{j=1} (\neg b_{j+1}^v \rightarrow \neg b_j^v ))
        \end{split}
    \end{equation}
\newcommand{\limply}{\rightarrow}
We have, for all $1\leq i \leq n$,
\begin{equation}
    \phi'_v \models (\lnot p_i^v \limply v)	\label{eq:implies_pos}
\end{equation}
In fact,  $\phi'_v \models (\lnot p_i^v \rightarrow \lnot a_i^v) \land (\lnot a_i^v \limply \lnot a_{i-1}^v) \land \dots (\lnot a_2^v \limply \lnot a_{1}^v) \land (\lnot a_{1}^v \limply v)$. 
Similarly, 
\begin{equation}
    \phi'_v \models (\lnot n_i^v \limply \lnot v) \label{eq:implies_negat}
\end{equation}
We now define $\phi'$ as follows.
\begin{equation}\label{eq:7}
    \phi' := \phiprimer \land \bigwedge_{v \in V} \phi'_v
\end{equation}

We first show that $\phi'$ is a linear literal-planar formula,
and that it is polynomial-time constructible,
and then establish the equisatisfiability of $\phi$ and $\phi'$,
which concludes the proof of the reduction.

\begin{claim}
    The formula $\phi'$ is linear.
\end{claim}

\begin{proof}
    Consider the clauses of Equation~\ref{eq:phi_v}. 
    There is an intersection between the clauses $(p_i^v \lor \neg a_i^v)$ and $(\neg a_i^v \lor a_{i+1}^v)$ with a unique literal $\neg a_i^v$.
    There is also an intersection between the clauses $(n_j^v \lor \neg b_j^v)$ and $(\neg b_j^v \lor b_{j+1}^v)$ that contains one literal~$\neg b_j^v$.
    There are no other intersecting clauses; so all intersections do contain at most one literal,
    and each $\phi'_v$ is linear.
    
    In $\phiprimer$, all the occurrences of $v$ and $\neg v$ in $\phi$ are replaced with a unique literal, respectively $\neg p^v_i$ and $\neg n^v_j$, thus there is no intersection between clauses.
    $\phi'_v$ contains only non-negated literals of variables $p_{i}^v$ and $n_{j}^v$ while $\phiprimer$ contains only negated literals of variables $p_{i}^v$ and $n_{j}^v$, thus, there is no intersection between $\bigwedge_{v \in V} \phi'_v$ and $\phiprimer$.
    We conclude that $\phi'$ is a linear 3-CNF formula.
\end{proof}

\begin{figure}[t]
    \centering

		\begin{subfigure}[t]{0.45\linewidth}
			\centering
			
			\begin{tikzpicture}[scale=0.8, every node/.style={fill=white, scale=0.8}]
				\pgfdeclarelayer{background}
				\pgfdeclarelayer{foreground}
				\pgfsetlayers{background,main,foreground}

				\node (v) at (0, 0) [litteral] {$v$};

				\node (c1) at (-1.5, 1.25) [positiveclause] {$c_1$};
				\node (c2) at (-0.5, 1.25) [positiveclause] {$c_2$};
				\node (cn) at (1.5, 1.25) [positiveclause] {$c_n$};
				\node (bc1) at (-1.5, -1.5) [clause] {$\neg c_1$};
				\node (bc2) at (-0.5, -1.5) [clause] {$\neg c_2$};
				\node (bcm) at (1, -1.5) [clause] {$\neg c_m$};
			
				\begin{pgfonlayer}{background}
					\draw[-] (v) -- (c1);
					\draw[-] (v) -- (c2);
					\draw[-] (v) -- (cn);
				
					\draw[-] (v) -- (bc1);
					\draw[-] (v) -- (bc2);
					\draw[-] (v) -- (bcm);
				
					\draw[variablecycle] (3, 0) -- (-3, 0);
					\draw[variablecycle] (3, 2) -- (-3, 2);
					\draw[variablecycle] (-3, 2) -- (-3, 0);
					\draw[variablecycle] (3, 2) -- (3, 0);

					\node (na1) at (2, 0) [] {$~~\dots~~$};
					\node (na1) at (-2, 0) [] {$~~\dots~~$};
				\end{pgfonlayer}
			\end{tikzpicture}

			\caption{The portion of the \inciGraph $\inciGraphSign{\phi, \pi}$ concerning variable $v$ of a monotone planar formula $\phi$.}
		\end{subfigure}
		\hfill
		\begin{subfigure}[t]{0.45\linewidth}
			\centering
			
			\begin{tikzpicture}[scale=0.8, every node/.style={fill=white, scale=0.8}]
				\pgfdeclarelayer{background}
				\pgfdeclarelayer{foreground}
				\pgfsetlayers{background,main,foreground}

				\node (v) at (0, 0.2) [litteral] {$G_v$};
				\node (vprime) at (0, -0.2) [litteral] {$\bar G_v$};

				\node (c1) at (-1.5, 1.5) [positiveclause] {$c'_1$};
				\node (c2) at (-0.5, 1.5) [positiveclause] {$c'_2$};
				\node (cn) at (1.5, 1.5) [positiveclause] {$c'_n$};
				\node (bc1) at (-1.5, -1.5) [clause] {$\neg c'_1$};
				\node (bc2) at (-0.5, -1.5) [clause] {$\neg c'_2$};
				\node (bcm) at (1, -1.5) [clause] {$\neg c'_m$};

				\draw[-] (v) -- (c1);
				\draw[-] (v) -- (c2);
				\draw[-] (v) -- (cn);

				\draw[-] (vprime) -- (bc1);
				\draw[-] (vprime) -- (bc2);
				\draw[-] (vprime) -- (bcm);

		\node (na1) at (2, 0.5) [] {$~~\dots~~$};
	\node (na1) at (2, -0.5) [] {$~~\dots~~$};
	\node (na1) at (-2, 0.5) [] {$~~\dots~~$};
	\node (na1) at (-2, -0.5) [] {$~~\dots~~$};
	
	\node[minimum height=1.5cm,draw,dotted] (GG) {$\litClauGraphSign{\phi'_v}$};
	
	\draw[variablecycle] (3, 0.5) -- (-3, 0.5);
	\draw[variablecycle] (3, -0.5) -- (-3, -0.5);
	\draw[variablecycle] (3, -0.5) -- (3, 0.5);
	\draw[variablecycle] (-3, -0.5) -- (-3, 0.5);

	\node[opacity=0.5,minimum height=1.5cm,draw,dotted] (GG) {$\litClauGraphSign{\phi'_v}$};
			\end{tikzpicture}

			\caption{
				The portion of the planar embedding of $\litClauGraphSign{\phi', \pi'}$ concerning the graph $\litClauGraphSign{\phi'_v}$.
			}
			\label{fig:lit_cycle_reduc}
		\end{subfigure}

		\vspace{1cm}

		\begin{subfigure}[b]{\linewidth}
			\centering

			\newcommand\gadgetci[3]{
				\node (a1) at (#1+1, 2) [litteral] {$a_#2^v$};
				\node (np1) at (#1+2, 2) [litteral] {$\lnot p_#2^v$};
				\node (p1) at (#1+3, 2) [litteral] {$p_#2^v$};
				\node (c1) at (#1+1, 3) [clause] {$#3 \lor a_#2^v$};	
				\draw[-] (na1) -- (c1);
				
				\node(na1) at (#1+4, 2) [litteral] {$\lnot a_#2^v$};

				\node (c2) at (#1+3, 3) [clause] {$p_#2^v \lor \lnot a_#2^v$};
				\node (cla1) at (#1+2, 3.75) [positiveclause] {$c'_#2$};
				
				\draw (np1) edge[paired, bend right=30] (p1);
				\draw (na1) edge[paired, bend left=30] (a1);

				\draw[-] (a1) -- (c1);
				\draw[-] (na1) -- (c2);
				\draw[-] (p1) -- (c2);
				\draw[-] (np1) -- (cla1);
			}

			\newcommand\gadgetnci[3]{
				\node (a1) at (#1+1, 0) [litteral] {$b_#2^v$};
				\node (np1) at (#1+2, 0) [litteral] {$\lnot n_#2^v$};
				\node (p1) at (#1+3, 0) [litteral] {$n_#2^v$};
				\node (c1) at (#1+1, -1) [clause] {$#3 \lor b_#2^v$};	
				\draw[-] (na1) -- (c1);
				
				\node(na1) at (#1+4, 0) [litteral] {$\lnot b_#2^v$};

				\node (c2) at (#1+3, -1) [clause] {$n_#2^v \lor \lnot b_#2^v$};
				\node (cla1) at (#1+2, -1.75) [clause] {$\lnot c'_#2$};
				
				\draw (np1) edge[paired, bend left=30] (p1);
				\draw (na1) edge[paired, bend right=30] (a1);

				\draw[-] (a1) -- (c1);
				\draw[-] (na1) -- (c2);
				\draw[-] (p1) -- (c2);
				\draw[-] (np1) -- (cla1);
			}
			
			\scalebox{0.85}{
			\begin{tikzpicture}[scale=0.85, every node/.style={fill=white, scale=0.7}]
				\draw[variablecycle] (-1,2) -- (15.5,2);
				\draw[variablecycle] (-1,0) -- (15.5,0);

				\draw (0,0) edge[paired] (0,2);
				
				\node (na1) at (0, 2) [litteral] {$v$};
				\gadgetci {0} {1} {v}
				\gadgetci {4} {2} {\lnot a_1^v}
				\draw (na1) -- (4+4.5, 2.5);
				
				\node (na1) at (9.75, 2) [] {$~~~\dots~~~$};
				\gadgetci {10.5} {n} {\lnot a_{n-1}^v}
				
				\node (na1) at (0, 0) [litteral] {$\lnot v$};
				
				\gadgetnci {0} {1} {\lnot v}
				\gadgetnci {4} {2} {\lnot b_1^v}
				\draw (na1) -- (4+4.5, -0.5);
				
				\node (na1) at (9.75, 0) [] {$~~~\dots~~~$};
				\gadgetnci {10} {m} {\lnot b_{m-1}^v}
				\node at (16,0) {$\dots$};
				\node at (16,2) {$\dots$};
				\node at (-1.5,0) {$\dots$};
				\node at (-1.5,2) {$\dots$};
		\end{tikzpicture}}

		\caption{
			A planar embedding of the graph $\litClauGraphSign{\phi'_v}$. 
			All  literals nodes $\literal$ have an edge with their opposite nodes $\bar \literal$. 
			The \tikz[baseline=-1mm]{\draw (0, 0) edge[variablecycle]  (0.5, 0);}-edges represent the literal cycle that pass through each literal while preserving the planarity of $\litClauGraphSign{\phi'_v}$.}
		\label{fig:gadget_lit_claus_graph_phi_v}
    \end{subfigure}

    \caption{Construction of $\litClauGraphSign{\phi', \pi'}$ from $\inciGraphSign{\phi, \pi}$, by replacing each variable $v$ from a Monotone Planar \tsat instance $\phi$ by the gadgets $G_v$ and $\bar G_v$.}

    \label{fig:gadget_planar_lsat}
\end{figure}
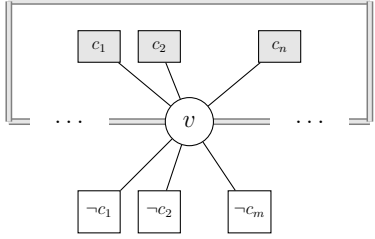
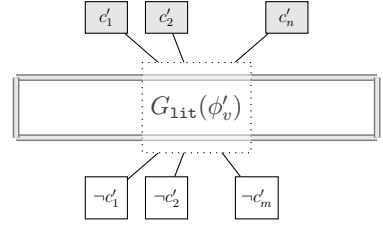
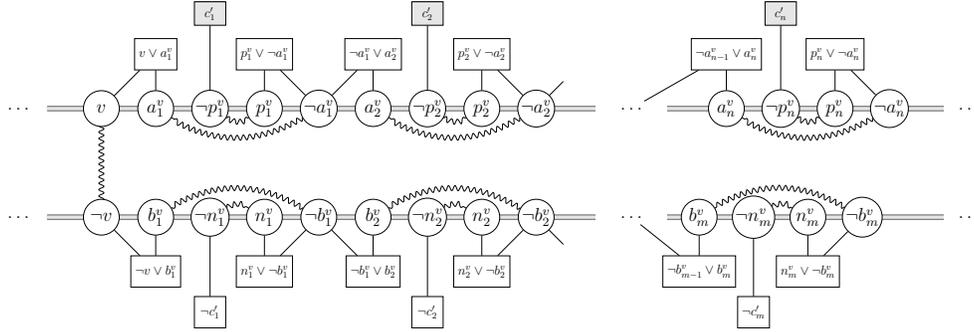

\begin{claim}
    $\litClauGraphSign{\phi',\pi'}$ is valid for some $\pi'$.
\end{claim}
\begin{proof}
    We know that $\phi$ is a monotone planar formula w.r.t. $\pi$.
    Thus the \inciGraph $\inciGraphSign{\phi,\pi}$ is planar with the variable cycle $\pi$ that separates positive and negative clauses in the considered embedding.
    We are going to construct the \litClausGraph $\litClauGraphSign{\phi', \pi'}$ for some $\pi'$, and prove that it is valid,
    that is, give a planar embedding in which paired edges are inside the literal cycle, and all clauses are outside of it.
    We proceed as follows.
    We modify $\inciGraphSign{\phi,\pi}$ by removing each variable node $v$
    and replacing it with $\litClauGraphSign{\phi'_v}$.
    Note that $\litClauGraphSign{\phi'_v}$ contains nodes~$v$ and $\lnot v$.
    The new node $v$ is linked to all positive clauses, while $\lnot v$ is linked to all negative clauses.
    Last, we add all paired edges.

    The graph $\litClauGraphSign{\phi'_v}$ is made of two disjoint parts,
    the first one containing all nodes connected to $v$, and the second one those connected to $\lnot v$.
    Figure~\ref{fig:lit_cycle_reduc} illustrates the replacement of node $v$ by $\litClauGraphSign{\phi'_v}$.
    A planar embedding for $\litClauGraphSign{\phi'_v}$ is given in
    Figure~\ref{fig:gadget_lit_claus_graph_phi_v}
    where the literals appear along two distinct lines.

    The literal cycle $\pi'$ is built as follows.
    Let $v_{\pi_1},\ldots,v_{\pi_n}$ denote the variables of~$\phi$ in the order they appear in $\inciGraphSign{\phi, \pi}$.
    $\pi'$ starts with the literals that appear in the upper part of Fig.~\ref{fig:gadget_lit_claus_graph_phi_v} from left to right, first for the variable
    $v_{\pi_1}$, then for $v_{\pi_2}$ etc. until $v_{\pi_n}$.
    Then we add literals from the lower part of Fig.~\ref{fig:gadget_lit_claus_graph_phi_v} from right to left for~$v_{\pi_n}$,
    then $v_{\pi_{n-1}}$, and downto $v_{\pi_1}$.

    In the presented embedding, the cycle $\pi'$ is such that all paired edges are inside, as seen in Fig.~\ref{fig:gadget_lit_claus_graph_phi_v},
    and all clauses are outside of it. Therefore, 
    $\litClauGraphSign{\phi',\pi'}$ is indeed valid.
\end{proof}

It follows that $\phi'$ is literal-planar.
Moreover $\phi'$ and $\pi'$ are constructible in polynomial time: we have in total $(2n + 2m)\times|V|+ |C|$ clauses.

\begin{claim}
    The formula $\phi$ is satisfiable if and only if $\phi'$ is satisfiable.    
\end{claim}
\begin{proof}
	\fbox{$\Rightarrow$}
    Assume $\phi$ is satisfiable, i.e., if there exists a truth assignment $\assignment : V \rightarrow \{0, 1\}$ of $V$ such that each clause $c \in C$ of $\phi$ is satisfied. 
    We show that $\phi'$ is satisfiable. We define a new assignment $\assignment'$ as follows:
    \begin{equation}
    	\begin{split}
            \assignment'(v) = \assignment(v) \\
    		\assignment'(n_j^{v}) =  \assignment'(b_j^{v}) = \assignment(v) \\
    		\assignment'(p_i^{v}) = \assignment'(a_i^{v}) = 1 - \assignment(v)
    	\end{split}
    \label{eq:assignment_paired_lit}
    \end{equation}
    Let us show that $\assignment' \models \phi'$. 
    First, $\assignment' \models \phiprimer$ since $\phiprimer$ is obtained from $\phi$ by replacing each $v$ by some $\lnot p_i^{v}$,
    and $\lnot v$ by some $n_i^{v}$.
    Second, $\assignment' \models \phi'_v$.
    In fact, $\phi'_v$ can be seen as the conjunction of two sequence of implications, namely,
    \[
        \phi'_v \models (\lnot p_i^v \rightarrow \lnot a_i^v) \land (\lnot a_i^v \limply \lnot a_{i-1}^v) \land \dots (\lnot a_2^v \limply \lnot a_{1}^v) \land   (\lnot a_{1}^v \limply v),
    \] 
    and
    \[
        \phi'_v \models (\lnot n_i^v \rightarrow \lnot b_i^v) \land (\lnot b_i^v \limply \lnot b_{i-1}^v) \land \dots (\lnot b_2^v \limply \lnot b_{1}^v) \land   (\lnot b_{1}^v \limply v).
    \] 
    The first set of implications holds because either $\assignment'(v) = 1$ and all variables in this chain are true, 
    or $\assignment'(v) = 0$ and they are all false; and similarly for the second sequence of implications.

    \fbox{$\Leftarrow$}
    Conversely, consider a truth assignment $\assignment'$ of $V'$ that satisfies $\phi'$. We define $\assignment$ as the restriction of $\assignment'$ to $V$.
    By \eqref{eq:implies_pos} and \eqref{eq:implies_negat}, if $\assignment'(\lnot p_i^v) = 1$ for some $i$, then 
    $\assignment'(\lnot n_j^v) = 0$ for all $j$. In this case, we have $\assignment'(v) = 1$ by \eqref{eq:implies_pos}.    
    Conversely, if $\assignment'(\lnot n_j^v) = 1$ for some $j$,
    then $\assignment'(\lnot p_i^v) = 0$ for all $i$,
    and $\assignment'(v) = 0$.

    Then because $\assignment' \models \phiprimer$, we have $\assignment \models \phi$ since satisfiability is preserved by replacing each 
    $\lnot p_i^v$ by $v$, and each $\lnot p_j^v$ by $\lnot v$.
\end{proof}

 We just established that \PLLPTSAT is NP-hard. Because \PLLPTSAT is a fragment of SAT, it is \npc.

\subsection{NP-completeness of \LPTSAT}

\newcommand{\MPLtwoOne}{Monotone Linear Planar\xspace}

In this section, we prove the NP-hardness for the class of \emph{monotone linear planar 3-CNF} formulas,
a subclass of linear planar 3-CNF formulas.

\begin{definition}
    \label{def:cmpl}
    A \emph{monotone linear planar 3-CNF} is a linear and monotone planar formula.
\end{definition}

We now prove the NP-hardness of \MPLtwoOne \tsat by a reduction from \PLLPTSAT. Again \MPLtwoOne \tsat takes both $\phi$ and $\pi$ in the input.

\newcommand\friendassignment[1]{{#1}^\pm}

\begin{theorem}
	\label{th:constrainedmonotoneplanarlinearthreesatnpcomplete}
	\MPLtwoOne \tsat is \npc.
    NP-hardness holds even when each variable appears negatively in at most one clause.
\end{theorem}

Let $\phi$ a \PLLPTformula with $V$ the set of variables, and $C$ the set of clauses. 
For each variable $v$ in $V$ we introduce a fresh variable~$v'$.
Define $\psi'$ as follows.
\begin{equation}
    \label{eq:reduction_planar_lsat}
    \psi' := \bigwedge_{v \in V} (\neg v \lor \neg v')
\end{equation}

The 3-CNF formula $\phi'$ is defined as the conjunction of $\psi'$
with the formula obtained from $\phi$ 
by replacing each $\neg v$ by $v'$, for all $v \in V$.

For instance, consider the following \PLLPTformula (Fig.~\ref{fig:planar_paired_lit_lsat_example}):

\begin{equation}
    \phi := (a \lor \neg b \lor d) \land (b \lor \neg c \lor d) \land (\neg d \lor c) \land (c \lor \neg a)
\end{equation}

We have

\begin{equation}
    \phi' := (a \lor b' \lor d) \land (b \lor c' \lor d) \land (d' \lor c) \land (c \lor a') \land \psi'
\end{equation}

\newcommand\pairededge{\protect\tikz[baseline=-1mm]{\protect\draw[paired] (0, 0) -- (0.7,0);}}
\begin{figure}[t]
    \centering
    
    $\litClauGraphSign{\phi,\pi}$
      \begin{tikzpicture}[scale=0.7, every node/.style={fill=white, scale=0.8}]
    	\pgfdeclarelayer{background}
    	\pgfdeclarelayer{foreground}
    	\pgfsetlayers{background,main,foreground}

    	\node (n1) at (0, 0) [litteral] {$a$};
    	\node (n2) at (1.5, 0) [litteral] {$\neg b$};
    	\node (n3) at (3, 0) [litteral] {$b$};
    	\node (n4) at (4.5, 0) [litteral] {$\neg c$};
    	\node (n5) at (6, 0) [litteral] {$d$};
    	\node (n6) at (7.5, 0) [litteral] {$\neg d$};
    	\node (n7) at (9, 0) [litteral] {$c$};
    	\node (n8) at (10.5, 0) [litteral] {$\neg a$};
    	
    	\node (c1) at (3, 2.25) [draw, rectangle, minimum width=4cm] {$a \lor \neg b \lor d$};
    	\node (c2) at (4.5, 1.5) [draw, rectangle] {$b \lor \neg c \lor d$};
    	\node (c3) at (7.5, 1.5) [draw, rectangle] {$\neg d \lor c$};
    	\node (c4) at (10.5, 1.5) [draw, rectangle] {$c \lor \neg a$};
    	
    	\begin{pgfonlayer}{background}
    		\draw[variablecycle] (n1) -- (n2);
    		\draw[variablecycle] (n2) -- (n3);
    		\draw[variablecycle] (n3) -- (n4);
    		\draw[variablecycle] (n4) -- (n5);
    		\draw[variablecycle] (n5) -- (n6);
    		\draw[variablecycle] (n6) -- (n7);
    		\draw[variablecycle] (n7) -- (n8);
    		\draw[variablecycle] (n1) -- (-1,0) -- (-1,-2) -- (11.5,-2) -- (11.5,0) -- (n8);
    		
    		\draw (n1) edge[paired, bend right=30] (n8);
    		\draw (n2) edge[paired, bend right=20] (n3);
    		\draw (n4) edge[paired, bend right=30] (n7);
    		\draw (n5) edge[paired, bend right=20] (n6);
    		
    		\draw[-] (n1) -- (c1);
    		\draw[-] (n2) -- (c1);
    		\draw[-] (n5) |- (c1);
    		
    		\draw[-] (n3) -- (c2);
    		\draw[-] (n4) -- (c2);
    		\draw[-] (n5) -- (c2);
    		
    		\draw[-] (n6) -- (c3);
    		\draw[-] (n7) -- (c3);
    		
    		\draw[-] (n7) -- (c4);
    		\draw[-] (n8) -- (c4);

    	\end{pgfonlayer}
    \end{tikzpicture}

\begin{tikzpicture}
	\draw (0, 0) edge[-latex, line width=2mm] (0, -1);
\end{tikzpicture}

$\inciGraphSign{\phi', \pi'}$ 
    \begin{tikzpicture}[scale=0.8, every node/.style={fill=white, scale=0.8}]
        \pgfdeclarelayer{background}
        \pgfdeclarelayer{foreground}
        \pgfsetlayers{background,main,foreground}

        \node (n1) at (0, 0) [draw, circle] {$a$};
        \node (n2) at (1.5, 0) [draw, circle] {$b'$};
        \node (n3) at (3, 0) [draw, circle] {$b$};
        \node (n4) at (4.5, 0) [draw, circle] {$c'$};
        \node (n5) at (6, 0) [draw, circle] {$d$};
        \node (n6) at (7.5, 0) [draw, circle] {$d'$};
        \node (n7) at (9, 0) [draw, circle] {$c$};
        \node (n8) at (10.5, 0) [draw, circle] {$a'$};

        \node[positiveclause] (c1) at (3, 2.25) [draw, rectangle, minimum width=4cm] {$a \lor b' \lor d$};
        \node[positiveclause] (c2) at (4.5, 1.5) [draw, rectangle] {$b \lor c' \lor d$};
        \node[positiveclause] (c3) at (7.5, 1.5) [draw, rectangle] {$d' \lor c$};
        \node[positiveclause] (c4) at (10.5, 1.5) [draw, rectangle] {$c \lor a'$};

        \node (c5) at (5.5,-2.5) [draw, rectangle] {$\neg a \lor \neg a'$};
        \node (c6) at (6.75,-1.75) [draw, rectangle] {$\neg c' \lor \neg c$};
        \node (c7) at (2.25,-1.0) [draw, rectangle] {$\neg b' \lor \neg b$};
        \node (c8) at (6.75,-1.0) [draw, rectangle] {$\neg d' \lor \neg d$};
    
        \begin{pgfonlayer}{background}
            \draw[variablecycle] (n1) -- (n2);
            \draw[variablecycle] (n2) -- (n3);
            \draw[variablecycle] (n3) -- (n4);
            \draw[variablecycle] (n4) -- (n5); 
            \draw[variablecycle] (n5) -- (n6);
            \draw[variablecycle] (n6) -- (n7);
            \draw[variablecycle] (n7) -- (n8);
            \draw[variablecycle] (n1) -- (-1,0) -- (-1,-3) -- (11.5,-3) -- (11.5,0) -- (n8);

            \draw[-] (n1) |- (c5);
            \draw[-] (n8) |- (c5);

            \draw[-] (n4) |- (c6);
            \draw[-] (n7) |- (c6);

            \draw[-] (n2) -- (c7);
            \draw[-] (n3) -- (c7); 

            \draw[-] (n5) -- (c8);
            \draw[-] (n6) -- (c8); 
            
            \draw[-] (n1) -- (c1);
            \draw[-] (n2) -- (c1);
            \draw[-] (n5) |- (c1);
    
            \draw[-] (n3) -- (c2);
            \draw[-] (n4) -- (c2);
            \draw[-] (n5) -- (c2);
    
            \draw[-] (n6) -- (c3);
            \draw[-] (n7) -- (c3);
    
            \draw[-] (n7) -- (c4);
            \draw[-] (n8) -- (c4);

        \end{pgfonlayer}
    \end{tikzpicture}
    
    \caption{
    	From the literal-clause graph $\litClauGraphSign{\phi,\pi}$, we construct the planar \inciGraph $\inciGraphSign{\phi', \pi'}$ 
    	obtained as follows: we replace each negative literals
    	$\lnot x$ by $x'$, and each paired edge $x \pairededge \lnot x$ a clause $\lnot x' \lor \lnot x'$. 
    	This is used in the reduction given in Theorem~\ref{th:constrainedmonotoneplanarlinearthreesatnpcomplete} from a \PLLPTformula $\phi$ to a monotone planar linear 3-CNF $\phi'$. %
    	}
    
    \label{fig:reduction_paired_lit_plsat_to_planar_lsat}
\end{figure}
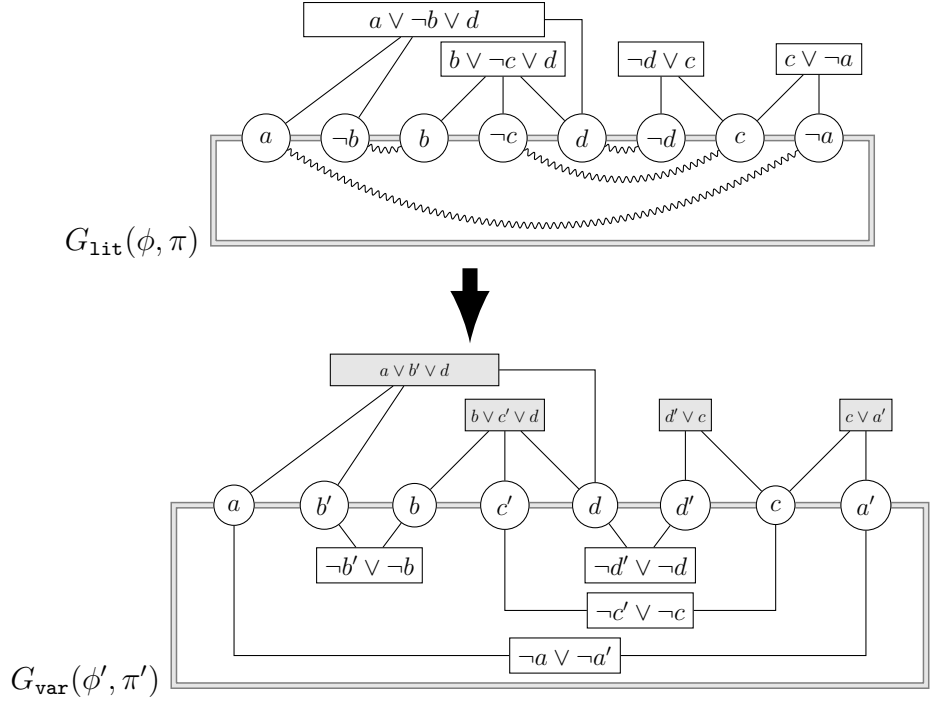

We can verify that $\phi'$ is a linear 3-CNF: $\phi$ is a linear 3-CNF and replacing each negated literal by new variables preserve linearity. 
Furthermore, clauses of $\psi'$ do not intersect each other,
and the clauses of $\phi'$ and $\psi'$ also do not intersect since the former contains only positive literals, and the latter only negative ones.

Recall that $\litClauGraphSign{\phi,\pi}$ is planar and valid, by hypothesis.
Consider the variable cycle $\pi'$ obtained from $\pi$ by replacing each $\lnot v$ by $v'$.
We argue that $\inciGraphSign{\phi',\pi'}$ is planar. 
In fact, $\inciGraphSign{\phi',\pi'}$ can be obtained from $\litClauGraphSign{\phi,\pi}$ 
by renaming each node $\lnot v$  into $v'$, and adding the new clauses $\lnot v \lor \lnot v'$.
Because $\litClauGraphSign{\phi,\pi}$ is valid, all clauses are outside of the literal cycle.
It follows that in $\inciGraphSign{\phi',\pi'}$, all positive clauses of $\pi'$ are outside the variable cycle $\pi$,
while the negative clauses of $\psi$ are placed inside the cycle.
Moreover, each negative clause $\lnot v \lor \lnot v'$ can be seen as replacing a paired edge between $v$ and $\lnot v$.
Therefore $\litClauGraphSign{\phi,\pi}$ is planar and valid.

An example of this transformation is provided in %
Fig.~\ref{fig:reduction_paired_lit_plsat_to_planar_lsat}.

Note that our construction is a monotone linear planar 3-CNF:
each clause contains only negated or only positive literals,
and the positive clauses are separated from the negative clauses by the variable cycle.

\begin{claim}
   Formula $\phi$ is satisfiable if and only if $\phi'$ is satisfiable.
    \label{lem:linear_literal_planar_sat_iff_linear_lanar_sat}
\end{claim}

\begin{proof}
    \fbox{$\Rightarrow$}
    Assume $\phi$ is satisfiable, i.e. there exists a truth assignment $\nu$ for $V$ such that each clause $c$ of $\phi$ is satisfied. 
    We define $\friendassignment\nu$ as the 
    \emph{friend assignment} obtained from $\nu$ as follows. 
    For each $v \in V$,

    \begin{equation}
        \begin{split}
            \friendassignment\nu(v) := \nu(v) \\
            \friendassignment\nu(v') := 1 - \nu(v)
        \end{split}
    \label{eq:assignment_paired_lit}
    \end{equation}

    We have $\friendassignment\nu \models (\neg v \lor \neg v')$ for all variable $v$. Furthermore, the transformed clauses are also made true by $\friendassignment\nu$. So $\friendassignment\nu \models \phi'$, and $\phi'$ is satisfiable.

    \fbox{$\Leftarrow$}
    Conversely, let $\nu'$ be a truth assignment for $V'$ that satisfies $\phi'$. 
    Let $\nu$ denote the restriction of $\nu'$ to $V$.
    
    All transformed clauses of $\phi'$ are satisfied by $\nu'$.
    Let $c$ be a clause of $\phi$ and $c'$ the corresponding clause of $\phi'$ obtained from $c$.
    Let $\ell \in c'$ such that $\nu' \models \ell$.
    If $\ell$ is in $V$, then it also appears in $c$, and $\nu(\ell) = \nu'(\ell)$, thus $\nu \models c$.
    Otherwise, $\ell = v'$ for some variable $v \in V$, and $c$ contains the literal $\lnot v$.
    But the clause $\lnot v \lor \lnot v'$ of $\phi'$ implies that in this case, $\nu'(v)=0$, that is, $\nu(v) = 0$.
    It follows that $\nu \models c$.
    
    Hence, $\phi$ is satisfiable.
\end{proof}

\newtheorem{corollary}{Corollary}[theorem]

In conclusion, we proved that \MPLtwoOne is NP-complete. 
As a corollary, we get our main result stated in the following theorem.

\begin{theorem}
	\LPTSAT is \npc.
	\label{th:planar_linear_tsat_npc}
\end{theorem}

\section{Cycle Augmentation for Planar Graphs}
\label{section:cycleaugmentation}
\newcommand{\genericcycle}{\acycle}
\newcommand{\vcycle}{$V'$-cycle\xspace}

In this section, we introduce the \emph{planar cycle augmentation} problem.
This consists, given a planar graph $G$ and a subset $V'$ of vertices, in computing a $V'$-cycle\footnote{i.e. a cycle among vertices in $V'$}~$\genericcycle$ that can be added to $G$ while preserving planarity. 

\begin{definition}[Planar Cycle Augmentation]
    Planar cycle augmentation is the following decision problem:
    \begin{itemize}
        \item Instance: an undirected planar graph $G = (V,E)$, a subset $V' \subseteq V$.
        \item Question: is there a $V'$-cycle $\genericcycle$ such that the multi-graph $G' = (V, E \uplus \genericcycle)$ is planar?
    \end{itemize}
    \label{def:planar_cycle_augment_decision}
\end{definition}

Planar cycle augmentation is the cornerstone of this article. 
In \Cref{sec:PLLPTSATreconfiguration}, a $V'$-cycle corresponds to a literal cycle in the \PLLPTSAT Reconfiguration problem.
The existence of a literal cycle will be crucial for proving the PSPACE-completeness of 2D CMAPF (\Cref{th:2D CMAPFPSPACEcomplete}). 
In fact, the exact complexity of 2D CMAPF was left open in \cite{cmapfIsse}
because existing planar 3-SAT reconfiguration problems in the literature~\cite{monotonNAEtSATRecon} are not sufficiently powerful as they do not mention any notion of variable or literal cycle.
The developments of this section allows us to define a reconfiguration problem powerful enough to fill this gap.

\begin{figure}[h]
    \centering

        \begin{tikzpicture}[
        	baseline=0mm,
            scale=0.9, 
            every node/.style={scale=0.7}
        ]
                \node [style={variable_node}] (0) at (-1.75, 0.25) {$a$};
                \node [style={variable_node},inVprime] (1) at (-1, -1) {$b$};
                \node [style={variable_node}] (2) at (0, 0) {$c$};
                \node [style={variable_node},inVprime] (3) at (-0.75, 0.75) {$d$};
                \node [style={variable_node},inVprime] (4) at (0.75, 1) {$e$};
                \node [style={variable_node},inVprime] (5) at (1.5, -1) {$f$};
                \node [style={variable_node}] (6) at (2.25, -0.25) {$g$};
                \draw (6) to (4);
                \draw (4) to (5);
                \draw (5) to (6);
                \draw (5) to (1);
                \draw (2) to (5);
                \draw (2) to (3);
                \draw (3) to (4);
                \draw (3) to (0);
                \draw (0) to (1);
        \end{tikzpicture}
	\fleche{}
        \begin{tikzpicture}[
        	baseline=0mm,
            scale=0.9, 
            every node/.style={scale=0.7}
        ]
            \node [style={variable_node}] (0) at (-1.75, 0.25) {$a$};
            \node [style={variable_node}, inVprime] (1) at (-1, -1) {$b$};
            \node [style={variable_node}] (2) at (0, 0) {$c$};
            \node [style={variable_node}, inVprime] (3) at (-0.75, 0.75) {$d$};
            \node [style={variable_node}, inVprime] (4) at (0.75, 1) {$e$};
            \node [style={variable_node}, inVprime] (5) at (1.5, -1) {$f$};
            \node [style={variable_node}] (6) at (2.25, -0.25) {$g$};
            \node [style=none] (7) at (-2.25, 1.25) {};
            \draw (6) to (4);
            \draw (4) to (5);
            \draw (5) to (6);
            \draw (5) to (1);
            \draw (2) to (5);
            \draw (2) to (3);
            \draw (3) to (4);
            \draw (3) to (0);
            \draw (0) to (1);
            \draw [style={variablecycle}, bend right=15] (3) to (4);
            \draw [style={variablecycle}, bend right=15, looseness=0.75] (5) to (1);
            \draw [style={variablecycle}, bend left=15] (3) to (5);
            \draw [style={variablecycle}, bend right=270, looseness=1.50] (1) to (7.center);
            \draw [style={variablecycle}, bend left] (7.center) to (4);
        \end{tikzpicture}

    \hfill
    	\footnotesize 
    \begin{tabular}{ll}
    	\tikz[baseline=0mm]{\node[variable_node,inVprime] {$x$};	}
    	& 	Vertices in $V'$ \\
    	\tikz[baseline=0mm]{\draw[variablecycle] (0, 0) -- (1, 0);} & $V'$-cycle
    \end{tabular}
    
    \caption{Positive instance of planar cycle augmentation with $V' = \set{b, d, e, f}$. A $V'$-cycle $\genericcycle = (b, e, d, f)$ is shown. Adding $\genericcycle$ preserves planarity.}
    \label{fig:cycle_augment_example}
\end{figure}
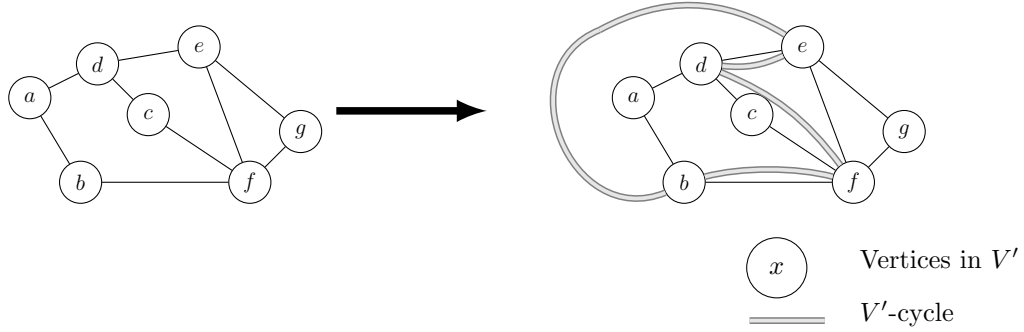

\Cref{fig:cycle_augment_example} shows an example of an instance of planar cycle augmentation. 
The graph augmented with a cycle $\genericcycle$ can be seen on the right in the figure.

\newcommand{\cycleaugment}{\emph{planar cycle augmentation}\xspace}

\subsection{NP-completeness of Cycle augmentation}

\begin{theorem}
    \emph{Planar cycle augmentation} is \npc.

    \label{th:general_planar_cycle_construction_npc}
\end{theorem}

\begin{proof}
    \emph{Planar cycle augmentation} is in NP by guessing a $V'$-cycle and checking the planarity of $G'$ in polynomial time \cite{hopcroft1974efficient}.    
    We prove the NP-hardness of \cycleaugment by reducing from the following problem: decide if an arbitrary maximal\footnote{maximal means that adding another edge to the graph would break planarity} planar graph is \emph{Hamiltonian} \cite{hamiltonianMaximalPlanarDecPbNPC}. 
    Consider a maximal planar graph $G = (V,E)$. The planar cycle augmentation instance is $(G, V)$.
    As $G$ is a maximal planar graph, for all cycle $\genericcycle$, $G' = (V, E \uplus \genericcycle)$ is planar iff $\genericcycle \subseteq E$.
    Hence, finding a $V'$-cycle $\genericcycle$ is equivalent to finding an \emph{Hamiltonian cycle} within the graph $G$.
    Hence \cycleaugment is NP-hard.
\end{proof}

\subsection{A sufficient condition for cycle existence}

\newcommand{\vfg}{\texttt{Kite}}
\newcommand{\vertexFaceGraph}{kite-graph\xspace}

\newcommand{\gvf}{K_{I}}
\newcommand{\eulerCycle}{\mathscr{C}}
\newcommand{\red}{\texttt{red}\xspace}
\newcommand{\blue}{\texttt{blue}\xspace}

\newcommand{\gvfface}{\mathfrak{f}}
\newcommand{\degree}[1]{\texttt{deg}_{#1}}

Given planar cycle augmentation instance $(G, V')$, we provide a sufficient condition on $(G, V')$ for the existence of a \vcycle $\genericcycle$.
We also provide a polynomial time algorithm to compute a \vcycle $\genericcycle$ if the condition is met. The condition is the existence of a \emph{kite-graph},
defined below.

\begin{definition}[Kite-Graph] %
	Consider a cycle augmentation instance $I = (G, V')$. 
    A \emph{\vertexFaceGraph} of $I$ denoted $\gvf = (V' \uplus F', E_{K})$ is a bipartite graph between $V'$ and some fresh set $F'$ of vertices
    with the following properties. %
        \begin{enumerate}
        \item $\gvf \cup G$ has a planar embedding,
    	\item $\gvf$ is Eulerian (see \Cref{prop:eulerian_graph_prop}),
    	\item for all $v \in V'$ in $\gvf$, the degree of $v$ in $K_I$ is 2, i.e. $\degree{K_I}(v) = 2$. 
    \end{enumerate}
    \label{def:kitegraph}
\end{definition}

Informally, each vertex of $F'$ represents a face of $G$, and \vertexFaceGraph links each vertex in $V'$ to the faces it touches.
A \vertexFaceGraph can be seen in \Cref{fig:face_vertex_graph_construction}.

In the following developments and \Cref{alg:constructingVprimecycle}, we use several
notions about planar graphs and their dual graphs. These are formally defined in \Cref{section:appendix-planar-graphs}.
Readers with a familiarity with planar graphs can read the following and only refer to the appendix to clarify certain points, 
while reading the appendix first might be useful if one is not familiar with planar graphs.

Let $I = (G = (V,E), V')$ be a \cycleaugment instance.
Let $\gvf$ be a \vertexFaceGraph of $I$.
By \Cref{prop:bipartite_dual_eulerian}, the dual graph $\gvf^*$ is bipartite 
i.e. 2-colorable. We fix a 2-coloring of faces of $\gvf$; the colors are called red\footnote{thick} and blue\footnote{thin}. As $\gvf^*$ is connected, the partition in two-color classes is unique, i.e. there is a unique 2-coloring of faces of $\gvf$, up to renaming. Edges of $G$ are colored accordingly as follows: an edge located (Definition \ref{def:graph_embedded_inside_face}) in a red (resp. blue) face is red (resp. blue). In other words, the set of edges is uniquely partitioned in two sets. Let $\Pi_{G,K_I}$ be this partition.

\newcommand{\blueface}{\draw[blueface, draw=white, line width=2mm] (-2.7, -1.8) rectangle (2.8, 2.4);}

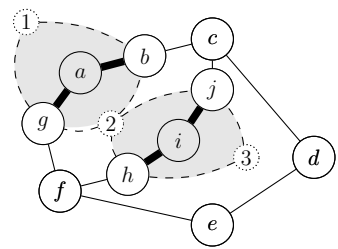
\begin{figure}[h]
    \centering
 
        \begin{tikzpicture}[
            scale=0.9, 
            every node/.style={scale=0.7}
        ]
			\blueface
			
            \node [style={variable_node}, inVprime] (b) at (-0.25, 1.25) {$b$};
            \node [style={variable_node}, inVprime] (j) at (0.75, 0.75) {$j$};
            \node [style={variable_node}] (a) at (-1.2, 1) {$a$};
            \node [style={variable_node}] (c) at (0.75, 1.5) {$c$};
            \node [style={variable_node}] (i) at (0.25, 0) {$i$};
            \node [style={variable_node}, inVprime] (h) at (-0.5, -0.5) {$h$};
            \node [style={variable_node}] (f) at (-1.5, -0.75) {$f$};
            \node [style={variable_node}, inVprime] (g) at (-1.75, 0.25) {$g$};
            \node [style={variable_node}] (d) at (2.25, -0.25) {$d$};
            \node [style={variable_node}] (e) at (0.75, -1.25) {$e$};
            \node [style={dual_g_face}] (3) at (1.25, -0.25) {$3$};
            \node [style={dual_g_face}] (2) at (-0.75, 0.25) {$2$};
            \node [style={dual_g_face}] (1) at (-2, 1.75) {$1$};

        	\draw[redface, draw=none] (1.center) to[bend left=20] (b.center) to[bend left=20] (2) to[bend left=20] (g.center) to[bend left=40] cycle;
        	
        	\draw[redface, draw=none] (2.center) to[bend left=20] (b.center) to[bend left=20] (2.center) to[bend left=20] (j.center) to[bend left=20] (3.center) to[bend left=20] (h.center) to[bend left=20] cycle;

        	\node [style={variable_node}, inVprime] (b) at (-0.25, 1.25) {$b$};
        	\node [style={variable_node}, inVprime] (j) at (0.75, 0.75) {$j$};
        	\node [style={variable_node}] (a) at (-1.2, 1) {$a$};
        	\node [style={variable_node}] (c) at (0.75, 1.5) {$c$};
        	\node [style={variable_node}] (i) at (0.25, 0) {$i$};
        	\node [style={variable_node}, inVprime] (h) at (-0.5, -0.5) {$h$};
        	\node [style={variable_node}] (f) at (-1.5, -0.75) {$f$};
        	\node [style={variable_node}, inVprime] (g) at (-1.75, 0.25) {$g$};
        	\node [style={variable_node}] (d) at (2.25, -0.25) {$d$};
        	\node [style={variable_node}] (e) at (0.75, -1.25) {$e$};
        	\node [style={dual_g_face}] (3) at (1.25, -0.25) {$3$};
        	\node [style={dual_g_face}] (2) at (-0.75, 0.25) {$2$};
        	\node [style={dual_g_face}] (1) at (-2, 1.75) {$1$};
        	
            \draw[rededge] (a) to (g);
            \draw[rededge] (a) to (b);
            \draw[blueedge] (b) to (c);
            \draw[blueedge] (c) to (j);
            \draw[rededge] (j) to (i);
            \draw[rededge] (i) to (h);
            \draw[blueedge] (h) to (f);
            \draw[blueedge] (f) to (g);
            \draw[blueedge] (c) to (d);
            \draw[blueedge] (e) to (d);
            \draw[blueedge] (e) to (f);
            \draw [style={paired_planar}, bend right=15] (3) to (j);
            \draw [style={paired_planar}, bend left=15] (3) to (h);
            \draw [style={paired_planar}, bend left=15, looseness=1.25] (h) to (2);
            \draw [style={paired_planar}, bend left=15] (2) to (j);
            \draw [style={paired_planar}, bend right=15] (2) to (b);
            \draw [style={paired_planar}, bend left=15, looseness=1.25] (2) to (g);
            \draw [style={paired_planar}, bend left] (g) to (1);
            \draw [style={paired_planar}, bend left=15] (1) to (b);

        \end{tikzpicture}

        \caption{A planar embedding of $G \cup \gvf$  with $I = (G, \{g,b,j,h\})$.  The graph $G$ is shown in plain edges, and the \vertexFaceGraph is shown in dashed edges.}
    \label{fig:face_vertex_graph_construction}
\end{figure}

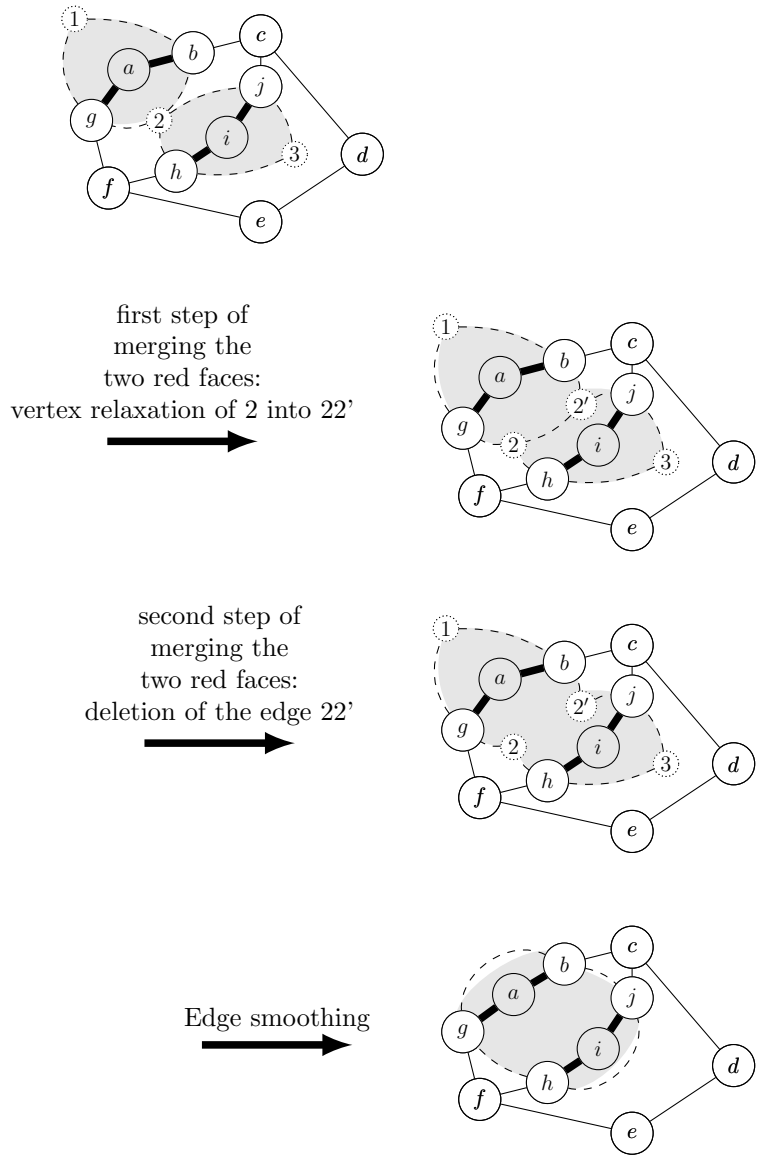
\begin{figure}
	\begin{center}
			\begin{tikzpicture}[
				baseline=0mm,
				scale=0.9, 
				every node/.style={scale=0.7}
				]
		
			\draw[blueface] (-3, -2) rectangle (3, 2.5);
			
				\node [style={variable_node}, inVprime] (b) at (-0.25, 1.25) {$b$};
				\node [style={variable_node}, inVprime] (j) at (0.75, 0.75) {$j$};
				\node [style={variable_node}] (a) at (-1.2, 1) {$a$};
				\node [style={variable_node}] (c) at (0.75, 1.5) {$c$};
				\node [style={variable_node}] (i) at (0.25, 0) {$i$};
				\node [style={variable_node}, inVprime] (h) at (-0.5, -0.5) {$h$};
				\node [style={variable_node}] (f) at (-1.5, -0.75) {$f$};
				\node [style={variable_node}, inVprime] (g) at (-1.75, 0.25) {$g$};
				\node [style={variable_node}] (d) at (2.25, -0.25) {$d$};
				\node [style={variable_node}] (e) at (0.75, -1.25) {$e$};
				\node [style={dual_g_face}] (3) at (1.25, -0.25) {$3$};
				\node [style={dual_g_face}] (2) at (-0.75, 0.25) {$2$};
				\node [style={dual_g_face}] (1) at (-2, 1.75) {$1$};

				\draw[redface] (1.center) to[bend left=20] (b.center) to[bend left=20] (2) to[bend left=20] (g.center) to[bend left=40] cycle;
				
				\draw[redface] (2.center) to[bend left=20] (b.center) to[bend left=20] (2.center) to[bend left=20] (j.center) to[bend left=20] (3.center) to[bend left=20] (h.center) to[bend left=20] cycle;

				\node [style={variable_node}, inVprime] (b) at (-0.25, 1.25) {$b$};
				\node [style={variable_node}, inVprime] (j) at (0.75, 0.75) {$j$};
				\node [style={variable_node}] (a) at (-1.2, 1) {$a$};
				\node [style={variable_node}] (c) at (0.75, 1.5) {$c$};
				\node [style={variable_node}] (i) at (0.25, 0) {$i$};
				\node [style={variable_node}, inVprime] (h) at (-0.5, -0.5) {$h$};
				\node [style={variable_node}] (f) at (-1.5, -0.75) {$f$};
				\node [style={variable_node}, inVprime] (g) at (-1.75, 0.25) {$g$};
				\node [style={variable_node}] (d) at (2.25, -0.25) {$d$};
				\node [style={variable_node}] (e) at (0.75, -1.25) {$e$};
				\node [style={dual_g_face}] (3) at (1.25, -0.25) {$3$};
				\node [style={dual_g_face}] (2) at (-0.75, 0.25) {$2$};
				\node [style={dual_g_face}] (1) at (-2, 1.75) {$1$};

				\draw[rededge] (a) to (g);
				\draw[rededge] (a) to (b);
				\draw[blueedge] (b) to (c);
				\draw[blueedge] (c) to (j);
				\draw[rededge] (j) to (i);
				\draw[rededge] (i) to (h);
				\draw[blueedge] (h) to (f);
				\draw[blueedge] (f) to (g);
				\draw[blueedge] (c) to (d);
				\draw[blueedge] (e) to (d);
				\draw[blueedge] (e) to (f);
				\draw [style={paired_planar}, bend right=15] (3) to (j);
				\draw [style={paired_planar}, bend left=15] (3) to (h);
				\draw [style={paired_planar}, bend left=15, looseness=1.25] (h) to (2);
				\draw [style={paired_planar}, bend left=15] (2) to (j);
				\draw [style={paired_planar}, bend right=15] (2) to (b);
				\draw [style={paired_planar}, bend left=15, looseness=1.25] (2) to (g);
				\draw [style={paired_planar}, bend left] (g) to (1);
				\draw [style={paired_planar}, bend left=15] (1) to (b);
			\end{tikzpicture}

						\hfill
						\fleche{\begin{tabular}{c}
						first step of \\
merging the \\
two red faces:
\\
vertex relaxation of 2 into 22'\end{tabular}}
				\begin{tikzpicture}[
					baseline=0mm,
				scale=0.9, 
				every node/.style={scale=0.7}
				]
				
				\blueface
				
				\node [style={variable_node}, inVprime] (b) at (-0.25, 1.25) {$b$};
				\node [style={variable_node}, inVprime] (j) at (0.75, 0.75) {$j$};
				\node [style={variable_node}] (a) at (-1.2, 1) {$a$};
				\node [style={variable_node}] (c) at (0.75, 1.5) {$c$};
				\node [style={variable_node}] (i) at (0.25, 0) {$i$};
				\node [style={variable_node}, inVprime] (h) at (-0.5, -0.5) {$h$};
				\node [style={variable_node}] (f) at (-1.5, -0.75) {$f$};
				\node [style={variable_node}, inVprime] (g) at (-1.75, 0.25) {$g$};
				\node [style={variable_node}] (d) at (2.25, -0.25) {$d$};
				\node [style={variable_node}] (e) at (0.75, -1.25) {$e$};
				\node [style={dual_g_face}] (3) at (1.25, -0.25) {$3$};
				\node [style={dual_g_face}] (2) at (-1, 0) {$2$};
				\node [style={dual_g_face}] (2') at (0, 0.6) {$2'$};
				\node [style={dual_g_face}] (1) at (-2, 1.75) {$1$};

				\draw[redface] (1.center) to[bend left=20] (b.center) to[bend left=20] (2') to[bend left=10] (j.center) to[bend left=20] (3.center) to[bend left=20] (h.center) to[bend left=20] (2.center) to[bend left=20] (g.center) to[bend left=30] cycle;

				\node [style={variable_node}, inVprime] (b) at (-0.25, 1.25) {$b$};
				\node [style={variable_node}, inVprime] (j) at (0.75, 0.75) {$j$};
				\node [style={variable_node}] (a) at (-1.2, 1) {$a$};
				\node [style={variable_node}] (c) at (0.75, 1.5) {$c$};
				\node [style={variable_node}] (i) at (0.25, 0) {$i$};
				\node [style={variable_node}, inVprime] (h) at (-0.5, -0.5) {$h$};
				\node [style={variable_node}] (f) at (-1.5, -0.75) {$f$};
				\node [style={variable_node}, inVprime] (g) at (-1.75, 0.25) {$g$};
				\node [style={variable_node}] (d) at (2.25, -0.25) {$d$};
				\node [style={variable_node}] (e) at (0.75, -1.25) {$e$};
				\node [style={dual_g_face}] (3) at (1.25, -0.25) {$3$};
				\node [style={dual_g_face}] (2) at (-1, 0) {$2$};
				\node [style={dual_g_face}] (2') at (0, 0.6) {$2'$};
				\node [style={dual_g_face}] (1) at (-2, 1.75) {$1$};

				\draw[rededge] (a) to (g);
				\draw[rededge] (a) to (b);
				\draw[blueedge] (b) to (c);
				\draw[blueedge] (c) to (j);
				\draw[rededge] (j) to (i);
				\draw[rededge] (i) to (h);
				\draw[blueedge] (h) to (f);
				\draw[blueedge] (f) to (g);
				\draw[blueedge] (c) to (d);
				\draw[blueedge] (e) to (d);
				\draw[blueedge] (e) to (f);
				\draw [style={paired_planar}, bend right=15] (3) to (j);
				\draw [style={paired_planar}, bend left=15] (3) to (h);
				\draw [style={paired_planar}, bend left=15, looseness=1.25] (h) to (2);
				\draw [style={paired_planar}, bend left=15] (2') to (j);
				\draw [style={paired_planar}, bend right=15] (2') to (b);
				\draw [style={paired_planar}, bend left=15, looseness=1.25] (2) to (g);
				\draw [style={paired_planar}, bend left] (g) to (1);
				\draw [style={paired_planar}, bend left=15] (1) to (b);
				
				\draw [style={paired_planar}, bend left=15] (2') to (2);
				
			\end{tikzpicture}

			\hfill
			\fleche{\begin{tabular}{c}second step of \\
									merging the \\
									 two red faces:
									 \\
									 deletion of the edge 22'
					\end{tabular}}
	\begin{tikzpicture}[
		baseline=0mm,
	scale=0.9, 
	every node/.style={scale=0.7}
	]
	
	\blueface

	\node [style={variable_node}, inVprime] (b) at (-0.25, 1.25) {$b$};
	\node [style={variable_node}, inVprime] (j) at (0.75, 0.75) {$j$};
	\node [style={variable_node}] (a) at (-1.2, 1) {$a$};
	\node [style={variable_node}] (c) at (0.75, 1.5) {$c$};
	\node [style={variable_node}] (i) at (0.25, 0) {$i$};
	\node [style={variable_node}, inVprime] (h) at (-0.5, -0.5) {$h$};
	\node [style={variable_node}] (f) at (-1.5, -0.75) {$f$};
	\node [style={variable_node}, inVprime] (g) at (-1.75, 0.25) {$g$};
	\node [style={variable_node}] (d) at (2.25, -0.25) {$d$};
	\node [style={variable_node}] (e) at (0.75, -1.25) {$e$};
	\node [style={dual_g_face}] (3) at (1.25, -0.25) {$3$};
	\node [style={dual_g_face}] (2) at (-1, 0) {$2$};
	\node [style={dual_g_face}] (2') at (0, 0.6) {$2'$};
	\node [style={dual_g_face}] (1) at (-2, 1.75) {$1$};

	\draw[redface] (1.center) to[bend left=20] (b.center) to[bend left=20] (2') to[bend left=10] (j.center) to[bend left=20] (3.center) to[bend left=20] (h.center) to[bend left=20] (2.center) to[bend left=20] (g.center) to[bend left=30] cycle;

	\node [style={variable_node}, inVprime] (b) at (-0.25, 1.25) {$b$};
	\node [style={variable_node}, inVprime] (j) at (0.75, 0.75) {$j$};
	\node [style={variable_node}] (a) at (-1.2, 1) {$a$};
	\node [style={variable_node}] (c) at (0.75, 1.5) {$c$};
	\node [style={variable_node}] (i) at (0.25, 0) {$i$};
	\node [style={variable_node}, inVprime] (h) at (-0.5, -0.5) {$h$};
	\node [style={variable_node}] (f) at (-1.5, -0.75) {$f$};
	\node [style={variable_node}, inVprime] (g) at (-1.75, 0.25) {$g$};
	\node [style={variable_node}] (d) at (2.25, -0.25) {$d$};
	\node [style={variable_node}] (e) at (0.75, -1.25) {$e$};
	\node [style={dual_g_face}] (3) at (1.25, -0.25) {$3$};
	\node [style={dual_g_face}] (2) at (-1, 0) {$2$};
	\node [style={dual_g_face}] (2') at (0, 0.6) {$2'$};
	\node [style={dual_g_face}] (1) at (-2, 1.75) {$1$};

	\draw[rededge] (a) to (g);
	\draw[rededge] (a) to (b);
	\draw[blueedge] (b) to (c);
	\draw[blueedge] (c) to (j);
	\draw[rededge] (j) to (i);
	\draw[rededge] (i) to (h);
	\draw[blueedge] (h) to (f);
	\draw[blueedge] (f) to (g);
	\draw[blueedge] (c) to (d);
	\draw[blueedge] (e) to (d);
	\draw[blueedge] (e) to (f);
	\draw [style={paired_planar}, bend right=15] (3) to (j);
	\draw [style={paired_planar}, bend left=15] (3) to (h);
	\draw [style={paired_planar}, bend left=15, looseness=1.25] (h) to (2);
	\draw [style={paired_planar}, bend left=15] (2') to (j);
	\draw [style={paired_planar}, bend right=15] (2') to (b);
	\draw [style={paired_planar}, bend left=15, looseness=1.25] (2) to (g);
	\draw [style={paired_planar}, bend left] (g) to (1);
	\draw [style={paired_planar}, bend left=15] (1) to (b);
\end{tikzpicture}

\hfill
\fleche{Edge smoothing}
	\begin{tikzpicture}[
	baseline=0mm,
	scale=0.9, 
	every node/.style={scale=0.7}
	]
	
	\blueface
	
	\node [style={variable_node}, inVprime] (b) at (-0.25, 1.25) {$b$};
	\node [style={variable_node}, inVprime] (j) at (0.75, 0.75) {$j$};
	\node [style={variable_node}] (a) at (-1, 0.8) {$a$};
	\node [style={variable_node}] (c) at (0.75, 1.5) {$c$};
	\node [style={variable_node}] (i) at (0.25, 0) {$i$};
	\node [style={variable_node}, inVprime] (h) at (-0.5, -0.5) {$h$};
	\node [style={variable_node}] (f) at (-1.5, -0.75) {$f$};
	\node [style={variable_node}, inVprime] (g) at (-1.75, 0.25) {$g$};
	\node [style={variable_node}] (d) at (2.25, -0.25) {$d$};
	\node [style={variable_node}] (e) at (0.75, -1.25) {$e$};

	\draw[redface] (g.center) to[bend left=110] (b.center)  to[bend left=10] (j.center) to[bend left=90] (h.center) to[bend left=20] (g.center) to[bend left=30] cycle;

	\node [style={variable_node}, inVprime] (b) at (-0.25, 1.25) {$b$};
	\node [style={variable_node}, inVprime] (j) at (0.75, 0.75) {$j$};
	\node [style={variable_node}] (a) at (-1, 0.8) {$a$};
	\node [style={variable_node}] (c) at (0.75, 1.5) {$c$};
	\node [style={variable_node}] (i) at (0.25, 0) {$i$};
	\node [style={variable_node}, inVprime] (h) at (-0.5, -0.5) {$h$};
	\node [style={variable_node}] (f) at (-1.5, -0.75) {$f$};
	\node [style={variable_node}, inVprime] (g) at (-1.75, 0.25) {$g$};
	\node [style={variable_node}] (d) at (2.25, -0.25) {$d$};
	\node [style={variable_node}] (e) at (0.75, -1.25) {$e$};

	\draw[rededge] (a) to (g);
	\draw[rededge] (a) to (b);
	\draw[blueedge] (b) to (c);
	\draw[blueedge] (c) to (j);
	\draw[rededge] (j) to (i);
	\draw[rededge] (i) to (h);
	\draw[blueedge] (h) to (f);
	\draw[blueedge] (f) to (g);
	\draw[blueedge] (c) to (d);
	\draw[blueedge] (e) to (d);
	\draw[blueedge] (e) to (f);
	\draw [style={paired_planar}, bend right=60] (h) to (j);
	\draw [style={paired_planar}, bend left=15] (h) to (g);
	\draw [style={paired_planar}, bend left=15] (b) to (j);
	\draw [style={paired_planar}, bend left=60] (g) to (b);
\end{tikzpicture}
	\end{center}
	\caption{Execution of the algorithm. The algorithm starts with an embedding of $G \cup K_I$. The algorithm keeps track of a graph $K_I(i)$ (dotted and dashed), and keeps merging $K_I(i)$-faces that are of the same color. Once there is exactly two remaining $K_I(i)$-faces, we perform edge smoothing.\label{figure:algorithmmergingfacesthensmoothing}}
\end{figure}

We say that a cycle $\pi$ \emph{separates} two sets of edges $E'$ and $E''$ in an embedding if all edges in $E'$ are in the interior of $\pi$
and all edges in $E''$ are in the exterior, or vice versa.

\begin{algorithm}
    \caption{$V'$-cycle computation}\label{alg:constructingVprimecycle}
    \begin{algorithmic}[1]
        \Require an instance $I = (G, V')$ with a graph $G$, a subset $V'$ of vertices, a  kite-graph $K_I$ for $I$
        \Ensure $V'$-cycle $\genericcycle$ such that $(V, E \uplus \genericcycle)$ is planar
        \State Compute a planar embedding of $G\cup\gvf$, and set $\gvf(0) = \gvf$ \label{line:planaremb}
        \State $i \gets 0$
        \While{$\gvf(i)$ has strictly more than two faces}
            \State Select two different $\gvf(i)$-faces $\gvfface_1$ and $\gvfface_2$ of the same color incident to some $\gvf(i)$-vertex~$f$. \label{line:f1f2}
            \State Obtain $G \cup \gvf(i+1)$ by applying vertex relaxation to $f$ 
                (see \Cref{def:vertex_relaxation}), thus merging $\gvfface_1$ with $\gvfface_2$ \label{line:merging}
            \State $i \gets i+1$ 	
        \EndWhile
    \State	\Return $V'$-cycle $\genericcycle$ obtained from $\gvf(i)$ by applying edge smoothing (\Cref{def:edge_smoothing}) at all vertices not in $V'$\label{line:edgesmoothing}
    \end{algorithmic}
\end{algorithm}

The algorithm to compute a $V'$-cycle is presented in \Cref{alg:constructingVprimecycle}. Here we refer to two operations defined using the dual graph. We explain these intuitively, and give the formal definitions in the appendix.
Vertex relaxation applies to two faces $\gvfface_1$ and $\gvfface_2$ incident to a common vertex~$f$, and consists in creating an edge incident to both faces.
This operation appears in the second step of \Cref{figure:algorithmmergingfacesthensmoothing}, and is defined formally 
in \Cref{def:vertex_relaxation}.
Furthermore, we use the edge smoothing operation which consists in removing a vertex with degree 2 while joining its two neighbors.
For example, applying edge smoothing to all white vertices below result in the graph on the right.

\begin{center}
    \begin{tikzpicture}[yscale=0.5]
        \foreach \a in {0,...,16} {
            \node[draw,circle, inner sep=0.5mm] (u\a) at ({\a*30}:1cm) {};
        }
            \foreach \a in {0,...,8} {
            \node[draw,circle, inner sep=0.5mm, inVprime] at ({\a*60}:1cm) {};
        }
        \foreach \a[evaluate={\ai=int(\a +1);}] in {0,...,15} {
            \draw (u\a) -- (u\ai);
        }
    \end{tikzpicture}
    \tikz[baseline=-6mm]{\draw[line width=1mm, -latex] (0, 0) -- (1, 0);}
    \begin{tikzpicture}[yscale=0.5]
    \foreach \a in {0,...,8} {
        \node[draw,circle, inner sep=0.5mm,inVprime] (u\a) at ({\a*60}:1cm) {};
    }
    \foreach \a[evaluate={\ai=int(\a+1);}] in {0,...,7} {
        \draw (u\a) -- (u\ai);
    }
\end{tikzpicture}
\end{center}
This is also applied in the last step of \Cref{figure:algorithmmergingfacesthensmoothing}.

\begin{theorem}
    Given a $I = (G,V')$ a planar augmentation instance. If there exists a {\vertexFaceGraph} $\gvf$ for $I$, then $G$ can be augmented with a cycle~$\acycle$ such that $G \cup \acycle$ is planar.
    Moreover, $\acycle$ separates the red and blue edges. If $\gvf$ is given, then the cycle $\acycle$ can be computed in polynomial time in $|G \cup \gvf|$.
    \label{th:construct_cycle_poly_time}
\end{theorem}
    We prove this theorem in rest of this section.

    First, note that since $\gvf$ is planar and Eulerian, by Property \ref{prop:bipartite_dual_eulerian}, its dual graph $\gvf^* = (F^*, E^*)$ is bipartite, i.e. $\gvf^*$ has a 2-coloring.

    We claim that Algorithm~\ref{alg:constructingVprimecycle} constructs a $V'$-cycle.
    The algorithm start by computing a planar embedding of $G\cup\gvf$ (and at the same time of $\gvf^*$).
    Then it merges faces, which produce the sequences $\gvf = \gvf(0), \gvf(1), \dots$, and $\gvf^* = \gvf^*(0), \gvf^*(1), \dots$.
    An example is given in Figure~\ref{figure:algorithmmergingfacesthensmoothing} where faces $\gvfface_1 = 1b2g$ and $\gvfface_2 = 2j3h$ are merged.
 
    The last step (Line~\ref{line:edgesmoothing}) consists in removing vertices that are not in $V'$
    as in the last step of Figure~\ref{figure:algorithmmergingfacesthensmoothing} where $1, 2, 3$ and $2'$ are removed.    
 \begin{claim}
 	\Cref{alg:constructingVprimecycle} runs in polynomial time in the size of $G$.
 	\label{fact:constructingVprimecyclecomplexity}
 \end{claim} 
 \begin{proof}
    At	\Cref{line:planaremb}  we compute a planar embedding of $\gvf$ (and at the same time of $\gvf^*$) in polynomial time (Property~\ref{prop:dual_given_embedding_poly_time}).
    Finding the two faces on \Cref{line:f1f2} can be done by inspecting all $\gvf(i)$-vertices $v$, in linear time.
    
    Let us explain how merging two faces $\gvfface_1$ and $\gvfface_2$ at \Cref{line:merging} can be performed in polynomial time. Technically merging two faces $\gvfface_1$ and $\gvfface_2$ consists in two operations.

    \begin{itemize}
    	\item 
        Apply vertex relaxation (\Cref{def:vertex_relaxation}) in $\gvf(i)$ (by \Cref{prop:vertex_relaxation_duality_edge_addition}),
        which consists in adding the edge $\gvfface_1\gvfface_2$ in $\gvf^*(i)$.
        Let $f$ be the common $\gvf(i)$-vertex of this vertex relaxation operation ($f=2$ in \Cref{figure:algorithmmergingfacesthensmoothing}),
        and note that $f$ corresponds to a face of $G$. 
        Let $f'$ be the new $\gvf(i)$-vertex from the vertex relaxation (in \Cref{figure:algorithmmergingfacesthensmoothing}, $f'$ is 2').
        We then obtain new embedding $\gvf(i)'$ where vertex $f$ have been relaxed into the edge $ff'$.
    	
    	\item Remove the edge $\gvf(i)'$-edge $ff'$ that corresponds to an edge contraction of $\gvfface_1\gvfface_2$ in $\gvf^*(i)$ by Property~\ref{prop:edge_contraction}.
    \end{itemize}
    
    Thanks to Property~\ref{prop:poly_time_embedding_modif}, the merging of two faces (\Cref{line:merging}) can be performed in polynomial time.
 \end{proof}

\begin{invariant}
	 $\gvf(i)$ is a kite-graph.
	 \label{invariant:kitegraph}
\end{invariant}
\begin{proof}
  Since $\gvf = \gvf(0)$ we know that $\gvf(0)$ is a kite-graph.
  
  The edge $ff'$ separates two faces $\gvfface_1$ and $\gvfface_2$ such that $\gvfface_1 \neq \gvfface_2$.
  Since we remove $ff'$, by Property \ref{prop:operations_preserve_planar_and_conn} $\gvf(i+1)$ is connected. 
  The faces $\gvfface_1$ and $\gvfface_2$ are of the same color. After merging $\gvf^*(i)$-vertices $\gvfface_1$ and $\gvfface_2$ into $\gvfface_{1,2}$, we get the dual $\gvf^*(i+1)$
  which is bipartite: in fact, $\gvfface_{1,2}$ has the same color as $\gvfface_1$ and $\gvfface_2$, while all other $\gvf^*(i+1)$-vertices keep their colors
  as in $\gvf^*(i)$.
  Hence by Property \ref{prop:bipartite_dual_eulerian}, $\gvf(i+1)$ is Eulerian.
  The vertex $f$ is in $F'$ and not in $V'$ meaning that all the vertices $v \in V'$ are of degree $2$ in $\gvf(i+1)$.
  Hence $\gvf(i+1)$ is a kite-graph.
\end{proof}

The following invariant says that the 2-partition of $G$-edges remains the same.

\begin{invariant}
    $\Pi_{G,\gvf(i)}$= $\Pi_{G,\gvf}$.
	\label{fact:vertexspliting}
\end{invariant}
\begin{proof}
    Initially $\Pi_{G,\gvf(0)}$= $\Pi_{G,\gvf}$.
Suppose $\Pi_{G,\gvf(i)}$= $\Pi_{G,\gvf}$.
	Now consider the merge of face $\gvfface_1$ and $\gvfface_2$. First we have a vertex relaxation of $f$ into $ff'$. By \Cref{prop:location_after_vertex_relaxation}, edges of $G$ in $\gvfface_1$ (resp. $\gvfface_2$) remain in the new face $\gvfface'_1$ (resp. $\gvfface'_2$). Then, an edge deletion of $ff'$ to get a face $\gvfface_{1, 2}$. Edges of $G$ in $\gvfface_1$ and $\gvfface_2$ are now in face $\gvfface_{1, 2}$ thanks to \Cref{prop:location_after_edge_deletion}. The other $G$-edges remain in the same faces. 

	As said, the graph $K^*(i+1)$ is obtained from  $K^*(i)$ by merging $K^*(i)$-vertices $\gvfface_1, \gvfface_2$ which were of the same color into a $K^*(i+1)$-vertex $\gvfface_{1, 2}$. We color $\gvfface_{1, 2}$ with the color of $\gvfface_1$ while the other vertices keep their colors.
	
	Finally, edges in $G$ keep their colors.
	Thus $\Pi_{G,\gvf(i+1)}$= $\Pi_{G,\gvf}$.
\end{proof}

\begin{claim}
	Algorithm \ref{alg:constructingVprimecycle} terminates.
\end{claim}
\begin{proof}
	The number of $K(i)$-faces is strictly decreasing.
\end{proof}

\begin{claim}
	The selection of faces at Line~\ref{line:f1f2} always succeed.
    \label{fact:algorithm_always_terminate_if_g_planar_eulerian}
\end{claim}
\begin{proof}
	\newcommand{\gvfi}{\gvf(i)}
	\newcommand{\gvfistar}{\gvf^*(i)}
    Suppose that $\gvfi$ has strictly more than two faces. Thus, $\gvfistar$ %
    has strictly more than two vertices.
    
    By contradiction, suppose that there are no pairs of faces in $\gvfi$ of the same color, sharing a vertex.
    In the dual graph, this means that there are no pairs of $\gvfistar$-vertices $\gvfface_1$ and $\gvfface_2$ of the same color and sharing a common face $f$ in $\gvfistar$.
    This implies that all the $\gvfistar$-faces $f$ contain only two $\gvfistar$-vertices (otherwise, there are two $\gvfistar$-vertices of the same color sharing a common face).
    It follows that all the faces $f$ of $\gvfistar$ contain two edges.
    Since $\gvfi$ is connected by Property \ref{prop:if_g_conn_then_g_dual_dual_is_g}, the dual of $\gvfistar$ is $\gvfi$.
    Hence all $\gvfi$-vertices are of degree 2.
    We conclude that $\gvfi$ is a cycle. As $\gvfi$ is Eulerian (because $\gvfi$ is a kite-graph, see \Cref{invariant:kitegraph}), by \Cref{prop:planar_eulerian_g_two_faces_is_cycle}, $\gvfi$ has two faces. Contradiction.
\end{proof}

\begin{claim}[Soundness]
   \Cref{alg:constructingVprimecycle} returns a $V'$-cycle $\genericcycle$ such that $(V, E \uplus \genericcycle)$ is planar. Moreover, $\acycle$ separates the red and blue edges.
\end{claim}

\begin{proof}
    At the end, $\gvf(i)$ has two faces. By \Cref{prop:planar_eulerian_g_two_faces_is_cycle}, $\gvf(i)$ is a cycle.
    The final step consists in edge smoothing (\Cref{def:edge_smoothing}) to remove vertices not in $V'$ from $\gvf(i)$: we get a $V'$-cycle $\genericcycle$. 
    
    By \Cref{fact:vertexspliting}, $\Pi_{G,\gvf(i)}= \Pi_{G,\gvf}$. The cycle $K(i)$ respects the partition $\Pi_{G,\gvf}$ meaning the red $G$-edges w.r.t. $K_I$ are inside while the blue ones are outside, up to color renaming. So the $V'$-cycle $\pi$ also respects this partition.
    As $G \cup \gvf(i)$ is planar so is $G \cup \genericcycle$.
\end{proof}

This concludes the proof of \Cref{th:construct_cycle_poly_time}.

\subsection{A Feasible Class of Instances}

\newcommand{\specialInst}{dually connected matching\xspace}
\newcommand{\edgedeletion}{{-}}

In this section we introduce a class of instances for which a $V'$-cycle  $\acycle$ exists and can be computed in polynomial time.
These instances are those which have a \emph{\specialInst}.

\begin{definition}[Dually Connected Matching]
    Let $I = (G = (V,E),V')$ be a cycle augmentation instance.
    A subset $E'\subseteq E \cap V' \times V'$ of edges is a \emph{dually connected matching} for $I$ if:
    \begin{enumerate}
    	\item $E'$ is a $V'$-perfect matching: for all $v \in V'$ there is an edge $e \in E'$ incident to $v$, and for all $e, e' \in E'$, if $e \neq e'$ then $e \cap e' = \emptyset$.
        \item $E'$ is dually connected: the graph $G_{E'}^*$ obtained from $G^*$ by only keeping the dual edges of edges in
$E'$ is connected\footnote{i.e. 
          $G^* \edgedeletion \bijectionedgetodualG(E \setminus E')$
           is connected where
           $\bijectionedgetodualG(E \setminus E') := \set{\bijectionedgetodualG(e) \mid e \in (E \setminus E')}$.
           See \Cref{def:edge_deletion} for the definition of the edge deletion operation $\edgedeletion$.}.
        
    \end{enumerate}
    \label{def:SepcialInst}
\end{definition}

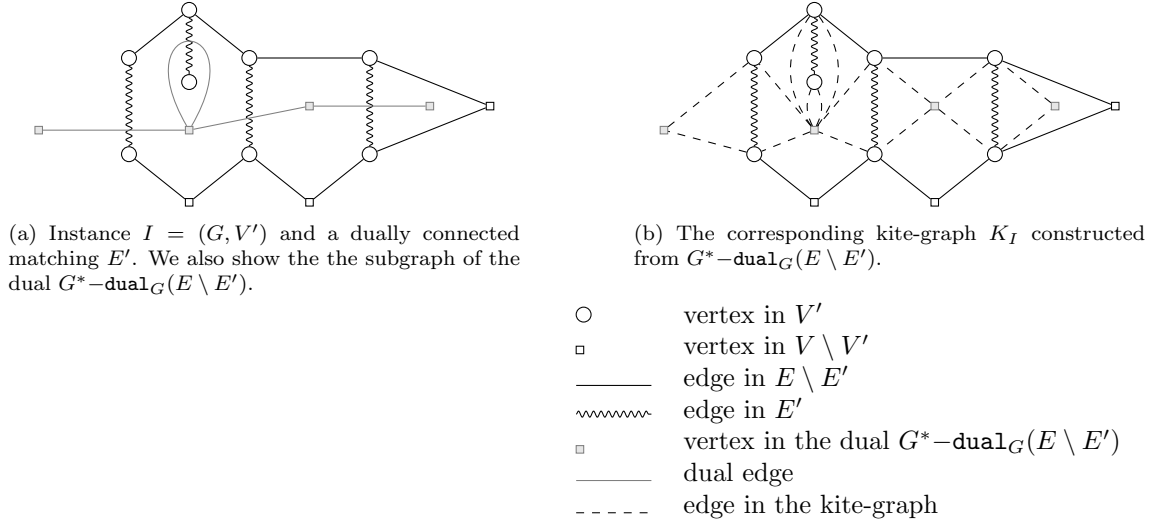
\begin{figure}[h]
    \centering
        \begin{subfigure}[t]{0.45\textwidth}
            \centering

            \begin{tikzpicture}[scale=0.8, yscale=0.8]
            \node[inVprime] (v0) at (-1, 2) {};
            \node[inVprime] (v1) at (-2, 1) {};
            \node[inVprime] (v2) at (-2, -1) {};
            \node[node] (v3) at (-1, -2) {};
            \node[inVprime] (v4) at (0, -1) {};
            \node[inVprime] (v5) at (0, 1) {};
            \node[inVprime] (v6) at (2, 1) {};
            \node[node] (v7) at (4, 0) {};
            \node[inVprime] (v8) at (2, -1) {};
            \node[node] (v9) at (1, -2) {};
            \node[inVprime] (v10) at (-1, 0.5) {};
            
            \node[dual] (v11) at (3, 0) {};
            \node[dual] (v12) at (1, 0) {};
            \node[dual] (v13) at (-1, -0.5) {};
            \node[dual] (v14) at (-3.5, -0.5) {};
            
            \draw (v1) edge (v0);
            \draw (v0) edge (v5);
            \draw (v2) edge (v3);
            \draw (v3) edge (v4);
            \draw (v4) edge (v9);
            \draw (v9) edge (v8);
            \draw (v8) edge (v7);
            \draw (v7) edge (v6);
            \draw (v6) edge (v5);
            
            \draw (v1) edge[paired] (v2);
            \draw (v5) edge[paired] (v4);
            \draw (v6) edge[paired] (v8);
            \draw (v0) edge[paired] (v10);
            
            \draw[dual] (v12) edge (v13);
            \draw[dual] (v13) edge (v14);
            \draw[dual] (v13) edge[loop,looseness=100,in=115, out=65] (v13);
            \draw[dual] (v11) edge (v12);
        \end{tikzpicture}

        \caption{Instance $I = (G, V')$ and a dually connected matching $E'$. We also show the the subgraph of the dual $G^* \edgedeletion \bijectionedgetodualG(E \setminus E')$.}
        \label{fig:dualg_edge_coll}
    \end{subfigure}
    \hfill
    \begin{subfigure}[t]{0.45\textwidth}
        \centering
        \begin{tikzpicture}[scale=0.8, yscale=0.8]
          \node[inVprime] (v0) at (-1, 2) {};
          \node[inVprime] (v1) at (-2, 1) {};
          \node[inVprime] (v2) at (-2, -1) {};
          \node[node] (v3) at (-1, -2) {};
          \node[inVprime] (v4) at (0, -1) {};
          \node[inVprime] (v5) at (0, 1) {};
          \node[inVprime] (v6) at (2, 1) {};
          \node[node] (v7) at (4, 0) {};
          \node[inVprime] (v8) at (2, -1) {};
          \node[node] (v9) at (1, -2) {};
          \node[inVprime] (v10) at (-1, 0.5) {};
           
           \node[dual] (v11) at (3, 0) {};
           \node[dual] (v12) at (1, 0) {};
           \node[dual] (v13) at (-1, -0.5) {};
           \node[dual] (v14) at (-3.5, -0.5) {};
          
           \draw (v1) edge (v0);
           \draw (v0) edge (v5);
           \draw (v2) edge (v3);
           \draw (v3) edge (v4);
           \draw (v4) edge (v9);
           \draw (v9) edge (v8);
           \draw (v8) edge (v7);
           \draw (v7) edge (v6);
           \draw (v6) edge (v5);
           
           \draw (v1) edge[paired] (v2);
           \draw (v5) edge[paired] (v4);
           \draw (v6) edge[paired] (v8);
           \draw (v0) edge[paired] (v10);

           \draw[kite] (v6) edge (v12);
           \draw[kite] (v12) edge (v8);
           \draw[kite] (v8) edge (v11);
           \draw[kite] (v12) edge (v5);
           \draw[kite] (v12) edge (v4);
           \draw[kite] (v4) edge (v13);
           \draw[kite] (v5) edge (v13);
           \draw[kite] (v13) edge[bend left=20] (v10);
           \draw[kite] (v13) edge[bend right=20] (v10);
           \draw[kite] (v13) edge[bend right=30] (v0);
           \draw[kite] (v13) edge[bend left=30] (v0);
           \draw[kite] (v11) edge (v6);
           \draw[kite] (v13) edge (v2);
           \draw[kite] (v2) edge (v14);
           \draw[kite] (v14) edge (v1);
           \draw[kite] (v13) edge (v1);
        \end{tikzpicture}
        \caption{The corresponding \vertexFaceGraph $K_I$ constructed from $G^* \edgedeletion \bijectionedgetodualG(E \setminus E')$.}
        \label{fig:face_embedded_graph_after_vertex_contraction}
    \end{subfigure}

\hfill
{\footnotesize
\begin{tabular}{ll}
\tikz{\node[inVprime] {};} & vertex in $V'$ \\
\tikz{\node[node] {};} & vertex in $V \setminus V'$ \\
\tikz{\draw (0, 0) edge (1, 0);} & edge in $E \setminus E'$ \\
\tikz{\draw (0, 0) edge[paired] (1, 0);} & edge in $E'$ \\
\tikz[baseline=0mm]{\node[dual] {};} & vertex in the dual $G^* \edgedeletion \bijectionedgetodualG(E \setminus E')$ \\
\tikz{\draw (0, 0) edge[dual] (1, 0);} & dual edge \\
\tikz[baseline=0mm]{\draw (0, 0) edge[kite] (1, 0);} & edge in the kite-graph
\end{tabular}
}
    \caption{An instance, a dually connected matching, and the corresponding \vertexFaceGraph.}
    \label{fig:edge_coll_example}
\end{figure}

In \Cref{def:SepcialInst}, point 1 says that $E'$ is a $V'$-perfect matching meaning that $E'$ touch all vertices in $V'$, and edges in $E'$ do not share common vertices. Point 2 says that when the dual graph $G^*$ remains connected even if we only keep the dual edges of edges in $E'$ (recall that a dual edge of $e$ is the edge linking the two faces touching $e$). \Cref{fig:dualg_edge_coll} gives an example of a dually connected matching.

\begin{theorem}
    Let $I = (G=(V,E),V')$. If $I$ has a \specialInst $E'$ %
     then a $V'$-cycle $\acycle$ can be computed in poly-time that separates 
      $E'$ from~$E \setminus E'$.
    \label{th:pi_constructible_for_edge_coll_instance}
\end{theorem}
\begin{proof}
The proof consists in building an embedding of a kite-graph $\gvf$.
    Let $G_{E'}^* := G^* \edgedeletion \bijectionedgetodualG(E \setminus E')$.
    Since $E'$ is a \specialInst, by \Cref{def:SepcialInst}, $G_{E'}^*$ is planar and connected. 
    Consider the planar 
    embeddings $\Gamma_{G_{E'}^*}$ and $\Gamma_G$ (Definition \ref{def:planar_embedding}).
    
    \paragraph{Definition of the graph $K_I$}
    The set of vertices of $K_I$ is $V' \cup \set{\text{vertices of $G_{E'}^*$}}$. Edges of $K_I$ are obtained as follows.
    Each edge $FF'$ in $G_{E'}^*$ corresponds to an edge $uv \in E'$. 
    For each edge $FF'$ in $G_{E'}^*$, the four edges $e_1 = Fu,  e_2 = uF',
        e_3 = Fv, e_4 = vF'$ are in $K_I$.

     \paragraph{Definition of the embedding $\Gamma_{K_I}$}   
        The embedding $\Gamma_{K_I}$ of $K_I$ is obtained from $\Gamma_{G_{E'}^*}$. We first add the edges $e_1, e_2, e_3, e_4$ for each edge $FF'$ 
    (see the operation of edge addition on embedding, \Cref{def:edge_addition}).
    Those edges $e_1, e_2, e_3, e_4$ form two faces in $\Gamma_{G_{E'}^*}'$:
    $\gvfface^1 = (e_1,e_2,FF')$ and $\gvfface^2 = (e_4,e_3,FF')$.
    Since $FF'$ touches two different faces $\gvfface^1$ and $\gvfface^2$, we can remove $FF'$ from $\Gamma_{G_{E'}^*}'$ (\Cref{def:edge_deletion}).
    This deletion implies that $\gvfface^1$ and $\gvfface^2$ are merged into the face $\gvfface_{uv} = (e_1,e_2,e_3,e_4)$
     (Property \ref{prop:edge_contraction}). We obtain an embedding of $\Gamma_{K_I}$.
    Since our edge addition and edge deletion preserve planarity and connectivity (Property \ref{prop:operations_preserve_planar_and_conn}), $\Gamma_{K_I}$ is planar and connected (see for instance the Figure~\ref{fig:face_embedded_graph_after_vertex_contraction}).

    \paragraph{Proof that $K_I$ is a kite-graph}
    The graph $K_I$ is bipartite: edges in $K_I$ are between a vertex in $V'$ and a vertex of $G^*_{E'}$. 
    
    Consider a vertex in $V'$. 
    By \Cref{def:SepcialInst}, each vertex $v$ in $V'$ either appears in some edge in $E'$ and then is linked to two $G$-faces $F$ and $F'$. So $\degree{K_I}(v) = 2$ (point 3 in \Cref{def:kitegraph}).
    
    Moreover a vertex $F$ is of even degree since $F$ is linked to one or multiple pairs of $u$ and $v$ such that $uv \in E'$. Thus $K_I$ is Eulerian (point 2 in \Cref{def:kitegraph}).

    $K_I \cup G$ is obtained from $G$ by adding fresh vertices $F, F'$ etc. located inside $G$-faces, and edges from vertices in $V'$ to these fresh vertices. So
    $K_I \cup G$ is planar.

    \paragraph{About colors}
    Recall that $K_I$ is a kite-graph. By~\Cref{prop:bipartite_dual_eulerian}, $K^*_I$ is bipartite hence 2-colorable. Consider a coloring of $K^*_I$. Let us prove that $E'$ are located in red faces while $E \setminus E'$ are in blue faces (up to renaming the colors).
    
    Given $uv \in E'$, let $\gvfface_{uv}$ be the $K_I$-face in which the edge $uv$ is located. Let us prove that $\gvfface_{uv}$ are all of the same color -- let say red.
    
    Suppose that there exists $K_I$-faces $\gvfface_{uv}$ and $\gvfface_{u'''v'''}$ of different colors with $uv, u'''v''' \in E'$. As $G^*_{E'}$ is connected (see~\Cref{fig:dualg_edge_coll}), there is a path linking the dual edge of $uv$ to the dual edge of $u'''v''''$:

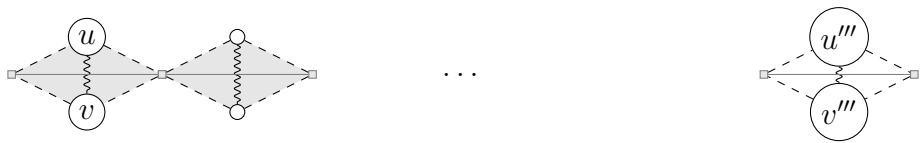
    \newcommand\drawdiamond[3]{
    	\draw[#3] (0, 0) -- (1, 0.5) -- (2, 0) -- (1, -0.5) -- cycle;
    	\node[inVprime] (u) at (1, 0.5) {$#1$};
    	\node[inVprime] (v) at (1, -0.5) {$#2$};
    	\node[dual] (a) at (0, 0) {};
    	\node[dual] (b) at (2, 0) {};
    	\draw[paired] (u) -- (v);
    	\draw[dual] (a) -- (b);
    	\draw[kite] (u) to (b) to (v) to (a) to (u);
    }
    
    \begin{center}
    \begin{tikzpicture}
    	\drawdiamond uv{redface}
    	\begin{scope}[xshift=2cm]
    	\drawdiamond {}{}{redface}
       	\node at (4, 0) {$\dots$};
       	\end{scope}
   		 	    	
    	\begin{scope}[xshift=10cm]
    	\drawdiamond {u'''}{v'''}{blueface}
       	\end{scope}
    \end{tikzpicture}
    \end{center}
     Along this path, there are two consecutive dual edges of $u'v'$ and $u''v''$ in $E'$ with $\gvfface_{u'v'}$ and $\gvfface_{u''v''}$ of different colors:
     
    \begin{center}
     \begin{tikzpicture}
     	\drawdiamond {u'}{v'}{redface}
     	\begin{scope}[xshift=2cm]
     	\drawdiamond {u''}{v''}{blueface}
       	\end{scope}
     \end{tikzpicture}
     \end{center}    
     
      The faces $\gvfface_{u'v'}$ and $\gvfface_{u''v''}$ are around a common vertex $F$. We now consider the faces around $F$ in clockwise-order. There are three consecutive faces around $F$ in the clockwise order of the form $\gvfface_{u^1v^1}$, $\gvfface_{other}$, $\gvfface_{u^2v^2}$ with $\gvfface_{u^1v^1}$ and $\gvfface_{u^2v^2}$ of different colors with $u^1v^1, u^2v^2 \in E'$, and $\gvfface_{other}$ is a $K_I$-face in which no edge of $E'$ is located\footnote{Consecutive $K_I$-faces of the form $\gvfface_{u^1v^1}$ and $\gvfface_{u^2v^2}$ in the clockwise order is impossible because $E'$ is a matching.}. W.l.o.g. $\gvfface_{u^1v^1}$ is red and $\gvfface_{u^2v^2}$ is blue:
      W.l.o.g. $\gvfface_{other}$ is red.
           
          
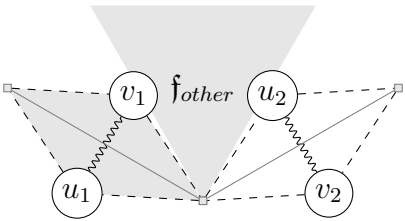
\begin{center}
           \begin{tikzpicture}[scale=1.5]
           \draw[redface] (0, 0) -- (120:2cm) -- (60:2cm) -- cycle;
           \node at (0, 1) {$\gvfface_{other}$};
           	\begin{scope}[rotate=150]
           	\drawdiamond {u_1}{v_1}{redface}
           	\end{scope}
           	\begin{scope}[rotate=30]
           	\drawdiamond {u_2}{v_2}{blueface}
           	\end{scope}
           \end{tikzpicture}
           \end{center}

        Contradiction because $\gvfface_{u^1v^1}$ and $\gvfface_{other}$ are of the same color.
    We conclude that all $\gvfface_{e}$ with $e \in E'$ are all red.

    Thus edges in $E'$ are red while those in $E \setminus E'$ are blue. By \Cref{th:construct_cycle_poly_time}, a $V'$-cycle~$\pi$ is constructible in polynomial time such that $\pi$ separates $E'$ from $E \setminus E'$.
\end{proof}

We are now ready to present our results on the hardness of reconfigurations problems in the next section.
We will prove the PSPACE-hardness of the linear-literal planar 3-SAT and linear planar 3-SAT reconfiguration problems
crucially applying \Cref{th:pi_constructible_for_edge_coll_instance}.

\section{Background on Reconfiguration Problems}
\label{sec:background-reconfiguration}

\subsection{\NCL (NCL)}
\newcommand{\edgeweighttwo}[2]{\tikz[baseline=-1mm]{
		\node[inner sep=0.1mm] (a) at (0.8, 0) {$#1$};
		\draw (0, 0) edge[two weig edge] node[above] {$#2$} (a);}
	}
	
\newcommand{\edgeweightone}[2]{\tikz[baseline=-1mm]{
		\node[inner sep=0.1mm] (a) at (0.8, 0) {$#1$};
		\draw (0, 0) edge[one weig edge] node[above] {$#2$} (a);}
}
	
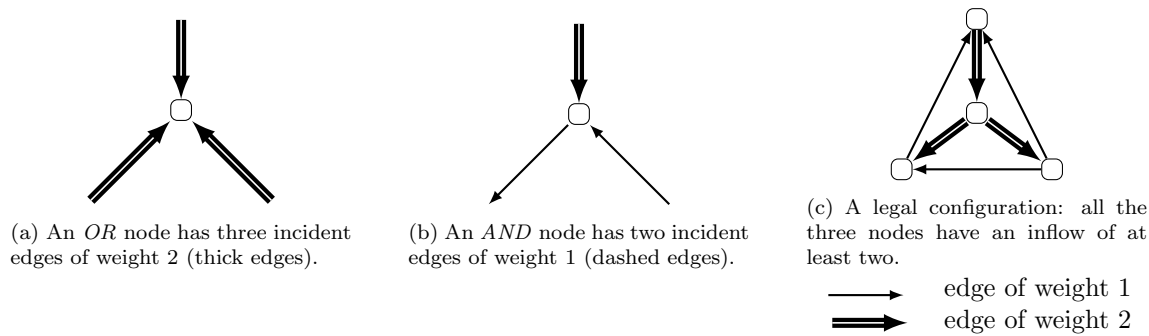
\begin{figure}[h]
    \centering

    \begin{subfigure}[b]{0.30\linewidth}
        \centering
        
        \begin{tikzpicture}[
            scale=0.8, 
            every node/.style={fill=white, scale=1.0},
        ]
            \pgfdeclarelayer{background}
            \pgfdeclarelayer{foreground}
            \pgfsetlayers{background,main,foreground}

            \node[nclnode] (n1) at (0, 0) {};

            \begin{pgfonlayer}{background}
                \draw[two weig edge] (0,1.5) -- (n1);
                \draw[two weig edge] (1.5,-1.5) -- (n1);
                \draw[two weig edge] (-1.5,-1.5) -- (n1);
            \end{pgfonlayer}
        \end{tikzpicture}

        \caption{An \emph{OR} node has three incident edges of weight 2 (thick edges).}
    \end{subfigure}
    \hfill
    \begin{subfigure}[b]{0.30\linewidth}
        \centering

        \begin{tikzpicture}[
            scale=0.8, 
            every node/.style={fill=white, scale=1.0},
        ]
            \pgfdeclarelayer{background}
            \pgfdeclarelayer{foreground}
            \pgfsetlayers{background,main,foreground}

            \node[nclnode] (n1) at (0, 0) {};

            \begin{pgfonlayer}{background}
                \draw[two weig edge] (0,1.5) -- (n1);
                \draw[one weig edge] (1.5,-1.5) -- (n1);
                \draw[one weig edge] (n1) -- (-1.5,-1.5) ;
            \end{pgfonlayer}
        \end{tikzpicture}
        \caption{An \emph{AND} node has two incident edges of weight 1 (dashed edges).}
    \end{subfigure}
    \hfill
    \begin{subfigure}[b]{0.30\linewidth}
        \centering

        \begin{tikzpicture}[
            scale=1.0, 
            every node/.style={fill=white, scale=1.0},
        ]
            \pgfdeclarelayer{background}
            \pgfdeclarelayer{foreground}
            \pgfsetlayers{background,main,foreground}

            \node[nclnode] (and_top) at (0, 1)  {};
            \node[nclnode] (and_bot_right) at (1, -1)  {};
            \node[nclnode] (and_bot_left) at (-1, -1) {};
            \node[nclnode] (or_center) at (0, -0.25) {};

            \begin{pgfonlayer}{background}
                \draw[one weig edge] (and_bot_right) -- (and_top);
                \draw[one weig edge] (and_bot_right) -- (and_bot_left);
                \draw[one weig edge] (and_bot_left) -- (and_top);

                \draw[two weig edge] (or_center) -- (and_bot_left);
                \draw[two weig edge] (or_center) -- (and_bot_right);
                \draw[two weig edge] (and_top) -- (or_center);
            \end{pgfonlayer}
        \end{tikzpicture}
   
        \caption{A legal configuration: all the three nodes have an inflow of at least two.}
    \end{subfigure}

\hfill
\footnotesize
    \begin{tabular}{ll}
	\tikz{\draw[one weig edge] (0, 0) -- (1, 0);} & edge of weight 1
	\\
	\tikz{\draw[two weig edge] (0, 0) -- (1, 0);} & edge of weight 2
\end{tabular}

    \caption{Nondeterministic constraint logic (NCL). A constraint graph contains OR nodes (a) and AND nodes (b). A constraint graph is shown in (c).}
    
    \label{fig:nc_example}
\end{figure}

In nondeterministic constraint logic, a \emph{constraint graph} is an undirected graph. W.l.o.g. (see \cite{gamesNCLthesis}) we suppose that a constraint graph is planar, cubic\footnote{The degree of any node is 3.} and that each edge has a weight of either 1 or 2, and that there are only two types of nodes: AND nodes that have two incident edges of weight 1 and one of weight 2, and OR with three incident edges of weight 2 (see Figure~\ref{fig:nc_example}, (a) and (b)). An edge of weight 1 (resp. 2) is denoted by $\edgeweightone{}{}$ (resp. $\edgeweighttwo{}{}$). 

A \emph{configuration} $c$ of such a graph gives an orientation to each edge. A configuration $c$ is \emph{legal} when it respects the following flow constraint: the sum of the weights of edges pointing to a node $v$ is at least $2$, for all nodes $v$ (see Figure~\ref{fig:nc_example}(c)).
More precisely if $u$ is an \emph{OR} node, then at least one edge must be pointing toward $u$: \edgeweighttwo{u}{} in $c$.
Else, if $u$ is an \emph{AND} node, either two edges of weight $1$ point towards $u$, or one edge of weight $2$ points toward $u$: (\edgeweightone{u}{} and \edgeweightone{u}{}) or \edgeweighttwo{u}{\xspace}.

\newcommand{\rightarrowflip}[1]{\rightsquigarrow^{\text{flip}}_{#1}}
\newcommand{\rightarrowflipvar}[2]
{\rightsquigarrow^{\text{flip(\ensuremath{#1})}}_{#2}}
\newcommand{\rightarrowflips}[1]{\rightsquigarrow^{\text{flips}}_{#1}}

Given a constraint graph $G$, we say that  $t$ is \emph{reachable} from $s$, denoted by $s \rightarrowflips G t$, if  there is a sequence of legal configurations $\langle c_1,\dots,c_n \rangle$ where each $c_{i+1}$ is obtained from $c_i$ by a single edge flip\footnote{i.e. $c_i$ and $c_{i+1}$ gives the same orientation to all edges, except one for which the orientation is flipped} and $c_1 = s$, $c_n = t$.

\begin{definition}[\ctoc Problem (C2C)]
     Given a constraint graph $G$ and two legal configurations $s$ and $t$ in $G$, do we have $s \rightarrowflips G t$? (\cite{gamesNCLthesis})
\end{definition}

\begin{theorem}\cite{gamesNCLthesis}
    C2C is PSPACE-complete.%
\end{theorem}

\subsection{Boolean Reconfiguration Problems}

The \emph{Boolean reconfiguration problem} takes as an input a Boolean formula $\phi$ and  two assignments $\assignment, \assignment'$ satisfying $\phi$. 
The problem consists in determining whether $\assignment$ can be transformed into $\assignment'$ by a sequence of flips of a single variable at each step, while ensuring that all intermediate assignments satisfy $\phi$. 
Formally, we write $\assignment \rightarrowflips{\phi} \assignment'$ when there is a sequence $\langle \assignment_1, \assignment_2 \dots,\assignment_n \rangle$ such that $\assignment_1 = \assignment$, $\assignment_n = \assignment'$ and for all $i=1..n-1$,  $\assignment_{i+1}$ is obtained by a single variable flip from $\assignment_i$ and 
 $\assignment_i \models \phi$. 
 Here, a variable flip refers to flipping the value of a variable from true to false, or from false to true.
 We also make use of the following notations: $\assignment \rightarrowflip{\phi} \assignment'$ means that $\assignment'$ is obtained by one variable flip from $\assignment$, and $\assignment \rightarrowflipvar v {\phi} \assignment'$ means that $\assignment'$ is obtained from $\assignment$ by flipping variable $v$.

\begin{definition}[Boolean Reconfiguration]
  Given a Boolean formula $\phi$, two assignments $ \assignment, \assignment'$ satisfying $\phi$, do we have $\assignment \rightarrowflips{\phi} \assignment'$? 
    \label{def:reconfigurable_bsat_instance}
\end{definition}

3-SAT Reconfiguration is the restriction of Boolean reconfiguration when the formulas $\phi$ is a 3-CNF.

\begin{theorem}[\cite{monotonNAEtSATRecon}]
	\tsat Reconfiguration is \pspacec even if the \inciGraph $\inciGraphSign{\phi}$ is planar.
    \label{th:tsat_recon_pspace}
\end{theorem}

\section{Complexity of \LPTSAT Reconfiguration}
\label{sec:complexity-reconfiguration}
\label{sec:PLLPTSATreconfiguration}
We are now interested in the complexity of the reconfiguration version of \LPTSAT and \PLLPTSAT. They are respectively the Boolean reconfiguration problems restricted to linear planar 3-CNFs, and linear literal-planar 3-CNF.
We prove that both problems are PSPACE-complete.

\subsection{PSPACE-Completeness of \PLLPTSAT Reconfiguration}
\label{subsection:reconfiguration}

It is important to note that the sole condition in the result of the literature (\Cref{th:tsat_recon_pspace} \cite{monotonNAEtSATRecon} is that $\inciGraphSign{\phi}$ is planar. Our contribution is to show that the constructed formula in the original reduction in \cite{monotonNAEtSATRecon} is indeed a 
\emph{linear literal-planar 3-CNF}.
In particular, we prove that there exists a literal-cycle $\pi$ such that $\litClauGraphSign{\phi, \pi}$ is planar. To do that, we use \Cref{th:pi_constructible_for_edge_coll_instance}.

Like the corresponding satisfiability problems, the \PLLPTSAT reconfiguration problem also take respectively some embedding of $\litClauGraphSign{\phi, \pi}$ in the input.

\begin{theorem}
	\PLLPTSAT Reconfiguration is PSPACE-complete.
    \label{th:proof_ltsat_reconfig_pspace}
\end{theorem}

We reduce planar C2C to \PLLPTSAT Reconfiguration. 
For completeness, we recall the construction given in \cite{monotonNAEtSATRecon}.

Consider a planar C2C instance $I = (G, s, t)$, where $G = (V,E)$ is a planar constraint graph and $s$ and $t$ are legal configurations in $G$.
We will construct an instance $I' = (\phi, \pi, \assignment, \assignment')$ of \PLLPTSAT Reconfiguration with $\phi$ a linear planar 3-CNF formula, $\pi$ a literal-cycle and $\assignment, \assignment'$ satisfying~$\phi$.

We represent each edge $e \in E$ by an ordered pair $uv$ where $u$ and $v$ are the endpoints of $e$. 
This fixes a \emph{referential orientation} of $e$ going from $u$ and pointing to~$v$.

\begin{figure}[t]
    \centering

    \begin{subfigure}[t]{0.45\linewidth}
        \centering
        
        \begin{tikzpicture}[
            xscale=0.8, 
            every node/.style={fill=white, scale=1.0},
        ]
            \pgfdeclarelayer{background}
            \pgfdeclarelayer{foreground}
            \pgfsetlayers{background,main,foreground}

            \node[nclnode] (n1) at (0, 0) [scale=0.7] {$v$};

            \node[nclnode] (n2) at (0,1.5) [scale=0.7] {$u$};
            
            \node[nclnode] (n3) at (1.5,-1.5) [scale=0.7] {$w$};

            \node[nclnode] (n4) at (-1.5,-1.5) [scale=0.7] {$x$};

            \begin{pgfonlayer}{background}
                \draw[two weig edge] (n2) -- node[midway, scale=0.6, inner sep=0.5mm] {$e$} (n1);
                \draw[two weig edge] (n3) -- node[midway, scale=0.6,inner sep=0.5mm] {$f$} (n1);
                \draw[two weig edge]  (n1) -- node[midway, scale=0.6,inner sep=0.5mm] {$g$} (n4);
            \end{pgfonlayer}
        \end{tikzpicture}
        \hfill
        \begin{tikzpicture}[
            xscale=0.65, 
            every node/.style={fill=white, scale=0.6}
        ]
            \pgfdeclarelayer{background}
            \pgfdeclarelayer{foreground}
            \pgfsetlayers{background,main,foreground}

            \node (n1) at (0, 0) [draw, rectangle] {$g^v \lor e^v \lor f^v$};

            \node (x) at (0,1) [draw, circle] {$e^v$};
            \node (x_p) at (0,2) [draw, circle] {$e^u$};

            \node (y) at (1,-1) [draw, circle] {$f^v$};
            \node (y_p) at (1.75,-1.75) [draw, circle] {$f^w$};
            
            \node (z) at (-1,-1) [draw, circle] {$g^v$};
            \node (z_p) at (-1.75,-1.75) [draw, circle] {$g^x$};

            \begin{pgfonlayer}{background}
                \draw[-] (x) -- (n1);
                \draw[-] (y) -- (n1);
                \draw[-] (z) -- (n1);

                \draw (x) edge[paired] (x_p);
                \draw (y) edge[paired] (y_p);
                \draw (z) edge[paired] (z_p);
            \end{pgfonlayer}
        \end{tikzpicture}

        \caption{An \emph{OR} node (left), and its gadget in the \litClausGraph of the 3-CNF formula (right).
        }
        \label{fig:incigraph_or_gadget}
    \end{subfigure}
    \hfill
    \begin{subfigure}[t]{0.45\linewidth}
        \centering

        \begin{tikzpicture}[
            xscale=0.8, 
            every node/.style={fill=white, scale=1.0},
        ]
            \pgfdeclarelayer{background}
            \pgfdeclarelayer{foreground}
            \pgfsetlayers{background,main,foreground}

            \node[nclnode] (n1) at (0, 0) [scale=0.7] {$v$};

            \node[nclnode] (n2) at (0,1.5) [scale=0.7] {$u$};
            
            \node[nclnode] (n3) at (1.5,-1.5) [scale=0.7] {$w$};

            \node[nclnode] (n4) at (-1.5,-1.5) [scale=0.7] {$x$};

            \begin{pgfonlayer}{background}
                \draw[two weig edge] (0,1.5) -- node[midway, scale=0.6] {$e$} (n1);
                \draw[one weig edge] (1.5,-1.5) -- node[midway, scale=0.6] {$f$} (n1);
                \draw[one weig edge] (n1) -- node[midway, scale=0.6] {$g$} (-1.5,-1.5);
            \end{pgfonlayer}
        \end{tikzpicture}
        \hfill
        \begin{tikzpicture}[
            xscale=0.6, 
            every node/.style={fill=white, scale=0.6}
        ]
            \pgfdeclarelayer{background}
            \pgfdeclarelayer{foreground}
            \pgfsetlayers{background,main,foreground}

            \node (n2) at (-1, 0) [draw, rectangle] {$g^v \lor e^v$};
            \node (n1) at (1, 0) [draw, rectangle] {$e^v \lor f^v$};

            \node (x) at (0,1) [draw, circle] {$e^v$};
            \node (xp) at (0,2) [draw, circle] {$e^u$};
            
            \node (y) at (1.5,-1) [draw, circle] {$f^v$};
            \node (yp) at (2,-2) [draw, circle] {$f^w$};
            
            \node (z) at (-1.5,-1) [draw, circle] {$g^v$};
            \node (zp) at (-2,-2) [draw, circle] {$g^x$};

            \begin{pgfonlayer}{background}
                \draw[-] (x) -- (n1);
                \draw[-] (x) -- (n2);
                \draw[-] (y) -- (n1);
                \draw[-] (z) -- (n2);

                \draw (x) edge[paired] (xp);
                \draw (y) edge[paired] (yp);
                \draw (z) edge[paired] (zp);
            \end{pgfonlayer}
        \end{tikzpicture}
        \caption{An \emph{AND} node (left), and its gadget \litClausGraph of the 3-CNF formula (right).}
        \label{fig:incigraph_and_gadget}
    \end{subfigure}

    \caption{Gadgets of the reduction from C2C to Linear Literal-Planar 3-SAT Reconfiguration (as in  \cite{monotonNAEtSATRecon}).
    \label{fig:NCLgadgets}
    These gadgets are building blocks of the literal-clause graph containing nodes for each literal and each clause.
    Recall that the squiggly edges are pair edges linking a variable and its negation.    
    }
    
    \label{fig:conversion_ncl_to_tsat}
\end{figure}
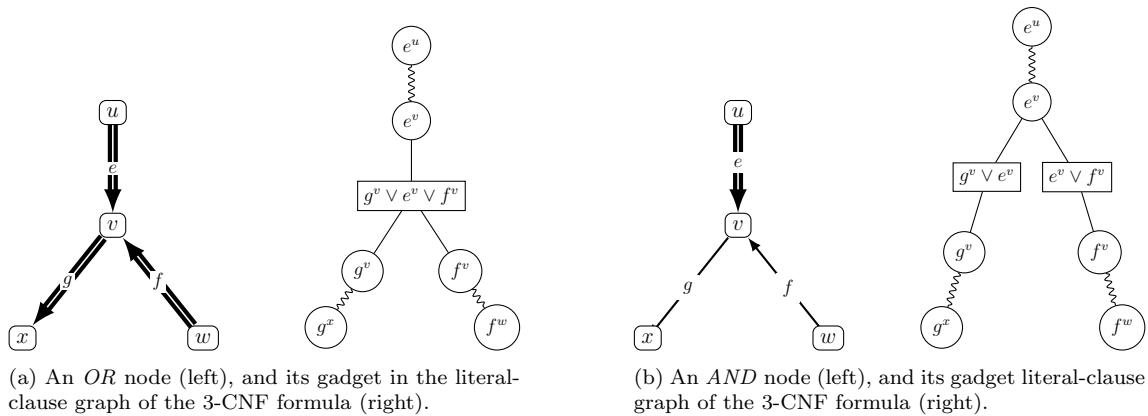

\paragraph*{Encoding configurations as assignments}
    An assignment encodes the orientations of the edges of $G$ in a given configuration.
    For each edge $e \in E$, we introduce a Boolean variable $\variableedge e$.
    Intuitively, $\variableedge e$ is true iff the orientation of edge $e = uv$ in the current configuration is the referential one, i.e. if $e$ points to $v$.
    Given an edge $e = uv$,  We will use the following notations: $e^v := \variableedge e$ and $e^u := \lnot \variableedge e$. 
    The intuitive meaning of the literal $e^v$ (resp. $e^u$) is that $e$ points to $v$ (to $u$).
    
    \paragraph*{Encoding a configuration into an assignment}
    Given a configuration $c$, we define the assignment $\assignment_c$ that encodes the configuration $c$ as follows. For each edge $e$,
    \begin{equation}
    	\label{equation:assignmentfromconfiguration}
    \assignment_c(\variableedge e) := \begin{cases}
    	1 \text{ if the orientation of $e$ is the referential one in $c$} \\
    	0 \text{ otherwise.}
    \end{cases}
    \end{equation} 
  
    \paragraph*{Initial and final configurations} We set $\assignment := \assignment_s$ and $\assignment := \assignment_t$.

\newcommand{\firstedge}{\alpha_v}
\newcommand{\secondedge}{\beta_v}
\newcommand{\thirdedge}{\gamma_v}

\newcommand{\sentencechelou}[2]{if the #1 edge $#2$ points to $v$ in the referential orientation, then $#2^v := \variableedge #2$. 
Otherwise, $#2^v := \neg \variableedge #2$. }

\paragraph*{Construction of $\phi$}
  Given an OR node $v$ in $G$, the corresponding formula $\psi_v$ says that at least one edge incident to $v$ points to $v$. As shown in Figure~\ref{fig:incigraph_or_gadget}, if the three\footnote{Remember that the degree of each node in $G$ is 3.} incident edges are $e$, $f$ and $g$, the formula $\psi_v$ is the clause $g^v \lor e^v \lor f^v$.
  
  Given an AND node $v$ in $G$, the corresponding formula $\psi_v$ says that if the 2-weighted edge does not point to $v$ then the two other 1-weighted edges should point to $v$. Formally, as shown in Figure~\ref{fig:incigraph_and_gadget}, if the incident edges are $e$, $f$ and $g$ and if $e$ is the 2-weighted edge, the formula $\psi_v$ is made of two clauses: $(g^v \lor e^v) \land (e^v\lor f^v)$.

\begin{figure}[t]
    \centering

    \begin{subfigure}[b]{0.25\linewidth}
        \centering

        \begin{tikzpicture}[
            scale=1.3, 
            every node/.style={fill=white, scale=0.9},
        ]
            \pgfdeclarelayer{background}
            \pgfdeclarelayer{foreground}
            \pgfsetlayers{background,main,foreground}

            \node[nclnode] (and_top) at (0, 1)  {$u$};
            \node[nclnode] (and_bot_right) at (1, -1) {$v$};
            \node[nclnode] (and_bot_left) at (-1, -1) {$w$};
            \node[nclnode] (or_center) at (0, -0.25) {$x$};

            \begin{pgfonlayer}{background}
                \draw[one weig edge] (and_bot_right) -- node[midway, scale=0.8] {$e$} (and_top);
                \draw[one weig edge] (and_bot_right) -- node[midway, scale=0.8] {$f$} (and_bot_left);
                \draw[one weig edge] (and_bot_left) -- node[midway, scale=0.8] {$g$} (and_top);

                \draw[two weig edge] (or_center) -- node[midway, scale=0.8] {$h$} (and_bot_left);
                \draw[two weig edge] (or_center) -- node[midway, scale=0.8] {$i$} (and_bot_right);
                \draw[two weig edge] (and_top) -- node[midway, scale=0.8] {$j$} (or_center);
            \end{pgfonlayer}
        \end{tikzpicture}
        \caption{A \ctoc instance.}
    \end{subfigure}
    \hfill
    \begin{subfigure}[b]{0.70\linewidth}
        \centering
    
        \begin{tikzpicture}[
            xscale=1.8, 
            every node/.style={fill=white, scale=0.7}
        ]
            \pgfdeclarelayer{nodelayer}
            \pgfdeclarelayer{edgelayer}
            \pgfsetlayers{nodelayer,edgelayer}

            \begin{pgfonlayer}{nodelayer}
                \node [style={clause_node}] (0) at (-0.5, 3.75) {$g^u \lor j^u$};
                \node [style={clause_node}] (1) at (-2.25, 0) {$h^w \lor f^w$};
                \node [style={clause_node}] (2) at (2.25, 0) {$i^v \lor f^v$};
                \node [style={litteral}] (4) at (1, 2.25) {$e^u$};
                \node [style={clause_node}] (6) at (0, 1.25) {$h^x \lor j^x \lor i^x$};
                \node [style={litteral}] (8) at (-0.75, 1) {$h^x$};
                \node [style={litteral}] (9) at (0.75, 1) {$i^x$};
                \node [style={litteral}] (10) at (1.5, 1.5) {$e^v$};
                \node [style={litteral}] (11) at (1.5, 0.5) {$i^v$};
                \node [style={litteral}] (12) at (-1.5, 0.5) {$h^w$};
                \node [style={litteral}] (13) at (0, 2.75) {$j^u$};
                \node [style={litteral}] (14) at (0, 2) {$j^x$};
                \node [style={litteral}] (15) at (-0.75, 0) {$f^w$};
                \node [style={litteral}] (16) at (0.75, 0) {$f^v$};
                \node [style={litteral}] (17) at (-1, 2.25) {$g^u$};
                \node [style={litteral}] (18) at (-1.5, 1.5) {$g^w$};
                \node [style={clause_node}] (19) at (2.25, 0.75) {$e^v \lor u^v$};
                \node [style={clause_node}] (20) at (-2.25, 0.75) {$g^w \lor h^w$};
                \node [style={clause_node}] (21) at (0.5, 3.75) {$j^u \lor e^u$};
            \end{pgfonlayer}
            \begin{pgfonlayer}{edgelayer}
                \draw (6) to (8);
                \draw (6) to (9);
                \draw (11) to (2);
                \draw (12) to (1);
                \draw (14) to (6);
                \draw (13) to (0);
                \draw (15) to (1);
                \draw (16) to (2);
                \draw (17) to (0);
                \draw [style={paired}] (9) to (11);
                \draw [style={paired}] (4) to (10);
                \draw [style={paired}] (17) to (18);
                \draw [style={paired}] (13) to (14);
                \draw [style={paired}] (15) to (16);
                \draw [style={paired}] (8) to (12);
                \draw (11) to (19);
                \draw (10) to (19);
                \draw (20) to (12);
                \draw (18) to (20);
                \draw (21) to (13);
                \draw (21) to (4);
            \end{pgfonlayer}
        \end{tikzpicture}
        \caption{Corresponding $\litClauGraphSign{\phi}$.}
        \label{fig:variableclausegraphforANDnode}
    \end{subfigure}

    \caption{Example of a transformation of a \ctoc instance to a Linear Literal-Planar 3-SAT instance.}
    
    \label{fig:conversion_ncl_instance_to_tsat_instance_embedding}
\end{figure}
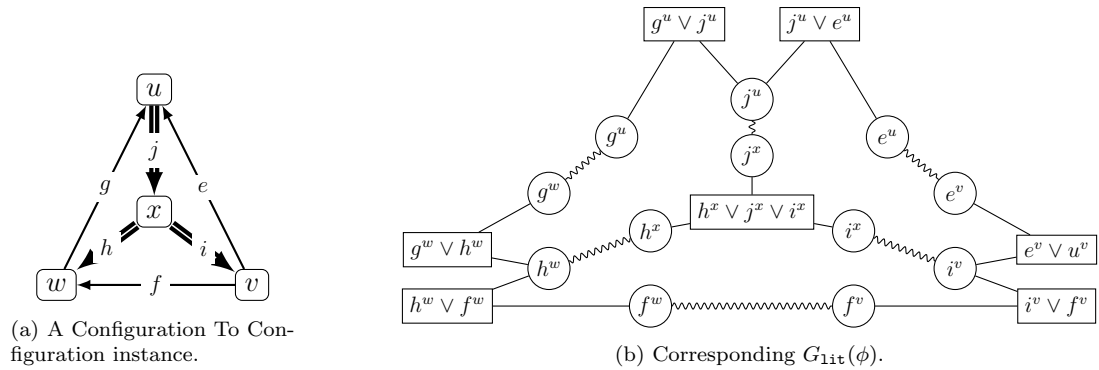

Let $\phi$ the following 3-CNF formula:

\begin{equation}
    \phi := \bigwedge_{v \in V} \psi_v
\end{equation}

\begin{lemma}
    \label{lemma:assignmentfromconfigurationlegalformula}
    Configuration $c$ is legal if and only if $\assignment_c \models \phi$, where $\assignment_c$ is defined in \Cref{equation:assignmentfromconfiguration}.
\end{lemma}
\begin{proof}
    $c$ is legal iff for all nodes $v$ in $G$, the in-flow in $v$ is greater than 2 iff for all nodes $v$ in $G$, $\assignment_c \models \psi_v$ iff $\assignment_c \models \phi$.
\end{proof}

\begin{lemma}
    Formula	$\phi$ is linear.
\end{lemma}
\begin{proof}
	If $e^v$ appears in $\psi_v$ as a literal, then it does not appear outside $\psi_v$.
    However, writing $e = uv$ or $vu$, literal $e^u = \lnot e^v$ only appears in $\psi_u$.
	Thus, if $v$ is an OR node, $e^v$ appears only in the unique clause of $\psi_v$. %
	If $v$ is an AND node, the two clauses in $\psi_v$ intersects in the single literal $e^v$ where $e$ is the 2-weighted edge incident to $v$. Any clause in $\psi_v$ does not intersect 
    any other clause outside of $\psi_v$ since $e^v$, $f^v$, $g^v$ do not appear elsewhere.
\end{proof} %

\Cref{fig:conversion_ncl_instance_to_tsat_instance_embedding} shows a constraint graph $G$ and the corresponding $\litClauGraphSign{\phi}$.

We say that a graph $H$ is a \emph{subdivision} of a graph $I$ if $H$ is obtained from $I$ by a (possibly empty) sequence of applications 
of \emph{edge subdivision} (see \Cref{def:edge_subdivsion}).
Intuitively, subdividing an edge consists in inserting a fresh node at the middle  of the edge. Said differently, edges of $I$ are replaced by finite paths to get $H$. 

\begin{lemma}
    The \litClausGraph $\litClauGraphSign{\phi}$ is a subdivision of $G$; thus $\litClauGraphSign{\phi}$ is planar.
    \label{lem:both_graph_planar}
\end{lemma}
\begin{proof}
    Consider the planar constraint graph $G$ of a C2C instance.
    The graph $\litClauGraphSign{\phi}$ is built by replacing each vertex of $G$ by its corresponding gadget (see \Cref{fig:NCLgadgets}). More precisely, we define a bijection $\lambda$ between vertices of degree 3 of $G$ (that is all vertices of $G$) and vertices of degree 3 in $\litClauGraphSign{\phi}$:
        \begin{itemize}
        \item To an OR node $v$, $\lambda(v) = g^v \lor e^v \lor f^v$ (see \Cref{fig:incigraph_or_gadget})
        \item To an AND node $v$, $\lambda(v) = e^v$ (see~\Cref{fig:incigraph_and_gadget}).
        \end{itemize}
    
    Now, for all edges $uv$ in $G$, there is a path between $\lambda(u)$ and $\lambda(v)$ where intermediate vertices (e.g. $e^v, g^v$, etc. in \Cref{fig:incigraph_or_gadget}, $e^u, g^v \lor e^v$, etc. in \Cref{fig:incigraph_and_gadget}) in the path are of degree 2.
    Note that vertices $\litClauGraphSign{\phi}$  are either of degree 3 or are intermediate vertices of some path.

    Therefore $\litClauGraphSign{\phi}$ is a subdivision of $G$.
\end{proof}

The previous fact can be observed in \Cref{fig:conversion_ncl_instance_to_tsat_instance_embedding}.

\begin{lemma}
    $\litClauGraphSign{\phi}$ can be augmented with a literal cycle $\pi$ such that $\litClauGraphSign{\phi, \pi}$ is planar. Moreover an embedding of $\litClauGraphSign{\phi, \pi}$ such that $\pi$ separates the paired literal edges from the other edges is constructible in polynomial-time.
    \label{lemma:linearsat_planar_cycle_augmentation}
\end{lemma}
\begin{proof}
    Consider $\litClauGraphSign{\phi} = (V,E)$ and $L \subset V$ such that $L$ is the set of all positive and negative literals.  
    Consider the cycle augmentation instance $I = (\litClauGraphSign{\phi}, L)$,
    and $E' := \{ (l,\neg l) \mid (l,\neg l) \in E \}$, the set of paired literals.
    
    We are going to show that $E'$ is a \specialInst (see \Cref{def:SepcialInst}).
    Then by \Cref{th:pi_constructible_for_edge_coll_instance}, a cycle $\pi$ between all literals is constructible in polynomial time and separates all the paired edges $(l, \neg l)$ from the rest of edges in $\litClauGraphSign{\phi, \pi}$.

    In this proof, we denote by $\litClauGraphSign{\phi}^*_{E'}$ the dual graph $\litClauGraphSign{\phi}^*$ restricted to the dual edges of $E'$.

    \begin{enumerate}
        \item $E'$ is a matching.

        \item It remains to show that $E'$ is dually connected.
        We show that for each vertex of degree $3$ in $\litClauGraphSign{\phi}$, there are $3$ different paired edges 
        between opposite literals $l, \neg l$.

        Consider the \emph{OR} gadget (see Figure \ref{fig:incigraph_or_gadget}), there are $3$ different paired edges around the central clause of degree $3$ (respectively $g^xg^v, f^wf^v$ and $e^ue^v$).
        
        Consider now the \emph{AND} gadget (see Figure \ref{fig:incigraph_and_gadget}), the central literal vertex of degree $3$ is connected with its opposite literal and its two clause vertices.
        Those two clauses of degree $2$ are neighbors with two different paired edges (respectively $g^xg^v, f^wf^v$ and $e^ue^v$).

        Since there are $3$ paired literal edges $e \in E'$ for each vertex of degree $3$ and since $\litClauGraphSign{\phi}$ is a subdivision of the cubic constraint graph $G$ (Fact \ref{lem:both_graph_planar}) the dual $\litClauGraphSign{\phi}^*$ restricted to $E'$ "$\litClauGraphSign{\phi}^*_{E'}$" is connected.
        Indeed, $\litClauGraphSign{\phi}^*_{E'}$ is the same graph as $G^*$ up to a renaming of the vertices.
        Since $G^*$ is connected (see \Cref{prop:dual_connected}), so is $\litClauGraphSign{\phi}^*_{E'}$.

    \end{enumerate}
\end{proof}

\begin{claim}
   $\assignment \rightarrowflips \phi \assignment'$  iff $s \rightarrowflips G t$.
\end{claim}
\begin{proof}
	\fbox{$\Leftarrow$} For a sequence of legal configurations $\langle c_1,\dots,c_n \rangle$, we construct $\langle \assignment_1,\dots,\assignment_n \rangle$ by $\assignment_i = \assignment_{c_i}$ (see Equation~\ref{equation:assignmentfromconfiguration}). By \Cref{lemma:assignmentfromconfigurationlegalformula}, $\assignment_i \models \phi$. The flip of $e$ between $c_i$ and $c_{i+1}$ becomes $\assignment_{i+1}(\variableedge{e}) = 1 - \assignment_i(\variableedge{e})$, while the other variables remain unchanged.
	
    \fbox{$\Rightarrow$} For $\langle \assignment_1,\dots,\assignment_n \rangle$, we construct each $c_i$ by reversing the edges $e$ that correspond to the variable flip of $\variableedge{e}$.
    By \Cref{lemma:assignmentfromconfigurationlegalformula}, as $\assignment_i \models \phi$, configuration $c_i$ is legal
    Furthermore, as $\assignment_{i+1}$ is obtained from $\assignment_{i}$ by flipping a variable $\variableedge{e}$, $c_{i+1}$ is obtained from $c_i$ by flipping edge $e$.
\end{proof}

This concludes the proof of \Cref{th:proof_ltsat_reconfig_pspace}.

\subsection{PSPACE-completeness of \MPLtwoOne \tsat Reconfiguration}
We now reduce \PLLPTSAT Reconfiguration, a PSPACE-complete problem by \Cref{th:proof_ltsat_reconfig_pspace} to \MPLtwoOne \tsat 
Reconfiguration.

\begin{theorem}
	\MPLtwoOne \tsat Reconfiguration is PSPACE-complete. 
    PSPACE-hardness holds even if each negative literal appears at most once.
    \label{th:linear_planar_tsat_pspace_c}
\end{theorem}

Consider $(\phi, \assignment, \assignment')$ a \PLLPTSAT Reconfiguration instance.
We construct $(\phi', \friendassignment\mu, \friendassignment{\mu'})$ a \MPLtwoOne \tsat Reconfiguration instance, where $\phi'$ is the formula defined in the proof of \Cref{th:constrainedmonotoneplanarlinearthreesatnpcomplete}, and $\friendassignment\mu$ and $\friendassignment{\mu'}$ are defined by 
\Cref{eq:assignment_paired_lit}.
We have $\mu \models \phi'$ and $\mu' \models \phi'$.

\begin{claim}
    $\assignment \rightarrowflips \phi \assignment'$  iff $\friendassignment{\assignment} \rightarrowflips {\phi'} \friendassignment{\assignment'}$.
\end{claim}

\begin{proof}
    \fbox{$\Rightarrow$} 
    Consider a sequence $\friendassignment{\assignment} = \nu_1 \rightarrowflip{\phi} \dots \rightarrowflip{\phi} \nu_n = \friendassignment{\assignment'}$. We construct a sequence  $\mu = \mu_1 \rightarrowflip{\phi'} \dots \rightarrowflip{\phi'} \mu_{2n} = \assignment'$ as follows.
        We replace each flip $\assignment_i \rightarrowflipvar{v}\phi \assignment_{i+1}$  by:
    \newcommand{\intermediateassignment}[1]{\xi_{#1}}
    \begin{enumerate}
        \item $\friendassignment{\assignment_i}
        \rightarrowflipvar{v}{\phi'}
            \intermediateassignment i \rightarrowflipvar{v'}{\phi'} \friendassignment{\assignment_{i+1}}$ if $\assignment_i \rightarrowflipvar{v}{\phi} \assignment_{i+1}$ flips $v$ from 1 to 0;
        \item $\friendassignment{\assignment_i}\rightarrowflipvar{v'}{\phi'} \intermediateassignment i \rightarrowflipvar{v}{\phi'} \friendassignment{\assignment_{i+1}}$ if $\assignment_i \rightarrowflipvar{v}{\phi} \assignment_{i+1}$ flips $v$ from 0 to 1.
    \end{enumerate}

    Remark that the choice of flipping $v$ first and then $v'$ or first $v'$ and then $v$ in the above cases 1 and 2 is chosen so that the intermediate assignment $\intermediateassignment i$ satisfies $\phi'$.

    \fbox{$\Leftarrow$} We transform a sequence $\friendassignment{\assignment} \rightarrowflips {\phi'} \friendassignment{\assignment'}$
    into a sequence $\assignment \rightarrowflips \phi \assignment'$ as follows: we keep the sequence of flips of variable $v$ but remove the flips of variables $v'$. For instance, from $\friendassignment{\assignment}
        \rightarrowflipvar{x_1} {\phi'}
        \rightarrowflipvar{x_3} {\phi'}
        \rightarrowflipvar{x_2' }{\phi'}
        \rightarrowflipvar{x_4} {\phi'} \friendassignment{\assignment'}$
        we get ${\assignment}
        \rightarrowflipvar{x_1} {\phi}
        \rightarrowflipvar{x_3} {\phi}
        \rightarrowflipvar{x_4} {\phi} 
        {\assignment'}$.
\end{proof}

This establishes that the \MPLtwoOne \tsat Reconfiguration problem is PSPACE-complete.

\section{Application to Connected Multi-Agent Pathfinding in 2D}
\label{sec:complexity-cmapf}

Connected multi-agent pathfinding (CMAPF) (\cite{DBLP:journals/trob/HollingerS12},\cite{cmapfTateo}) is the problem of finding collision-free paths for multiple agents in a graph while ensuring that the agents remain connected at all times. This connectivity constraint is critical in applications requiring continuous communication, such as robotic coordination in exploration or disaster response scenarios.
This problem is PSPACE-complete in general (\cite{cmapfTateo}, \cite{DBLP:journals/aamas/CharrierQSS20}). Recently, it has been proven PSPACE-complete even if the environment is included in a 3D grid and the connectivity is given with a radius $\radius$ (two agents can communicate if their distance is at most $\rho$) \cite{cmapfIsse}. 

Here we prove that the problem is PSPACE-complete in 2D grids using reductions from the \LPTSAT problem developed in the previous sections,
while also establishing that the bounded version of the problem is NP-complete.

\begin{figure}
	\begin{tikzpicture}
		contenu...
	\end{tikzpicture}
\end{figure}

\subsection{Our 2D setting}

A 2D CMAPF environment is a tuple $G = (V, E_M, \radius)$ where $V$ is a non-empty finite subset of $\setZ^2$, $E_M \subseteq \set{(u, v) \in V^2 \mid d_m(u, v) \leq 1}$ is the set of movement edges, representing the possible movements of agents, where $d_m$ is the Manhattan distance. 
In other words, $G = (V, E_M)$ is a finite subgraph of the grid $\setZ^2$. The quantity $\radius$ is the connectivity radius.
The connectivity relation $E_C \subseteq V \times V$ is implicitly defined by the radius: $E_C = \set{(u, v) \in V^2 \mid d_{m}(u, v) \leq \rho}$.

A \emph{configuration} $c$ assigns a position $c_i \in V$ to each agent $i \in \{1, \ldots, n\}$ such that $c_i \neq c_j$ for $i \neq j$. 
In other words, agents occupy pairwise distinct locations; configurations have \emph{no vertex conflicts}.
Furthermore, we suppose that a configuration is \emph{connected}: the induced subgraph of $(V, E_C)$ on vertices $\set{c_i \mid i \in \set{1, \dots, n}}$ is connected. 
This means that agents can always communicate either directly or indirectly via multi-hop.
The initial and target configurations are denoted by $s$ and~$t$, respectively. 

We write $c \rightarrow c'$ if  $(c_i, c'_i) \in E_M$ for all agents $i$, and if $(c,c')$ has \emph{no swapping conflict}:  
for all $i \neq j$, $(c_i, c_j) \neq (c'_j, c'_i)$ .
Intuitively, $c \rightarrow c'$ means that the group of agents move from configuration $c$ to configuration $c'$ all together in one step.

An \emph{execution} from $s$ to $t$ is a sequence of configurations $\langle c^0, c^1, \ldots, c^\ell \rangle$ such that $s = c^0 \rightarrow \dots \rightarrow c^\ell = t$, and the \emph{length} of this execution is $\ell$.
We write $s \rightarrowcmapf{G} t$ if there is an execution from $s$ to $t$.

\begin{definition}[2D CMAPF Problem]
	Given a CMAPF environment $G$, two configurations $s$ and $t$, determine if $s \rightarrowcmapf{G} t$.
\end{definition}

We also consider the bounded version of this problem where the witness execution is required to have respect a given bound.
\begin{definition}[Bounded 2D CMAPF Problem]
	Given a CMAPF environment $G$, two configurations $s$ and $t$, and a bound $b$ given in unary, 
	determine if there is an execution from $s$ to $t$ of length at most $b$, denoted by $s \rightarrowcmapfbounded G b t$.
\end{definition}

\subsection{Notations Used for the Gadgets}

\newcolumntype{Y}{>{\raggedright\arraybackslash}X}

\begin{table}[h]
		\begin{tabularx}{\textwidth}{b{1cm} X} 
			\begin{tikzpicture}[baseline=-1mm]
					\pgfsetlayers{edgelayer,nodelayer}
					\begin{pgfonlayer}{nodelayer}
							\node [style={blue_agent}] (1) at (0, 0) {};
					\end{pgfonlayer}
			\end{tikzpicture} 
			&
			\textbf{Provider}: Immobile agent that provides connectivity to moving agents. \\ 

			\begin{tikzpicture}[baseline=-1mm]
					\pgfsetlayers{edgelayer,nodelayer}
					\begin{pgfonlayer}{nodelayer}
							\node [style={green_agent}] (1) at (0, 0) {};
					\end{pgfonlayer}
			\end{tikzpicture} &
			\textbf{Requester}: Immobile agent that constrains some mobile agents to provide connectivity to them. The position of mobile agents that provide connectivity to a connectivity requester encodes  the truth value of some literal. \\ 
			\begin{tikzpicture}[baseline=3.5mm]
					\pgfsetlayers{edgelayer,nodelayer}
					\begin{pgfonlayer}{nodelayer}
							\node [style={movement_node}] (1) at (0, 0.5) {$x$};
					\end{pgfonlayer}
			\end{tikzpicture} &
			
			\textbf{Mobile agent}: A mobile agent, denoted by a letter $x$. \\
			\begin{tikzpicture}[baseline=3.5mm]
				\node [style={target}] (1) at (0, 0.5) {};
			\end{tikzpicture} &
			\textbf{Target node}: A target node represents the target position of some mobile agent. Mobile agents will move in parts of the graph isolated from each other, so each target node will belong to a unique mobile agent without ambiguity. Note that since \emph{immobile agents} (connectivity providers/requesters) cannot move, their targets are not specified in the figures as they are already at their target positions.  
	\end{tabularx}
\caption{Notations used in the description of the gadgets.\label{table:gadgetsnotation}}
\end{table}

\begin{table}[h]
	\begin{center}
	\footnotesize
		\begin{tabular}{lp{8cm}l}
			\begin{tikzpicture}
				\pgfsetlayers{edgelayer,nodelayer}
				\begin{pgfonlayer}{edgelayer}
					\draw [style={connectivity_provider_wire}] (0,0) to (1,0);
				\end{pgfonlayer}
			\end{tikzpicture}
			& A wire of connectivity providers & \wireproviders
		\\[2mm]
			\hline \\[-2mm]
			\begin{tikzpicture}
				\pgfsetlayers{edgelayer,nodelayer}
				\begin{pgfonlayer}{edgelayer}
					\draw [style={connectivity_requester_wire}] (0,0) to (1,0);
				\end{pgfonlayer}
			\end{tikzpicture} 
			& A wire of connectivity requesters & 
			\wirerequesters \\[2mm]
				\hline \\[-2mm]
			\begin{tikzpicture}[baseline=0mm]
				\pgfsetlayers{edgelayer,nodelayer}
				\begin{pgfonlayer}{edgelayer}
					\draw [style={movement_wire}] (0,0) to (1,0);
				\end{pgfonlayer}
			\end{tikzpicture}
			& a line of empty cells on which a mobile agent can move surrounded by connectivity providers & 
			\begin{tikzpicture}[baseline=2mm]
				\draw (0, 0) -- (2, 0);
						\foreach \x in {0, 0.5, ..., 2} {
						\node [style={blue_agent}] at (\x, 0.5) {};
						\node [style={blue_agent}] at (\x, -0.5) {};
						\node [style={movement_node}] (\x) at (\x, 0) {};
						}
			\end{tikzpicture}
		\end{tabular}
	\end{center}
	\caption{Notation for the wires.\label{table:wires}}
\end{table}

In the rest of the section, we will describe reductions to prove the NP-hardness of the bounded 2D CMAPF (where the length of the execution is bounded), 
and the PSPACE-hardness of 2D CMAPF.
In these reductions, we will transform a graph~$\inciGraphSign{\phi,\pi}$ into an instance of (bounded)  2D CMAPF instance.

\Cref{table:gadgetsnotation} gives the notations for specific cells in a grid which we will use to build our gadgets. 
\Cref{table:wires} gives the notations used in \Cref{fig:variable_gadget}-\ref{fig:reduction_unbounded_example} for \emph{wires}, which are consecutive specific cells. 
\emph{Wires of connectivity providers} can be thought of as live wires carrying a current, while 
\emph{wires of connectivity requesters} are used to connect agents to the current.
Connectivity is achieved if all agents are connected to the current.

All the gadgets we construct further are composed by the nodes described above.

\subsection{NP-Completeness of Bounded 2D CMAPF}
\label{subsection:np_hardness_bcmapf_proof}

\begin{theorem}
	Bounded 2D CMAPF is NP-complete, even if the bound is 2 and the radius is 1.
	\label{th:bounded2DCMAPFNPcomplete}
\end{theorem}

	We prove the NP-completeness of Bounded 2D CMAPF by a reduction from \MPLtwoOne \tsat. 
	Consider an instance of \MPLtwoOne \tsat, that is an embedding of $\inciGraphSign{\phi, \pi}$ for some cycle $\pi$.

	We construct a bounded 2D CMAPF instance $(G, s, t, b)$ with $b=2$ and a radius of 1.
	To construct~$G$ we replace the vertices of $\inciGraphSign{\phi, \pi}$ by gadgets that we will describe now.

\begin{figure}[h]
    \centering
    \pgfsetlayers{edgelayer,nodelayer}
    
    \begin{subfigure}[b]{0.45\linewidth}
        \centering
        \begin{tikzpicture}[
            scale=1, 
            every node/.style={scale=1}
        ]
            \begin{pgfonlayer}{nodelayer}
                \node (l) at (0, 0) [draw, circle, fill=white] {$x$};
            \end{pgfonlayer}
            \begin{pgfonlayer}{edgelayer}
                \draw[-] (0.5,1) -- (l);
                \draw[-] (-0.5,1) -- (l);
                \draw[-] (0,-1) -- (l);
                \draw[variablecycle] (1,0) -- (l);
                \draw[variablecycle] (-1,0) -- (l);
     \node [explanation] at (0, 1.4) {\begin{tabular}{c}
                (to clauses containing~$x$)
                \end{tabular}};
                \node [explanation] at (0, -1.3) {\begin{tabular}{c} 
                (to clause containing~$\lnot x$)
                 \end{tabular}};
            \end{pgfonlayer}
        \end{tikzpicture}

        \caption{A variable $x$ in $\inciGraphSign{\phi,\pi}$. Thick edges on each side are part of the variable cycle; the edges above are linked to positive clauses (0, 1 or 2 positive clauses); the edge below to a negative one (0 or 1 negative clause).}
    \end{subfigure}
    \hfill
    \begin{subfigure}[b]{0.45\linewidth}
        \centering
        \begin{tikzpicture}[
            scale=0.9, 
            every node/.style={scale=0.8}
        ]
        \begin{pgfonlayer}{nodelayer}
            \node [explanation] at (0, 1.4) {\begin{tabular}{c}
            positive output\\(to clause gadgets containing~$x$)
            \end{tabular}};
            \node [explanation] at (0, -2.3) {\begin{tabular}{c} 
            negative output\\ 
            (to clause gadgets containing~$\lnot x$)
             \end{tabular}};
             \node [style=none] (2) at (0, 1) {};
             \node [style=none] (3) at (-1, 0) {};
             \node [style=none] (4) at (1, 0) {};
             \node [style=none] (6) at (0, 0) {};
             \node [style=gadget] (7) at (0, 0) {$\splitgadget{x}$};
             \node [style=none] (8) at (-1, 1) {};
             \node [style=gadget] (9) at (0, -1) {\switch{x}};
             \node [style=none] (10) at (0, -2) {};
             \node [style=none] (11) at (1, -1) {};
             \node [style=none] (12) at (1.75, -1) {};
             \node [style=none] (13) at (-1.75, -1) {};
             \node [style=none] (15) at (1.5, 0.75) {};
             \node [style=none] (16) at (-1.5, 0.75) {};
             \node [style=none] (17) at (-1.5, -1.75) {};
             \node [style=none] (18) at (1.5, -1.75) {};
             \node [style=none] (19) at (1.25, 0.5) {$\texttt{x}$};
         \end{pgfonlayer}
         \begin{pgfonlayer}{edgelayer}
             \draw [style={connectivity_provider_wire}] (4.center) to (7);
             \draw [style={connectivity_provider_wire}] (11.center) to (4.center);
             \draw [style={connectivity_provider_wire}] (11.center) to (9);
             \draw [style={connectivity_provider_wire}] (11.center) to (12.center);
             \draw [style={connectivity_provider_wire}] (13.center) to (9);
             \draw [style={connectivity_requester_wire}] (9) to (10.center);
             \draw [style={connectivity_requester_wire}] (9) to (7);
             \draw [style={connectivity_requester_wire}] (7) to (3.center);
             \draw [style={connectivity_requester_wire}] (3.center) to (8.center);
             \draw (15.center) to (18.center);
             \draw (18.center) to (17.center);
             \draw (17.center) to (16.center);
             \draw (16.center) to (15.center);
             \draw [style={connectivity_requester_wire}] (2.center) to (7);
         \end{pgfonlayer}
        \end{tikzpicture}

        \caption{Abstract representation of the corresponding gadget. 
        See \Cref{fig:switch_np_hardness} (resp. \Cref{fig:split_np_hardness})
         for the content of $\switch{x}$ (resp. $\splitgadget{x}$).
          \label{figure:gadgetvariable}
          }
    \end{subfigure}

    \begin{subfigure}[b]{0.45\linewidth}
        \centering
            \begin{tikzpicture}[
                scale=0.9, 
                every node/.style={scale=0.8}
            ]
                \begin{pgfonlayer}{nodelayer}
                    \node [style={target}] (0) at (0, 0) {$x$};
                    \node [style={movement_node}] (1) at (0, 0.5) {};
                    \node [style={movement_node}] (2) at (0, -0.5) {};
                    \node [style={green_agent}] (17) at (0, 1.5) {};
                    \node [style={green_agent}] (31) at (0, -1) {};
                    \node [style={green_agent}] (32) at (0, -1.5) {};
                    \node [style={blue_agent}] (82) at (-0.5, 0) {};
                    \node [style={blue_agent}] (142) at (2, 0) {};
                    \node [style={blue_agent}] (144) at (1.5, 0) {};
                    \node [style={blue_agent}] (145) at (0.5, 0.5) {};
                    \node [style={blue_agent}] (146) at (1, 0) {};
                    \node [style={blue_agent}] (147) at (0.5, 0) {};
                    \node [style={green_agent}] (149) at (0, 1) {};
                    \node [style={blue_agent}] (153) at (0.5, -0.5) {};
                    \node [style=none] (157) at (-1.75, 1.25) {};
                    \node [style=none] (158) at (1.75, 1.25) {};
                    \node [style=none] (159) at (-1.75, -1.25) {};
                    \node [style=none] (160) at (1.75, -1.25) {};
                    \node [style=none] (161) at (1.07, 1) {\texttt{\switch{x}}};
                    \node [style={blue_agent}] (162) at (-0.5, 0.5) {};
                    \node [style={blue_agent}] (163) at (-0.5, -0.5) {};
                    \node [style={blue_agent}] (164) at (-1, 0) {};
                    \node [style={blue_agent}] (165) at (-1.5, 0) {};
                    \node [style={blue_agent}] (166) at (-2, 0) {};
                \end{pgfonlayer}
                \begin{pgfonlayer}{edgelayer}
                    \draw (2) to (0);
                    \draw (0) to (1);
                    \draw (158.center) to (157.center);
                    \draw (159.center) to (157.center);
                    \draw (160.center) to (159.center);
                    \draw (160.center) to (158.center);
                \end{pgfonlayer}
            \end{tikzpicture}
    
            \caption{The content of $\switch{x}$. Moving the agent $x$ up corresponds to setting variable $x$ to true.  Moving the agent $x$ down corresponds to setting variable $x$ to false.}
            \label{fig:switch_np_hardness}
        \end{subfigure}
        \hfill
        \begin{subfigure}[b]{0.45\linewidth}
            \centering
            \begin{tikzpicture}[
                scale=0.9, 
                every node/.style={scale=0.8}
            ]
                \begin{pgfonlayer}{nodelayer}
                    \node [style={green_agent}] (0) at (0, 1) {};
                    \node [style={target}] (1) at (0.5, 0.5) {$\splitagent x$};
                    \node [style={movement_node}] (3) at (0.5, 1) {};
                    \node [style={green_agent}] (4) at (0.5, 1.5) {};
                    \node [style={blue_agent}] (23) at (1, 0.5) {};
                    \node [style={blue_agent}] (24) at (1, 1) {};
                    \node [style={green_agent}] (41) at (0.5, 0) {};
                    \node [style={green_agent}] (42) at (0.5, -0.5) {};
                    \node [style={blue_agent}] (54) at (1.5, 0.5) {};
                    \node [style={blue_agent}] (55) at (2, 0.5) {};
                    \node [style={green_agent}] (58) at (-0.5, 1) {};
                    \node [style={green_agent}] (59) at (-1, 1) {};
                    \node [style={green_agent}] (91) at (-1.5, 1) {};
                    \node [style={green_agent}] (60) at (0.5, 2) {};
                    \node [style=none] (92) at (-1.25, 1.75) {};
                    \node [style=none] (93) at (2.25, 1.75) {};
                    \node [style=none] (94) at (-1.25, -0.25) {};
                    \node [style=none] (95) at (2.25, -0.25) {};
                    \node [style=none] (96) at (1.65, 1.5) {$\splitgadget x$};
                    \node [style={blue_agent}] (97) at (2.5, 0.5) {};
                \end{pgfonlayer}
                \begin{pgfonlayer}{edgelayer}
                    \draw (93.center) to (92.center);
                    \draw (94.center) to (92.center);
                    \draw (95.center) to (93.center);
                    \draw (95.center) to (94.center);
                    \draw (3) to (1);
                \end{pgfonlayer}
            \end{tikzpicture}
        
            \caption{The content of $\splitgadget{x}$. For the NP-hardness, the figure indicates the target of agent $\ell$. For the PSPACE-hardness, the position of the target of agent $\ell$ may be down or up.}
            \label{fig:split_np_hardness}
        \end{subfigure}
    
    \caption{NP-hardness: Gadget for a variable.}
    \label{fig:variable_gadget}
\end{figure}
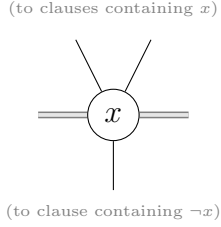
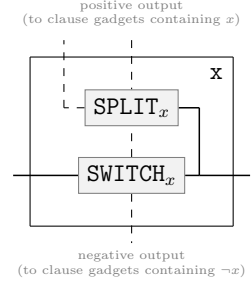
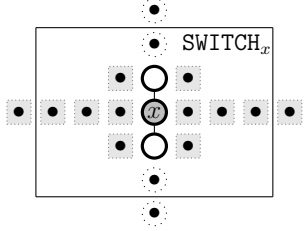
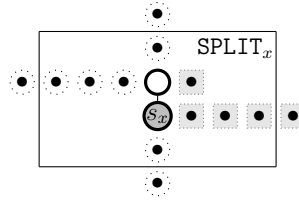

	\smallskip
	\noindent\textbf{Variable Gadget.}
	As shown in \Cref{figure:gadgetvariable}, the variable gadget $\texttt{x}$ is composed of two parts: $\switch x$ managing the choice of the Boolean value of variable $x$, and $\splitgadget{x}$ for splitting the wire connecting the variable to the clauses.
		Remember that $\phi$ is monotone linear, so one can assume, by \Cref{th:constrainedmonotoneplanarlinearthreesatnpcomplete}, that there is at most one occurrence of $\lnot x$ and at most \emph{two} occurrences for $x$.
	Hence, the gadget shown in \Cref{figure:gadgetvariable} has the appropriate number of outputs for connecting it to the clauses.

	\emph{Switch Part.}
	The switch gadget $\switch{x}$ is given in \Cref{fig:switch_np_hardness}. 
	It has a ``positive output'' and a ``negative output'': the positive (resp. negative) output is linked to clause gadgets whose clauses contain the positive literal $x$ (resp. the negative literal~$\lnot x$).
	This gadget encodes a valuation of $x$ and acts as a switch: at step 1, agent $x$ can either move up and make $x$ true, or move down and make $x$ false.
	It provides connectivity to the top requester wire \wirerequesters\ if $x$ is true, and to the bottom requester wire otherwise.
		
\begin{figure}[H]
    \pgfsetlayers{edgelayer,nodelayer}
    
    \centering
    
    \begin{subfigure}[b]{0.3\linewidth}
        \centering
        \begin{tikzpicture}[
            scale=1, 
            every node/.style={scale=1.2}
        ]
            \begin{pgfonlayer}{nodelayer}
                \node (c1) at (0, 0) [draw, rectangle, fill=white] {$\literal_1 \lor \literal_2 \lor \literal_3$};
                \node[explanation] at (-1, -1.3){to $\literal_1$};
                \node[explanation] at (0, -1.3){to $\literal_2$};
                \node[explanation] at (1, -1.3){to $\literal_3$};
            \end{pgfonlayer}
            \begin{pgfonlayer}{edgelayer}
                \draw[-] (0,-1) -| (c1);
                \draw[-] (-1,-1) |- (c1.west);
                \draw[-] (1,-1) |- (c1.east);
            \end{pgfonlayer}
        \end{tikzpicture}

        \caption{A clause in $\inciGraphSign{\phi,\pi}$ where $\literal_1, \literal_2, \literal_3$ are literals.
        Three edges shown below connect to literal nodes.
        }
    \end{subfigure}
    \hfill
    \begin{subfigure}[b]{0.3\linewidth}
        \centering
        \begin{tikzpicture}[
            scale=0.9, 
            every node/.style={scale=0.8}
        ]
        \begin{pgfonlayer}{nodelayer}
            \node [style=none] (1) at (0, -1.5) {};
            \node [style=none] (2) at (0, 1) {};
            \node [style=none] (3) at (-1.5, 0) {};
            \node [style=none] (4) at (1.5, 0) {};
            \node [style=none] (6) at (0, 0) {};
            \node [style=gadget] (7) at (0, 0) {$\literal_1 \lor \literal_2 \lor \literal_3$};
            \node [style=none] (8) at (-0.5, -1) {};
            \node [style=none] (9) at (-0.5, -0.25) {};
            \node [style=none] (10) at (0.5, -1) {};
            \node [style=none] (11) at (0.5, -0.25) {};
            
             \node[explanation,rotate=90] at (-2, 0){
             \begin{tabular}{c}
             to variable
              gadget\\ 
             corresponding \\
             to $\literal_1$
             \end{tabular}};

\node[explanation,rotate=90] at (2, 0){
             \begin{tabular}{c}
             to variable 
              gadget\\ 
             corresponding \\
             to $\literal_3$
             \end{tabular}};

\node[explanation] at (0, -2.2){
             \begin{tabular}{c}
             to variable 
              gadget\\ 
             corresponding \\
             to $\literal_2$
             \end{tabular}};
        \end{pgfonlayer}
        \begin{pgfonlayer}{edgelayer}
            \draw [style={connectivity_provider_wire}] (7) to (2.center);
            \draw [style={connectivity_requester_wire}] (7) to (4.center);
            \draw [style={connectivity_requester_wire}] (7) to (3.center);
            \draw [style={connectivity_requester_wire}] (7) to (1.center);
            \draw [style={connectivity_provider_wire}] (9.center) to (8.center);
            \draw [style={connectivity_provider_wire}] (11.center) to (10.center);
        \end{pgfonlayer}
        \end{tikzpicture}

        \caption{Abstract representation of the gadget.}
    \end{subfigure}
    \hfill
    \begin{subfigure}[b]{0.3\linewidth}
        \centering
        \begin{tikzpicture}[
            scale=0.9, 
            every node/.style={scale=0.8}
        ]
        \begin{pgfonlayer}{nodelayer}
            \node [style={movement_node}] (0) at (0, 2) {};
            \node [style={movement_node}] (1) at (-0.5, 2) {$c$};
            \node [style={green_agent}] (5) at (0, -1) {};
            \node [style={green_agent}] (12) at (0, 0.5) {};
            \node [style=none] (14) at (-1.75, 2.75) {};
            \node [style=none] (15) at (1.75, 2.75) {};
            \node [style=none] (16) at (-1.75, -1.25) {};
            \node [style=none] (17) at (1.75, -1.25) {};
            \node [style=none] (18) at (1.05, 2.5) {\texttt{CLAUSE}};
            \node [style={blue_agent}] (19) at (-1, 1) {};
            \node [style={blue_agent}] (21) at (1, 0) {};
            \node [style={blue_agent}] (22) at (1, -0.5) {};
            \node [style={movement_node}] (28) at (-0.5, 0.5) {};
            \node [style={movement_node}] (29) at (0.5, 0.5) {};
            \node [style={movement_node}] (31) at (0, 0) {};
            \node [style={blue_agent}] (35) at (0.5, -0.5) {};
            \node [style={blue_agent}] (37) at (-1, -0.5) {};
            \node [style={blue_agent}] (38) at (-1, -1.5) {};
            \node [style={blue_agent}] (39) at (-0.5, -0.5) {};
            \node [style={blue_agent}] (40) at (-0.5, 0) {};
            \node [style={green_agent}] (44) at (-1.5, 0.5) {};
            \node [style={blue_agent}] (46) at (-0.5, 1.5) {};
            \node [style={blue_agent}] (50) at (0.5, 1) {};
            \node [style={blue_agent}] (51) at (1, 1) {};
            \node [style={blue_agent}] (53) at (0.5, 1.5) {};
            \node [style={blue_agent}] (54) at (-0.5, 1) {};
            \node [style={movement_node}] (57) at (0, 1.5) {};
            \node [style={blue_agent}] (59) at (-1, -1) {};
            \node [style={blue_agent}] (60) at (1, -1) {};
            \node [style={blue_agent}] (61) at (0.5, 0) {};
            \node [style={blue_agent}] (62) at (-1, 0) {};
            \node [style={blue_agent}] (63) at (1, -1.5) {};
            \node [style={blue_agent}] (67) at (-1, 1.5) {};
            \node [style={blue_agent}] (68) at (-1.5, 2) {};
            \node [style={blue_agent}] (69) at (-1.5, 2.5) {};
            \node [style={blue_agent}] (70) at (-1.5, 3) {};
            \node [style={green_agent}] (74) at (1.5, 0.5) {};
            \node [style=target] (75) at (0.5, 2) {};
            \node [style=target] (76) at (0, 1) {$t$};
            \node [style=target] (77) at (1, 0.5) {$\literal_3$};
            \node [style=target] (78) at (-1, 0.5) {$\literal_1$};
            \node [style=target] (79) at (0, -0.5) {$\literal_2$};
            \node [style={green_agent}] (80) at (-2, 0.5) {};
            \node [style={green_agent}] (81) at (2, 0.5) {};
            \node [style={green_agent}] (82) at (0, -1.5) {};
            \node [style={blue_agent}] (83) at (-1.5, 1.5) {};
        \end{pgfonlayer}
        \begin{pgfonlayer}{edgelayer}
            \draw (0) to (1);
            \draw (15.center) to (14.center);
            \draw (16.center) to (14.center);
            \draw (17.center) to (16.center);
            \draw (17.center) to (15.center);
            \draw (75) to (0);
            \draw (57) to (76);
            \draw (77) to (29);
            \draw (78) to (28);
            \draw (79) to (31);
        \end{pgfonlayer}
        \end{tikzpicture}
    
        \caption{The content of the gadget.}
        \label{fig:clause_content_np_hardness}
    \end{subfigure}
    
    \caption{NP-hardness: gadget for a clause.}
    \label{fig:clause_gadget}
\end{figure}
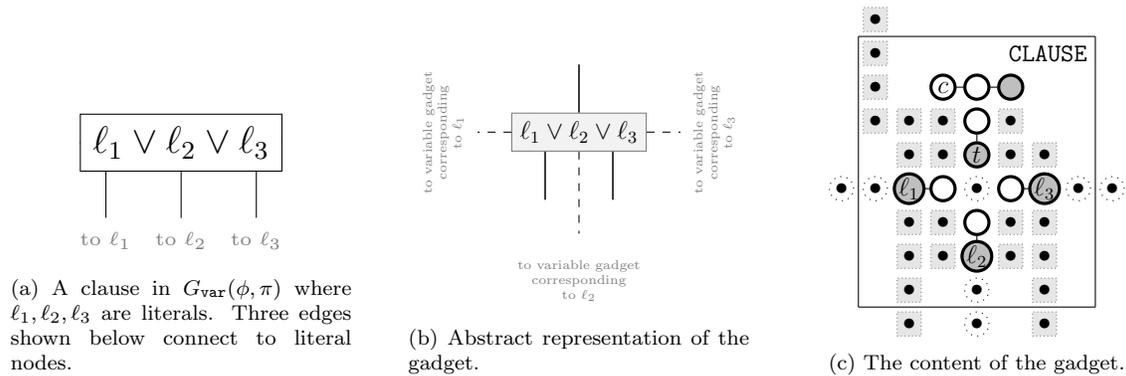

	\emph{Split Part.}
	The split gadget $\splitgadget{x}$ is presented in \Cref{fig:split_np_hardness}.
	As shown in \Cref{figure:gadgetvariable}, it is placed above the switch gadget.
	Initially, agent $\splitagent x$ ensures the connection to the requesters below,
	while requesters going to the clauses receive connectivity from agents $\ell_i$ in the clause gadget (see \Cref{fig:clause_gadget}) described later.
	Notice that if $x$ is false, agent $\splitagent{x}$ cannot move up since the requesters below would be disconnected.
	As soon as variable $x$ is set to true, the requesters below are connected to agent $x$ in $\switch{x}$, so $\splitagent{x}$ is free to move up at step 1.

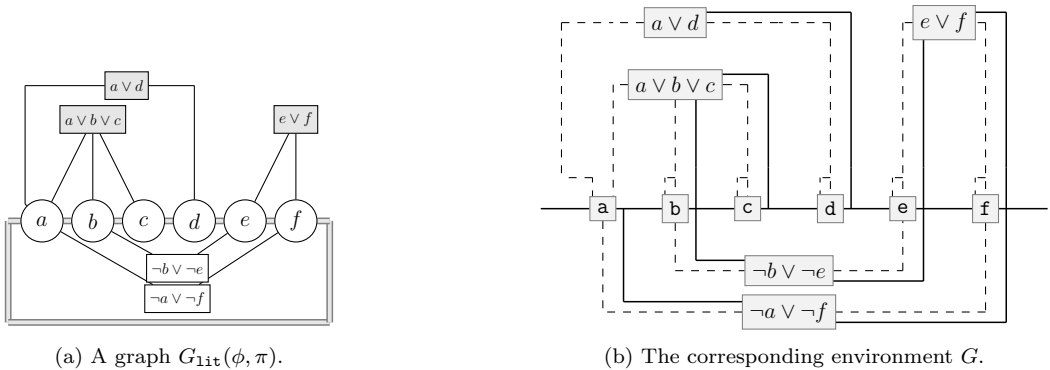
\begin{figure}[H]
    \centering
    \pgfsetlayers{edgelayer,nodelayer}
    
    \begin{subfigure}[b]{0.4\linewidth}
        \centering
        \begin{tikzpicture}[
            scale=0.9, 
            every node/.style={scale=0.7}
        ]
        \begin{pgfonlayer}{nodelayer}
            \node [style={positiveclause}] (0) at (-0.5, 2) {$a \lor b \lor c$};
            \node [style={variable_node}] (3) at (0.25, 0.5) {$c$};
            \node [style={positiveclause}] (4) at (2.5, 2) {$e \lor f$};
            \node [style={variable_node}] (5) at (1.75, 0.5) {$e$};
            \node [style={variable_node}] (6) at (2.5, 0.5) {$f$};
            \node [style={variable_node}] (7) at (-0.5, 0.5) {$b$};
            \node [style=none] (8) at (3, 0.5) {};
            \node [style=none] (9) at (3, -1) {};
            \node [style=none] (10) at (-1.75, -1) {};
            \node [style=none] (11) at (-1.75, 0.5) {};
            \node [style={positiveclause}] (13) at (0, 2.5) {$a \lor d$};
            \node [style={variable_node}] (14) at (1, 0.5) {$d$};
            \node [style=none] (15) at (1, 2.5) {};
            \node [style=none] (16) at (-1.5, 2.5) {};
            \node [style=none] (17) at (-1.5, 0.75) {};
            \node [style={variable_node}] (18) at (-1.25, 0.5) {$a$};
            \node [style={negativeclause}] (19) at (0.75, -0.65) {$\neg a \lor \neg f$};
            \node [style={negativeclause}] (20) at (0.75, -0.2) {$\neg b \lor \neg e$};
        \end{pgfonlayer}
        \begin{pgfonlayer}{edgelayer}
            \draw (4) to (5);
            \draw (4) to (6);
            \draw (0) to (7);
            \draw (0) to (3);
            \draw [style={variablecycle}] (7) to (3);
            \draw [style={variablecycle}] (5) to (6);
            \draw [style={variablecycle}] (11.center) to (10.center);
            \draw [style={variablecycle}] (10.center) to (9.center);
            \draw [style={variablecycle}] (9.center) to (8.center);
            \draw [style={variablecycle}] (8.center) to (6);
            \draw (15.center) to (14);
            \draw (15.center) to (13);
            \draw [style={variablecycle}] (3) to (14);
            \draw [style={variablecycle}] (14) to (5);
            \draw (17.center) to (16.center);
            \draw (16.center) to (13);
            \draw [style={variablecycle}] (18) to (7);
            \draw [style={variablecycle}] (18) to (11.center);
            \draw (18) to (0);
            \draw (17.center) to (18);
            \draw (18) to (19);
            \draw (19) to (6);
            \draw (5) to (20);
            \draw (20) to (7);
        \end{pgfonlayer}
        \end{tikzpicture}

        \caption{A graph $\litClauGraphSign{\phi, \pi}$.}
    \end{subfigure}
    \hfill
    \begin{subfigure}[b]{0.5\linewidth}
        \centering
        \begin{tikzpicture}[
            scale=0.55, 
            every node/.style={scale=0.7}
        ]
        \begin{pgfonlayer}{nodelayer}
            \node [style=none] (2) at (-2.5, 1.75) {};
            \node [style=none] (3) at (0.25, 1.75) {};
            \node [style=none] (4) at (-1, 1.75) {};
            \node [style=gadget] (5) at (-1, 1.75) {$a \lor b \lor c$};
            \node [style=none] (9) at (-0.5, 1.5) {};
            \node [style=none] (12) at (-3.75, -0.5) {};
            \node [style=none] (16) at (-3.75, 3.25) {};
            \node [style=none] (17) at (-2.75, -3.75) {};
            \node [style=none] (21) at (-2.75, -1.25) {};
            \node [style=gadget] (22) at (-2.75, -1.25) {$\texttt{a}$};
            \node [style=none] (27) at (0.75, -1.25) {};
            \node [style=gadget] (28) at (0.75, -1.25) {$\texttt{c}$};
            \node [style=none] (31) at (-2.5, 3.25) {};
            \node [style=none] (32) at (1.5, 3.25) {};
            \node [style=gadget] (34) at (-1, 3.25) {$a \lor d$};
            \node [style=none] (41) at (0.75, 1.75) {};
            \node [style=none] (49) at (4.5, -1.25) {};
            \node [style=gadget] (50) at (4.5, -1.25) {$\texttt{e}$};
            \node [style=none] (52) at (4.5, -2.75) {};
            \node [style=none] (53) at (-1, -2.75) {};
            \node [style=none] (56) at (2.75, 3.25) {};
            \node [style=none] (58) at (6.5, -3.75) {};
            \node [style=gadget] (67) at (5.5, 3.25) {$e \lor f$};
            \node [style=none] (72) at (4.5, 3.25) {};
            \node [style=none] (73) at (6.5, 3.25) {};
            \node [style=gadget] (74) at (-1, -1.25) {$\texttt{b}$};
            \node [style=gadget] (75) at (2.75, -1.25) {$\texttt{d}$};
            \node [style=gadget] (76) at (6.5, -1.25) {$\texttt{f}$};
            \node [style=none] (78) at (5, 3) {};
            \node [style=none] (79) at (-4.25, -1.25) {};
            \node [style=none] (80) at (8, -1.25) {};
            \node [style=none] (81) at (-0.5, -1.25) {};
            \node [style=none] (85) at (-2.25, -1.25) {};
            \node [style=none] (87) at (-0.5, -0.75) {};
            \node [style=none] (90) at (1.25, -1.25) {};
            \node [style=none] (92) at (1.25, 2) {};
            \node [style=none] (93) at (-0.25, 2) {};
            \node [style=none] (97) at (3.25, -1.25) {};
            \node [style=none] (98) at (3.25, 3.5) {};
            \node [style=none] (99) at (-0.5, 3.5) {};
            \node [style=none] (100) at (5, -1.25) {};
            \node [style=none] (103) at (7, -1.25) {};
            \node [style=none] (104) at (7, 3.5) {};
            \node [style=none] (105) at (6, 3.5) {};
            \node [style=gadget] (116) at (1.75, -2.75) {$\neg b \lor \neg e$};
            \node [style=gadget] (117) at (1.75, -3.75) {$\neg a \lor \neg f$};
            \node [style=none] (129) at (1.5, -2.5) {};
            \node [style=none] (130) at (-0.5, -2.5) {};
            \node [style=none] (132) at (5, -3) {};
            \node [style=none] (133) at (2, -3) {};
            \node [style=none] (134) at (2, -4) {};
            \node [style=none] (135) at (7, -4) {};
            \node [style=none] (136) at (-2.25, -3.5) {};
            \node [style=none] (137) at (1.5, -3.5) {};
            \node [style=none] (141) at (-0.5, -0.25) {};
            \node [style=none] (142) at (1.25, -0.75) {};
            \node [style=none] (145) at (1.25, -0.25) {};
            \node [style=none] (146) at (3.25, -0.75) {};
            \node [style=none] (149) at (3.25, -0.25) {};
            \node [style=none] (150) at (5, -0.75) {};
            \node [style=none] (153) at (5, -0.25) {};
            \node [style=none] (154) at (7, -0.75) {};
            \node [style=none] (157) at (7, -0.25) {};
            \node [style=none] (158) at (-3, -0.5) {};
            \node [style=none] (159) at (-3, -1) {};
            \node [style=none] (160) at (-2.5, -1) {};
            \node [style=none] (161) at (-1.25, -1) {};
            \node [style=none] (162) at (-1.25, -0.5) {};
            \node [style=none] (163) at (-1, -0.5) {};
            \node [style=none] (164) at (0.5, -1) {};
            \node [style=none] (165) at (0.5, -0.5) {};
            \node [style=none] (166) at (0.75, -0.5) {};
            \node [style=none] (167) at (2.5, -1) {};
            \node [style=none] (168) at (2.5, -0.5) {};
            \node [style=none] (169) at (2.75, -0.5) {};
            \node [style=none] (170) at (4.25, -1) {};
            \node [style=none] (171) at (4.25, -0.5) {};
            \node [style=none] (172) at (4.5, -0.5) {};
            \node [style=none] (173) at (6.25, -1) {};
            \node [style=none] (174) at (6.25, -0.5) {};
            \node [style=none] (175) at (6.5, -0.5) {};
        \end{pgfonlayer}
        \begin{pgfonlayer}{edgelayer}
            \draw [style={connectivity_requester_wire}] (5) to (3.center);
            \draw [style={connectivity_requester_wire}] (5) to (2.center);
            \draw [style={connectivity_requester_wire}] (16.center) to (12.center);
            \draw [style={connectivity_requester_wire}] (22) to (17.center);
            \draw [style={connectivity_requester_wire}] (34) to (32.center);
            \draw [style={connectivity_requester_wire}] (34) to (31.center);
            \draw [style={connectivity_requester_wire}] (41.center) to (3.center);
            \draw [style={connectivity_requester_wire}] (16.center) to (31.center);
            \draw [style={connectivity_requester_wire}] (56.center) to (32.center);
            \draw [style={connectivity_provider_wire}] (75) to (28);
            \draw [style={connectivity_provider_wire}] (28) to (74);
            \draw [style={connectivity_provider_wire}] (74) to (22);
            \draw [style={connectivity_provider_wire}] (75) to (50);
            \draw [style={connectivity_provider_wire}] (50) to (76);
            \draw [style={connectivity_provider_wire}] (79.center) to (22);
            \draw [style={connectivity_provider_wire}] (80.center) to (76);
            \draw [style={connectivity_provider_wire}] (93.center) to (92.center);
            \draw [style={connectivity_provider_wire}] (99.center) to (98.center);
            \draw [style={connectivity_provider_wire}] (105.center) to (104.center);
            \draw [style={connectivity_requester_wire}] (74) to (53.center);
            \draw [style={connectivity_requester_wire}] (58.center) to (117);
            \draw [style={connectivity_requester_wire}] (117) to (17.center);
            \draw [style={connectivity_requester_wire}] (53.center) to (116);
            \draw [style={connectivity_requester_wire}] (116) to (52.center);
            \draw [style={connectivity_requester_wire}] (52.center) to (50);
            \draw [style={connectivity_requester_wire}] (73.center) to (67);
            \draw [style={connectivity_requester_wire}] (67) to (72.center);
            \draw [style={connectivity_requester_wire}] (76) to (58.center);
            \draw [style={connectivity_provider_wire}] (130.center) to (81.center);
            \draw [style={connectivity_provider_wire}] (130.center) to (129.center);
            \draw [style={connectivity_provider_wire}] (132.center) to (100.center);
            \draw [style={connectivity_provider_wire}] (133.center) to (132.center);
            \draw [style={connectivity_provider_wire}] (135.center) to (103.center);
            \draw [style={connectivity_provider_wire}] (135.center) to (134.center);
            \draw [style={connectivity_provider_wire}] (85.center) to (136.center);
            \draw [style={connectivity_provider_wire}] (137.center) to (136.center);
            \draw [style={connectivity_provider_wire}] (81.center) to (87.center);
            \draw [style={connectivity_provider_wire}] (141.center) to (9.center);
            \draw [style={connectivity_provider_wire}] (100.center) to (150.center);
            \draw [style={connectivity_provider_wire}] (153.center) to (78.center);
            \draw [style={connectivity_provider_wire}] (104.center) to (157.center);
            \draw [style={connectivity_provider_wire}] (154.center) to (103.center);
            \draw [style={connectivity_provider_wire}] (146.center) to (97.center);
            \draw [style={connectivity_provider_wire}] (149.center) to (98.center);
            \draw [style={connectivity_provider_wire}] (142.center) to (90.center);
            \draw [style={connectivity_provider_wire}] (145.center) to (92.center);
            \draw [style={connectivity_provider_wire}] (157.center) to (154.center);
            \draw [style={connectivity_provider_wire}] (153.center) to (150.center);
            \draw [style={connectivity_provider_wire}] (149.center) to (146.center);
            \draw [style={connectivity_provider_wire}] (145.center) to (142.center);
            \draw [style={connectivity_provider_wire}] (141.center) to (87.center);
            \draw [style={connectivity_requester_wire}] (73.center) to (76);
            \draw [style={connectivity_requester_wire}] (72.center) to (50);
            \draw [style={connectivity_requester_wire}] (41.center) to (28);
            \draw [style={connectivity_requester_wire}] (5) to (74);
            \draw [style={connectivity_requester_wire}] (56.center) to (75);
            \draw [style={connectivity_requester_wire}] (2.center) to (160.center);
            \draw [style={connectivity_requester_wire}] (159.center) to (158.center);
            \draw [style={connectivity_requester_wire}] (158.center) to (12.center);
            \draw [style={connectivity_requester_wire}] (161.center) to (162.center);
            \draw [style={connectivity_requester_wire}] (163.center) to (162.center);
            \draw [style={connectivity_requester_wire}] (164.center) to (165.center);
            \draw [style={connectivity_requester_wire}] (166.center) to (165.center);
            \draw [style={connectivity_requester_wire}] (167.center) to (168.center);
            \draw [style={connectivity_requester_wire}] (169.center) to (168.center);
            \draw [style={connectivity_requester_wire}] (170.center) to (171.center);
            \draw [style={connectivity_requester_wire}] (172.center) to (171.center);
            \draw [style={connectivity_requester_wire}] (173.center) to (174.center);
            \draw [style={connectivity_requester_wire}] (175.center) to (174.center);
        \end{pgfonlayer}
        \end{tikzpicture}

        \caption{The corresponding environment $G$.}
        \label{fig:grid_graph}
    \end{subfigure}
    
    \caption{NP-hardness: overall idea of the reduction.}
    \label{fig:reduction_bouned_example}
\end{figure}

\begin{remark}
	If a variable $x$ appears only positively (resp. negatively), we connect the top (resp. bottom) requester wire to the nearest provider wire in $\switch{x}$. 
	If a variable $x$ appears positively in a single clause we connect the left requester wire of~$\splitgadget{x}$ to the right requester wire (see for example $b, c, d, e, f$ in \Cref{fig:grid_graph}).
\end{remark}

	\noindent\textbf{Clause Gadget.}
\Cref{fig:clause_gadget} shows a clause gadget. Here, the wire of requesters touching $\ell_i$ is linked to the positive output of the variable gadget for the variable $\ell_i$ when $\ell_i$ is positive, and to the negative output if $\ell_i$ is negative.
	In a clause gadget, the single requester \singlerequester at the center, called the
	\emph{isolated requester}, plays an important role.
	Suppose that the valuation chosen in the clause gadget satisfies the clause.
	This means that some literal $\ell_i$ is satisfied, so the requester wire \wirerequesters for this literal is connected to its variable gadget. 
	Therefore the corresponding agent $\ell_i$ is free to move towards the isolated requester \singlerequester at step 1. 
	Because $\ell_i$ will ensure connectivity to the isolated requester at step $2$, 
	So $t$ is free to move up at step $1$. Agent $c$ can thus move right at step 1 and reach its target at step 2 while all other agents return to their respective targets.
	In fact, $c$ can reach its target only when the valuation satisfies the clause following the described scenario.

\begin{remark}
	For a clause containing 2 literals instead of 3, the wire of requesters of~$\ell_2$ is never connected to any variable gadget.
	It follows that the agent $\ell_2$ cannot move and must give connectivity to the two requesters agent below them (see \Cref{fig:clause_content_np_hardness}).
	Therefore the agents $\ell_1$ and $\ell_3$ are the only ones capable of providing the connectivity to the isolated requester of the clause gadget.
	\label{remark:clause_two_lit}
\end{remark}

	\noindent\textbf{CMAPF environment construction.} 
	We now describe  a polynomial time algorithm to transform the \MPLtwoOne \tsat instance graph $\inciGraphSign{\phi, \pi}$ into our CMAPF environment $G$. 
	The intuitive idea is to replace each vertex (variable or clause) in the embedding of $\inciGraphSign{\phi, \pi}$ by its corresponding gadget (see \Cref{fig:reduction_bouned_example}).
	However, we work with a slightly modified graph $G'$ obtained from $\inciGraphSign{\phi, \pi}$ by adding some new edges as follows.
	Since formula $\phi$ is linear, for each clause~$c$ in $\phi$, there is at least one literal $\ell_c$ that appears only in that clause (otherwise the clause $c$ would intersect at least two other clauses). 
	For each clause $c$, we duplicate in $G'$ the edge $vc$ where $v$ is the variable in $\ell_c$.
	The graph $G'$ is planar and by \Cref{prop:operations_preserve_planar_and_conn} and by \Cref{prop:poly_time_embedding_modif}, $G'$ is constructible from $\inciGraphSign{\phi, \pi}$ in poly-time.
	Moreover, by splitting newly added edges\footnote{Splitting such edge have no consequences since the newly added vertices in $G'$ are replaced by some provider agent (corresponding to a node) within a provider wire in $G$.} the multigraph $G'$ is transformed into a simple graph such that a rectilinear embedding is constructible in poly-time~\cite{planarToRectilinear}.

	To construct $G$, each vertex of $G'$ is replaced by its corresponding gadget.
	Concerning the provider wires, we proceed as follows. 
	First, the variable gadgets are connected in a line, following the variable cycle $\pi$ ($\texttt{a}$ connected to $\texttt{b}$ connected to $\texttt{c}$, etc. in \Cref{fig:reduction_bouned_example}).
	Notice that this line forms a connected component in the communication graph since the $\switch{x}$ gadgets pass connectivity from left providers to right providers 
	(see \Cref{fig:switch_np_hardness}).
	
	Now we connect the clause gadgets to the variable gadget line.
	If there is a single edge between a variable $x$ and a clause $c$, then it is replaced by a wire of requesters \wirerequesters.
	If there are two, then we replace one with a wire of requesters, and the other one with a wire of providers \wireproviders\ .
	The wire of providers is positioned such that it does not give connectivity to the wire of requesters\footnote{Take for instance the edges from the variable gadget $b$ to the clause gadget $a \lor b \lor c$ in \Cref{fig:grid_graph}, here the dashed line representing the wire of requesters is far enough from the plain line representing the wire of providers avoiding giving connectivity to requesters for ``free''.}.
	As explained in the paragraph dedicated to the construction of $G'$, by linearity of the formula $\phi$ each clause contains a literal with a single occurrence.
	This implies by construction that each clause is connected to the line of variable gadget since the edge of this variable gadget $xc$ have been duplicated and replaced by a provider wire (see for instance \Cref{fig:grid_graph}, where the line of variable gadget is connected to each clause gadget).
	Therefore, in our construction, all provider wires are connected to each other.
	
	We set the radius $\radius = 1$.
	We just described an environment $G = (V, E_M, \radius)$. We define an instance $(G, s, t, b)$ of Bounded 2D CMAPF where $s$ and $t$ are respectively the initial and final configuration where the positions of the agents in $s$ (resp. in~$t$) are given by the very positions of the agents as described in the gadgets (resp. the nodes \begin{tikzpicture}
	\node [style={target}] (1) at (0, 0.5) {};
	\end{tikzpicture})
	and we set $b = 2$. In particular, the variable agents $x$ in the variable gadgets start in the middle vertex. As said previously, we have:
	
	\begin{claim}
		The obtained bounded 2D CMAPF instance $(G, s, t, b)$ is computable in poly-time in $|\phi|$. 
		\label{claim:bounded2DCMAPFpolytime}
	\end{claim}	
	\begin{proof}
		Let $k$ denote the size of a clause gadget, and $n$ the size of a variable gadget.
		Recall that $\inciGraphSign{\phi, \pi} = (V \cup C, E)$ with $C$ being the set of clause vertices and $V$ the set of variable vertices.
		We can select a constant $c$ to produce a grid large enough to contain all the gadget and all the wires produced from $E$: $S = c \cdot (|C|\cdot k + |V| \cdot n)$ with $S$ being the size of the grid. 
	\end{proof}

	\begin{claim}
		$\phi$ is satisfiable iff $s \rightarrowcmapfbounded G 2 t$.
		\label{claim:bounded2DCMAPFreductioncorrection}
	\end{claim}

\begin{proof}
	\fbox{$\Rightarrow$} If $\phi$ is satisfiable, there is an assignment $\assignment$ making $\phi$ true. At the first step, we move each agent $x$ in the variable gadget to its value in $\assignment$, that is, $x$ moves up if $\nu(x)=1$ and down otherwise in \Cref{fig:variable_gadget}. When $x$ moves up (resp. down), this gives connectivity to the wire of requesters towards agents $\ell_i$ in clause gadgets with $\ell_i = x$ (resp. $\ell_i = \lnot x$). 

	If $x$ is set to true, 
	then agent $s_x$ in the split gadgets moves up.
	This is possible since the requesters below the split gadget are now connected to agent $x$.
	Note that at this step, $s_x$ inside split gadgets become connected to the requesters reaching the clause gadgets.
	Recall that if $x$ is set to false, then at step $2$, $x$ provides connectivity to all requesters connecting the variable gadget to the clause containing $\lnot x$.

	Finally, consider a clause gadget for clause $c$, and let a literal $\ell_i$ that is satisfied in the current valuation. Then because all requesters touching agent $\ell_i$ are now connected,	agent $\ell_i$ can move towards the isolated requester.
	At the same time, agent $t$ moves up, and agent $c$ moves right towards its target.
	At the next step, $c$ reaches its target, and all other agents move back to their initial positions.
	Because $\nu \models \phi$, the move of agent $c$ occurs in \emph{all} clause gadgets. So \emph{all} agents reach their targets and do remain connected.

	\fbox{$\Leftarrow$} 
	Suppose  $s \rightarrowcmapfbounded G 2 t$.
	In particular all agents $c$ in clauses reach their targets in configuration $t$. This means that the agent $t$ moved up at the first step. 
	For this, some agent $\literal_i$ must have moved towards the center of the clause gadget away from the wire of requesters touching it.
	But this is only possible if these requesters remain connected.
	This means that, for a positive literal, agent $s_x$ inside the split gadget between this clause gadget and the variable gadget for $x$ has moved up (where $x$ is the variable appearing in $\ell_i$).
	In turn, this is only possible if agent $x$ has moved up so as to satisfy $\ell_i$ in the variable gadget.
	For a negative literal, there is no split gadget, and this means that $x$ has moved down in the variable gadget.

	Let $\nu$ be the valuation given by 
	$$\nu(x) = \begin{cases}
		1 \text{ if agent $x$ in the variable gadget for variable $x$ moved up} \\
		0 \text{ otherwise.}
	\end{cases}.$$ 
	Note that $\nu$ is well-defined since it is impossible for an agent $x$ in a variable gadget to have been up and down, due to the time limit of 2 steps. We get that $\nu \models \phi$.
\end{proof}

\subsection{PSPACE-completeness of CMAPF}

\begin{theorem}
	2D CMAPF is PSPACE-complete.
	\label{th:2D CMAPFPSPACEcomplete}
\end{theorem}

We reduce \MPLtwoOne \tsat Reconfiguration to 2D CMAPF.
Consider  an instance of \MPLtwoOne \tsat Reconfiguration $(\phi, \pi, \nu, \nu')$, and a valid planar embedding of $\inciGraphSign{\phi,\pi}$ (this embedding is implicitly given in the instance). 
Note that by linearity of $\phi$, there are at most \emph{two} occurrences of each positive literal, and 
by \Cref{th:linear_planar_tsat_pspace_c}, 
we can assume that there is at most one occurrence of each negative literal.

We construct a 2D CMAPF instance $(G, s, t)$ by replacing nodes of $\inciGraphSign{\phi,\pi}$ by gadgets as follows.

{
Let $\pi'$ denote an ordering of the variables obtained by removing one edge of the variable cycle $\pi$. So we consider a line of variables.
}

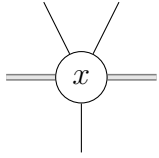
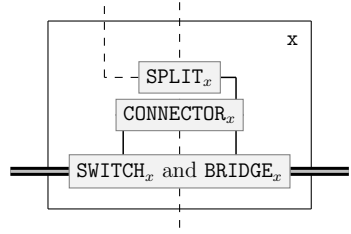
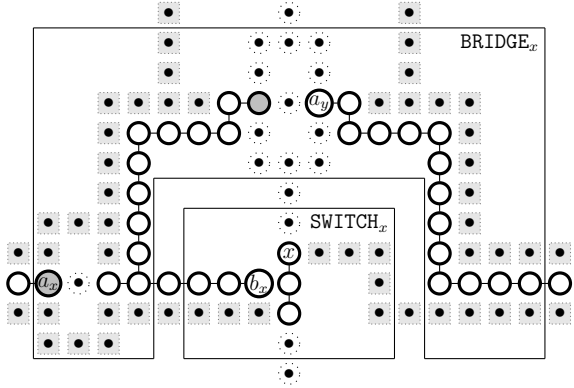
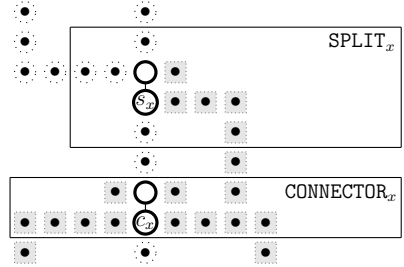
\begin{figure}[H]
    \centering
    \pgfsetlayers{edgelayer,nodelayer}
    
    \begin{subfigure}[t]{0.45\linewidth}
        \centering
        \begin{tikzpicture}[
            scale=1, 
            every node/.style={scale=1}
        ]
        \begin{pgfonlayer}{nodelayer}
            \node (l) at (0, 0) [draw, circle, fill=white] {$x$};
        \end{pgfonlayer}
        \begin{pgfonlayer}{edgelayer}
            \draw[-] (0.5,1) -- (l);
            \draw[-] (-0.5,1) -- (l);
            \draw[-] (0,-1) -- (l);
            \draw[variablecycle] (1,0) -- (l);
            \draw[variablecycle] (-1,0) -- (l);
        \end{pgfonlayer}
        \end{tikzpicture}

        \caption{A variable $x$ in $\inciGraphSign{\phi,\pi}$.}
    \end{subfigure}
    \hfill
    \begin{subfigure}[t]{0.45\linewidth}
        \centering
        \begin{tikzpicture}[
            scale=1, 
            every node/.style={scale=0.7}
        ]
        \begin{pgfonlayer}{nodelayer}
            \node [style=none] (0) at (0, 1.75) {};
            \node [style=none] (1) at (-1, 0.75) {};
            \node [style=none] (3) at (0, 0.75) {};
            \node [style=gadget] (4) at (0, 0.75) {$\splitgadget{x}$};
            \node [style=none] (5) at (-1, 1.75) {};
            \node [style=gadget] (6) at (0, -0.5) {$\switch{x}$ and $\bridge{x}$};
            \node [style=none] (7) at (0, -1.25) {};
            \node [style=none] (9) at (2.25, -0.5) {};
            \node [style=none] (10) at (-2.25, -0.5) {};
            \node [style=none] (11) at (1.75, 1.5) {};
            \node [style=none] (12) at (-1.75, 1.5) {};
            \node [style=none] (13) at (-1.75, -1) {};
            \node [style=none] (14) at (1.75, -1) {};
            \node [style=none] (15) at (1.5, 1.25) {$\texttt{x}$};
            \node [style=gadget] (16) at (0, 0.25) {$\texttt{CONNECTOR}_x$};
            \node [style=none] (17) at (0.75, 0.25) {};
            \node [style=none] (18) at (0.75, -0.5) {};
            \node [style=none] (19) at (-0.75, -0.5) {};
            \node [style=none] (20) at (-0.75, 0.25) {};
            \node [style=none] (21) at (0.75, 0.75) {};
        \end{pgfonlayer}
        \begin{pgfonlayer}{edgelayer}
            \draw [style={connectivity_requester_wire}] (6) to (7.center);
            \draw [style={connectivity_requester_wire}] (6) to (4);
            \draw [style={connectivity_requester_wire}] (4) to (1.center);
            \draw [style={connectivity_requester_wire}] (1.center) to (5.center);
            \draw (11.center) to (14.center);
            \draw (14.center) to (13.center);
            \draw (13.center) to (12.center);
            \draw (12.center) to (11.center);
            \draw [style={connectivity_requester_wire}] (0.center) to (4);
            \draw [style={connectivity_provider_wire}] (20.center) to (19.center);
            \draw [style={connectivity_provider_wire}] (20.center) to (16);
            \draw [style={connectivity_provider_wire}] (17.center) to (16);
            \draw [style={connectivity_provider_wire}] (17.center) to (18.center);
            \draw [style={connectivity_provider_wire}] (21.center) to (4);
            \draw [style={connectivity_provider_wire}] (21.center) to (17.center);
            \draw [style={movement_wire}] (6) to (10.center);
            \draw [style={movement_wire}] (6) to (9.center);
        \end{pgfonlayer}
        \end{tikzpicture}
        \caption{Abstract representation of the corresponding gadget. 
        See \Cref{fig:switch_and_bridge_pspace_hardness} for the content of 
        $\switch{x}$ and $\bridge{x}$ and \Cref{fig:connecter_split_gadget} 
         for $\texttt{CONNECTOR}_x$ and $\splitgadget{x}$.}
        \label{fig:scheme_var_gadget_pspace}
    \end{subfigure}

    \begin{subfigure}[t]{0.52\linewidth}
        \centering
        \begin{tikzpicture}[
            scale=0.8, 
            every node/.style={scale=0.7}
        ]
        \begin{pgfonlayer}{nodelayer}
            \node [style={movement_node}] (0) at (0, 0) {};
            \node [style={movement_node}] (1) at (0, 0.5) {$x$};
            \node [style={movement_node}] (2) at (0, -0.5) {};
            \node [style={movement_node}] (3) at (-0.5, 0) {$b_x$};
            \node [style={movement_node}] (4) at (-1, 0) {};
            \node [style={movement_node}] (5) at (-1.5, 0) {};
            \node [style={movement_node}] (6) at (-2, 0) {};
            \node [style={movement_node}] (10) at (-2.5, 0) {};
            \node [style={green_agent}] (17) at (0, 1.5) {};
            \node [style={green_agent}] (25) at (-0.5, 4) {};
            \node [style={green_agent}] (28) at (0.5, 4) {};
            \node [style={green_agent}] (29) at (0, 4) {};
            \node [style={green_agent}] (31) at (0, -1) {};
            \node [style={green_agent}] (32) at (0, -1.5) {};
            \node [style={blue_agent}] (54) at (-4, 1) {};
            \node [style={movement_node}] (57) at (-2.5, 1.5) {};
            \node [style={movement_node}] (63) at (-1, 2.5) {};
            \node [style={blue_agent}] (67) at (-1.5, 3) {};
            \node [style={blue_agent}] (68) at (-2, 3) {};
            \node [style={blue_agent}] (69) at (-2.5, 3) {};
            \node [style={blue_agent}] (71) at (-3, 3) {};
            \node [style={blue_agent}] (72) at (-3, 2.5) {};
            \node [style={blue_agent}] (73) at (-3, 2) {};
            \node [style={blue_agent}] (74) at (-3, 1.5) {};
            \node [style={movement_node}] (77) at (1, 2.5) {};
            \node [style={blue_agent}] (79) at (-0.5, -0.5) {};
            \node [style={blue_agent}] (80) at (-1, -0.5) {};
            \node [style={blue_agent}] (81) at (-1.5, -0.5) {};
            \node [style={blue_agent}] (82) at (-2, -0.5) {};
            \node [style={blue_agent}] (83) at (-2.5, -0.5) {};
            \node [style={blue_agent}] (85) at (-4, -0.5) {};
            \node [style={movement_node}] (87) at (-4.5, 0) {};
            \node [style={blue_agent}] (88) at (-4, 0.5) {};
            \node [style={blue_agent}] (89) at (-4.5, 0.5) {};
            \node [style={movement_node}] (93) at (2.5, 2) {};
            \node [style={movement_node}] (94) at (2.5, 1.5) {};
            \node [style={movement_node}] (95) at (2.5, 1) {};
            \node [style={movement_node}] (97) at (2.5, 0) {};
            \node [style={movement_node}] (98) at (3, 0) {};
            \node [style={movement_node}] (99) at (3.5, 0) {};
            \node [style={blue_agent}] (104) at (1.5, 3) {};
            \node [style={blue_agent}] (105) at (2, 3) {};
            \node [style={blue_agent}] (107) at (2.5, 3) {};
            \node [style={blue_agent}] (108) at (3, 3) {};
            \node [style={blue_agent}] (109) at (3, 2.5) {};
            \node [style={blue_agent}] (110) at (3, 2) {};
            \node [style={blue_agent}] (112) at (3, 1.5) {};
            \node [style={blue_agent}] (113) at (3, 0.5) {};
            \node [style={blue_agent}] (114) at (3.5, 0.5) {};
            \node [style={movement_node}] (116) at (-1.5, 2.5) {};
            \node [style={movement_node}] (118) at (1.5, 2.5) {};
            \node [style={blue_agent}] (119) at (-2, 3.5) {};
            \node [style={blue_agent}] (120) at (-2, 4) {};
            \node [style={green_agent}] (121) at (0.5, 2.5) {};
            \node [style={movement_node}] (122) at (-2.5, 0.5) {};
            \node [style={green_agent}] (123) at (-3.5, 0) {};
            \node [style={blue_agent}] (124) at (-4, -1) {};
            \node [style={blue_agent}] (131) at (-3.5, -1) {};
            \node [style={movement_node}] (132) at (-3, 0) {};
            \node [style={blue_agent}] (133) at (-3, -0.5) {};
            \node [style={blue_agent}] (134) at (-3, -1) {};
            \node [style={blue_agent}] (135) at (-3.5, 1) {};
            \node [style={blue_agent}] (136) at (-3, 0.5) {};
            \node [style={movement_node}] (137) at (-2, 2.5) {};
            \node [style={blue_agent}] (138) at (-4.5, -0.5) {};
            \node [style={blue_agent}] (139) at (3.5, -0.5) {};
            \node [style={blue_agent}] (140) at (3, -0.5) {};
            \node [style={blue_agent}] (141) at (2.5, -0.5) {};
            \node [style={blue_agent}] (142) at (2, -0.5) {};
            \node [style={blue_agent}] (144) at (1.5, 0) {};
            \node [style={blue_agent}] (145) at (1.5, 0.5) {};
            \node [style={blue_agent}] (146) at (1, 0.5) {};
            \node [style={blue_agent}] (147) at (0.5, 0.5) {};
            \node [style={blue_agent}] (148) at (-3, 1) {};
            \node [style={green_agent}] (149) at (0, 1) {};
            \node [style={green_agent}] (150) at (-0.5, 2) {};
            \node [style={green_agent}] (151) at (-0.5, 2.5) {};
            \node [style={green_agent}] (152) at (0, 2) {};
            \node [style={blue_agent}] (153) at (1.5, -0.5) {};
            \node [style={movement_node}] (154) at (2.5, 0.5) {};
            \node [style={blue_agent}] (155) at (3, 1) {};
            \node [style={movement_node}] (156) at (-2.5, 1) {};
            \node [style=none] (157) at (-1.75, 1.25) {};
            \node [style=none] (158) at (1.75, 1.25) {};
            \node [style=none] (159) at (-1.75, -1.25) {};
            \node [style=none] (160) at (1.75, -1.25) {};
            \node [style=none] (161) at (1, 1) {$\texttt{SWITCH}_x$};
            \node [style=none] (162) at (3.5, 4) {$\texttt{BRIDGE}_x$};
            \node [style=none] (163) at (-2.25, 1.75) {};
            \node [style=none] (164) at (2.25, 1.75) {};
            \node [style=none] (165) at (2.25, -1.25) {};
            \node [style=none] (166) at (4.25, -1.25) {};
            \node [style=none] (167) at (4.25, 4.25) {};
            \node [style=none] (168) at (-4.25, 4.25) {};
            \node [style=none] (169) at (-4.25, -1.25) {};
            \node [style=none] (170) at (-2.25, -1.25) {};
            \node [style={movement_node}] (171) at (4, 0) {};
            \node [style={movement_node}] (172) at (4.5, 0) {};
            \node [style={blue_agent}] (173) at (4, 0.5) {};
            \node [style={blue_agent}] (174) at (4.5, 0.5) {};
            \node [style={blue_agent}] (175) at (4, -0.5) {};
            \node [style={blue_agent}] (176) at (4.5, -0.5) {};
            \node [style={green_agent}] (179) at (0, 4.5) {};
            \node [style={green_agent}] (183) at (0.5, 2) {};
            \node [style={green_agent}] (184) at (0, 3) {};
            \node [style={green_agent}] (185) at (-0.5, 3.5) {};
            \node [style={green_agent}] (186) at (0.5, 3.5) {};
            \node [style=target] (187) at (-4, 0) {$a_x$};
            \node [style=target] (188) at (-0.5, 3) {};
            \node [style={blue_agent}] (191) at (-2, 4.5) {};
            \node [style={movement_node}] (193) at (1, 3) {};
            \node [style={movement_node}] (194) at (0.5, 3) {$a_y$};
            \node [style={movement_node}] (195) at (-1, 3) {};
            \node [style={movement_node}] (196) at (-2.5, 2) {};
            \node [style={movement_node}] (197) at (-2.5, 2.5) {};
            \node [style={movement_node}] (198) at (2, 2.5) {};
            \node [style={movement_node}] (199) at (2.5, 2.5) {};
            \node [style={blue_agent}] (224) at (2, 3.5) {};
            \node [style={blue_agent}] (225) at (2, 4) {};
            \node [style={blue_agent}] (226) at (2, 4.5) {};
        \end{pgfonlayer}
        \begin{pgfonlayer}{edgelayer}
            \draw (2) to (0);
            \draw (0) to (1);
            \draw (6) to (5);
            \draw (5) to (4);
            \draw (4) to (3);
            \draw (10) to (6);
            \draw (93) to (94);
            \draw (94) to (95);
            \draw (97) to (98);
            \draw (98) to (99);
            \draw (116) to (63);
            \draw (118) to (77);
            \draw (10) to (132);
            \draw (122) to (10);
            \draw (116) to (137);
            \draw (95) to (154);
            \draw (154) to (97);
            \draw (57) to (156);
            \draw (122) to (156);
            \draw (171) to (99);
            \draw (172) to (171);
            \draw (187) to (87);
            \draw (195) to (63);
            \draw (195) to (188);
            \draw (194) to (193);
            \draw (193) to (77);
            \draw (167.center) to (166.center);
            \draw (165.center) to (166.center);
            \draw (165.center) to (164.center);
            \draw (164.center) to (163.center);
            \draw (163.center) to (170.center);
            \draw (157.center) to (158.center);
            \draw (158.center) to (160.center);
            \draw (160.center) to (159.center);
            \draw (157.center) to (159.center);
            \draw (167.center) to (168.center);
            \draw (170.center) to (169.center);
            \draw (169.center) to (168.center);
            \draw (57) to (196);
            \draw (196) to (197);
            \draw (197) to (137);
            \draw (118) to (198);
            \draw (198) to (199);
            \draw (199) to (93);
        \end{pgfonlayer}
        \end{tikzpicture}
    
        \caption{Corresponding gadgets $\bridge x$ and $\switch x$ for variable~$x$. Agent $x$ determines the value of $x$ (up = true, down = false).
         Agent $x$ can move and switch its value only if $b_x$ is at the shown position. 
        Agent $a_y$ can also enter in $\bridge y$ where $y$ is the variable on the right of $x$.
        If $x$ is the left-most variable, then $a_x$ is replaced by a provider. If $x$ is the right-most variable, then $a_y$ is replaced by requester and we add providers to connect the bottom-right providers to the top-right providers
        (for example by replacing the empty cell on the extreme right by a provider).
        }
      \label{fig:switch_and_bridge_pspace_hardness}
    \end{subfigure}
    \hfill
    \begin{subfigure}[t]{0.45\linewidth}
        \centering
        \begin{tikzpicture}[
            scale=0.8, 
            every node/.style={scale=0.7}
        ]
        \begin{pgfonlayer}{nodelayer}
            \node [style={green_agent}] (0) at (0, -0.75) {};
            \node [style={blue_agent}] (1) at (-2, -0.75) {};
            \node [style={green_agent}] (2) at (0, 0.75) {};
            \node [style={movement_node}] (3) at (0, -0.25) {$c_x$};
            \node [style={movement_node}] (4) at (0, 0.25) {};
            \node [style=none] (5) at (4.25, -0.5) {};
            \node [style=none] (6) at (-2.25, -0.5) {};
            \node [style=none] (7) at (-2.25, 0.5) {};
            \node [style=none] (8) at (4.25, 0.5) {};
            \node [style=none] (9) at (3.25, 0.25) {$\texttt{CONNECTOR}_x$};
            \node [style={blue_agent}] (10) at (-0.5, 0.25) {};
            \node [style={blue_agent}] (11) at (-0.5, -0.25) {};
            \node [style={blue_agent}] (12) at (-2, -0.25) {};
            \node [style={blue_agent}] (13) at (-1.5, -0.25) {};
            \node [style={blue_agent}] (14) at (-1, -0.25) {};
            \node [style={blue_agent}] (15) at (0.5, -0.25) {};
            \node [style={blue_agent}] (16) at (1, -0.25) {};
            \node [style={blue_agent}] (17) at (1.5, -0.25) {};
            \node [style={blue_agent}] (18) at (2, -0.25) {};
            \node [style={blue_agent}] (19) at (0.5, 0.25) {};
            \node [style={blue_agent}] (20) at (2, -0.75) {};
            \node [style={blue_agent}] (21) at (1.5, 0.25) {};
            \node [style={blue_agent}] (22) at (1.5, 0.75) {};
            \node [style={green_agent}] (25) at (-0.5, 2.25) {};
            \node [style={movement_node}] (26) at (0, 1.75) {$s_x$};
            \node [style={movement_node}] (27) at (0, 2.25) {};
            \node [style={green_agent}] (28) at (0, 2.75) {};
            \node [style={blue_agent}] (29) at (0.5, 1.75) {};
            \node [style={blue_agent}] (30) at (0.5, 2.25) {};
            \node [style={green_agent}] (31) at (0, 1.25) {};
            \node [style={blue_agent}] (33) at (1, 1.75) {};
            \node [style={blue_agent}] (34) at (1.5, 1.75) {};
            \node [style={green_agent}] (35) at (-1, 2.25) {};
            \node [style={green_agent}] (35) at (-1.5, 2.25) {};
            \node [style={green_agent}] (36) at (-2, 2.25) {};
            \node [style={green_agent}] (37) at (-2, 2.75) {};
            \node [style={green_agent}] (38) at (-2, 3.25) {};
            \node [style={green_agent}] (39) at (0, 3.25) {};
            \node [style=none] (40) at (-1.25, 3) {};
            \node [style=none] (41) at (4.25, 3) {};
            \node [style=none] (42) at (-1.25, 1) {};
            \node [style=none] (43) at (4.25, 1) {};
            \node [style=none] (44) at (3.6, 2.75) {$\texttt{SPLIT}_x$};
            \node [style={blue_agent}] (46) at (1.5, 1.25) {};
        \end{pgfonlayer}
        \begin{pgfonlayer}{edgelayer}
            \draw (4) to (3);
            \draw (5.center) to (6.center);
            \draw (8.center) to (7.center);
            \draw (7.center) to (6.center);
            \draw (8.center) to (5.center);
            \draw (27) to (26);
            \draw (42.center) to (43.center);
            \draw (43.center) to (41.center);
            \draw (41.center) to (40.center);
            \draw (40.center) to (42.center);
        \end{pgfonlayer}
        \end{tikzpicture}
        \caption{Corresponding $\connectorgadget x$ and $\splitgadget{x}$ for variable $x$. $\connectorgadget x$ part transmits connectivity from the providers on the left to the providers
        on the right.
        Agent $c_x$ can move if agent $x$ in $\switch{x}$ is on top position. The source and target positions of agents $c_x$ and $s_x$ are determined by $\nu$ and $\nu'$:
        they are at the bottom position when $x$ is true, and on the top position otherwise.
        }
       \label{fig:connecter_split_gadget}
    \end{subfigure}

    \caption{A variable $x$ in $\inciGraphSign{\phi,\pi}$ is replaced by $\bridge x$ and $\switch x$.}
    \label{fig:variable_gadget_bridge_switch}
\end{figure}

\noindent\textbf{Variable Gadget.}
Consider a variable $x$, and let $y$ be the variable that is the right neighbor of $x$ in the order $\pi'$.
The variable vertex $x$ in $\inciGraphSign{\phi, \pi}$ is replaced by a variable gadget composed by multiple parts given in \Cref{fig:scheme_var_gadget_pspace}.

\emph{Switch Part.}
$\switch x$ (see \Cref{fig:switch_and_bridge_pspace_hardness}) works similarly as the variable gadget in the proof of \Cref{th:bounded2DCMAPFNPcomplete} (given in \Cref{fig:switch_np_hardness}): 
it has a positive (top) and negative (bottom) output depending on the position of the agent $x$.
The positive requester wire (leaving above, crossing through \bridge\xspace) and the negative requester wire (leaving below the gadget) are linked to respectively positive and negative literal agents $x$ and $\neg x$ that appear in the clause gadgets described below.
The initial and target positions of the agent $x$ in $s,t$ are given by valuations $\nu,\nu'$: 
$x$ starts at the top position if $\nu(x) = 1$, and on the bottom position otherwise, and similarly for $\nu'$ and the target position.

\emph{Bridge Part.}
In $\bridge x$, there are two particular requesters that need to be connected by positioning the agents $a_x$, $b_x$ or $a_y$.
First, the \emph{left requester} (to the right of $a_x$ in \Cref{fig:switch_and_bridge_pspace_hardness}) can be connected either
$a_x$ (as shown in the figure), or by~$b_x$.
Note here that $a_x$ also transmits connectivity from the top providers to the bottom providers.
Second, the \emph{top requester} on top of the gadget (to the left of $a_y$ in \Cref{fig:switch_and_bridge_pspace_hardness})
must be connected either by $a_y$ (as shown in the figure), or by $b_x$. 
On the other hand, $b_x$ is the only agent that can provide connectivity to $x$ when $x$ moves from positive to negative, or vice versa.
In fact, for $x$ to flip its value, it is necessary that the left requester is connected via $a_x$, and the top requester connected via $a_y$.

This guarantees that \emph{at most one} agent $x$ can switch its value at any step, simulating a flip of variable $x$.
In fact, for~$x$ to switch its value, $b_x$ must be placed on its left; which means that both $a_x$ and $a_y$ appear in the positions
shown in \Cref{fig:switch_and_bridge_pspace_hardness}. But if $a_y$ is connected to the top requester,
 then $b_y$ is next to the left requester in~$\bridge{y}$, far from~$y$; so $y$ cannot flip its value at this step. 
Let $z$ be the variable that appears on the right of $y$ in $\pi'$.
The position of $a_y$, in turn, means that $z$ cannot flip its value at this step (since $b_z$ must provide connectivity to the left requester). And so on.
A similar argument holds for the variables that are on the left of $x$ in~$\pi'$: because $a_x$ is next to the left requester in $\bridge{x}$,
the variable to the left of $x$ cannot switch values at that step.

In the initial configuration, all agents $a_x,b_x$ are at their target positions (their precise initial positions are actually not important as long as all the requesters have connectivity).

\emph{Split Part.}
The gadget 
$\splitgadget{x}$ is given in \Cref{fig:connecter_split_gadget} and is nearly the same as the one used in th proof of \Cref{th:bounded2DCMAPFNPcomplete} (see \Cref{fig:split_np_hardness}).
The only difference resides in the fact that now the source and target positions of agent $s_x$ are given by $\nu$ and~$\nu'$: the source position is up (resp. down) if $\nu(x) = 1$ (resp. $\nu = 0$). Same for~$\nu'$.

\emph{Connector Part.}
The gadget $\connectorgadget{x}$ in \Cref{fig:connecter_split_gadget} is only used to ensure connectivity from right to left providers.
This is necessary to construct a connected line of providers as in the proof of \Cref{th:bounded2DCMAPFNPcomplete}, while in the bounded case, the connectivity could be directly given in the switch gadget (see~\Cref{fig:switch_np_hardness}).
As in~$\splitgadget{x}$, the source and target position of agent $c_x$ are determined by $\nu$ and $\nu'$.

\begin{remark}
	We recall that an agent can move even after reaching its target: hence agent $b_x$ could provide connectivity to $x$ for multiple flips even if it reached its target to let move $a_y$ multiple times.
	At the end of the reconfiguration, for each variable~$x$, the agents $a_x$ and $b_x$ move one by one to the right until reaching their targets~\begin{tikzpicture}
	\node [style={target}] (1) at (0, 0.5) {};
	\end{tikzpicture}.
\end{remark}

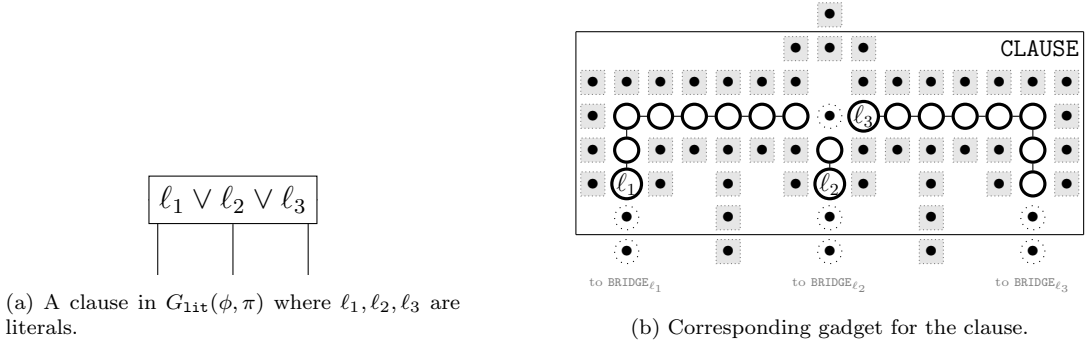
\begin{figure}[H]
    \centering
    \pgfsetlayers{edgelayer,nodelayer}
    
    \begin{subfigure}[b]{0.4\linewidth}
        \centering
        \begin{tikzpicture}[
            scale=1
        ]
            \begin{pgfonlayer}{nodelayer}
                \node (c1) at (0, 0) [draw, rectangle, fill=white] {$\literal_1 \lor \literal_2 \lor \literal_3$};
            \end{pgfonlayer}
            \begin{pgfonlayer}{edgelayer}
                \draw[-] (0,-1) -| (c1);
                \draw[-] (-1,-1) |- (c1.west);
                \draw[-] (1,-1) |- (c1.east);
            \end{pgfonlayer}
        \end{tikzpicture}

        \caption{A clause in $\litClauGraphSign{\phi,\pi}$ where $\literal_1, \literal_2, \literal_3$ are literals.}
    \end{subfigure}
    \hfill
    \begin{subfigure}[b]{0.55\linewidth}
        \centering
        \begin{tikzpicture}[
            scale=0.9, 
            every node/.style={scale=0.8}
        ]
            \begin{pgfonlayer}{nodelayer}
                \node [style={movement_node}] (2) at (-3, 3) {};
                \node [style={movement_node}] (3) at (-2.5, 3) {};
                \node [style={movement_node}] (4) at (-2, 3) {};
                \node [style={movement_node}] (5) at (-1.5, 2.5) {};
                \node [style={movement_node}] (6) at (-1.5, 2) {$\literal_2$};
                \node [style={movement_node}] (7) at (-1, 3) {$\literal_3$};
                \node [style={movement_node}] (8) at (-0.5, 3) {};
                \node [style={movement_node}] (12) at (-4.5, 3) {};
                \node [style={movement_node}] (13) at (-4.5, 2.5) {};
                \node [style={movement_node}] (14) at (-4.5, 2) {$\literal_1$};
                \node [style={green_agent}] (20) at (-1.5, 1.5) {};
                \node [style={green_agent}] (21) at (-1.5, 1) {};
                \node [style={green_agent}] (22) at (-4.5, 1.5) {};
                \node [style={green_agent}] (23) at (-4.5, 1) {};
                \node [style={movement_node}] (30) at (1.5, 3) {};
                \node [style={movement_node}] (31) at (1.5, 2.5) {};
                \node [style={movement_node}] (32) at (1.5, 2) {};
                \node [style={movement_node}] (33) at (1, 3) {};
                \node [style={green_agent}] (34) at (1.5, 1.5) {};
                \node [style={green_agent}] (35) at (1.5, 1) {};
                \node [style={movement_node}] (37) at (0, 3) {};
                \node [style={movement_node}] (38) at (0.5, 3) {};
                \node [style={blue_agent}] (45) at (-2, 2) {};
                \node [style={blue_agent}] (46) at (-2, 2.5) {};
                \node [style={green_agent}] (47) at (-1.5, 3) {};
                \node [style={movement_node}] (48) at (-3.5, 3) {};
                \node [style={movement_node}] (49) at (-4, 3) {};
                \node [style={blue_agent}] (50) at (-2.5, 2.5) {};
                \node [style={blue_agent}] (51) at (-3, 2.5) {};
                \node [style={blue_agent}] (52) at (-3.5, 2.5) {};
                \node [style={blue_agent}] (53) at (-4, 2.5) {};
                \node [style={blue_agent}] (54) at (-4, 2) {};
                \node [style={blue_agent}] (55) at (-3, 2) {};
                \node [style={blue_agent}] (56) at (-3, 1.5) {};
                \node [style={blue_agent}] (57) at (-3, 1) {};
                \node [style={blue_agent}] (58) at (-4.5, 3.5) {};
                \node [style={blue_agent}] (59) at (-4, 3.5) {};
                \node [style={blue_agent}] (60) at (-3.5, 3.5) {};
                \node [style={blue_agent}] (61) at (-3, 3.5) {};
                \node [style={blue_agent}] (62) at (-2.5, 3.5) {};
                \node [style={blue_agent}] (63) at (-2, 3.5) {};
                \node [style={blue_agent}] (64) at (-2, 4) {};
                \node [style={blue_agent}] (65) at (-1.5, 4) {};
                \node [style={blue_agent}] (66) at (-1, 4) {};
                \node [style={blue_agent}] (67) at (-1, 3.5) {};
                \node [style={blue_agent}] (68) at (-0.5, 3.5) {};
                \node [style={blue_agent}] (69) at (0, 3.5) {};
                \node [style={blue_agent}] (70) at (0.5, 3.5) {};
                \node [style={blue_agent}] (71) at (1, 3.5) {};
                \node [style={blue_agent}] (72) at (1.5, 3.5) {};
                \node [style={blue_agent}] (73) at (2, 3.5) {};
                \node [style={blue_agent}] (74) at (2, 3) {};
                \node [style={blue_agent}] (75) at (2, 2.5) {};
                \node [style={blue_agent}] (76) at (2, 2) {};
                \node [style={blue_agent}] (77) at (-1.5, 4.5) {};
                \node [style={blue_agent}] (78) at (-5, 3.5) {};
                \node [style={blue_agent}] (79) at (-5, 3) {};
                \node [style={blue_agent}] (80) at (-5, 2.5) {};
                \node [style={blue_agent}] (81) at (-5, 2) {};
                \node [style={blue_agent}] (82) at (1, 2) {};
                \node [style={blue_agent}] (83) at (1, 2.5) {};
                \node [style={blue_agent}] (84) at (0.5, 2.5) {};
                \node [style={blue_agent}] (85) at (0, 2.5) {};
                \node [style={blue_agent}] (86) at (-0.5, 2.5) {};
                \node [style={blue_agent}] (87) at (-1, 2.5) {};
                \node [style={blue_agent}] (88) at (-1, 2) {};
                \node [style={blue_agent}] (90) at (0, 1.5) {};
                \node [style={blue_agent}] (91) at (0, 1) {};
                \node [style={blue_agent}] (92) at (0, 2) {};
                \node [style=none] (93) at (-5.25, 4.25) {};
                \node [style=none] (94) at (2.25, 4.25) {};
                \node [style=none] (95) at (-5.25, 1.25) {};
                \node [style=none] (96) at (2.25, 1.25) {};
                \node [style=none] (97) at (1.6, 4) {\texttt{CLAUSE}};
                \node [explanation] at (-4.5, 0.5) {to $\bridge {\literal_1}$};
                \node [explanation] at (-1.5, 0.5) {to $\bridge {\literal_2}$};
                \node [explanation] at (1.5, 0.5) {to $\bridge {\literal_3}$};
                           
            \end{pgfonlayer}
            \begin{pgfonlayer}{edgelayer}
                \draw (6) to (5);
                \draw (8) to (7);
                \draw (4) to (3);
                \draw (3) to (2);
                \draw (12) to (13);
                \draw (13) to (14);
                \draw (32) to (31);
                \draw (31) to (30);
                \draw (30) to (33);
                \draw (12) to (49);
                \draw (49) to (48);
                \draw (48) to (2);
                \draw (8) to (37);
                \draw (37) to (38);
                \draw (38) to (33);
                \draw (94.center) to (93.center);
                \draw (95.center) to (93.center);
                \draw (96.center) to (94.center);
                \draw (96.center) to (95.center);
            \end{pgfonlayer}
        \end{tikzpicture}
    
        \caption{Corresponding gadget for the clause.}
        \label{fig:gadget_clause_pspace_detail}
    \end{subfigure}
    
    \caption{Clause gadget for showing the PSPACE-hardness.\label{fig:clause_gadget_PSPACE}}
\end{figure}

\noindent\textbf{Clause Gadget.}
Each clause vertex in $\inciGraphSign{\phi, \pi}$ is replaced by a clause gadget described in Figure~\ref{fig:clause_gadget_PSPACE}.
The clause gadget contains $2$ or $3$ literal agents depending on the size of the clause. %
Let the single requester~\singlerequester at the middle of \Cref{fig:clause_gadget_PSPACE} be called the \emph{central requester} %
(to the left of $\ell_3$ in the figure).
There is a requester wire leaving the clause gadget for each literal $\ell_i$, and reaching the variable gadget (of the variable in $\ell_i$);
thus connectivity to these requesters are ensured either by the corresponding variable agent $x$ (if the valuation of $x$ satisfies the literal $\ell_i$), 
or by the agent $\ell_i$ in the clause gadget if it is on the bottom position (see $\ell_1$ and $\ell_2$ in \Cref{fig:clause_gadget_PSPACE}).
The initial and target positions of each literal agent $\ell_i$ in $s,t$ are given by valuations $\nu,\nu'$ as follows.
If $\nu$ satisfies $\ell_i$, then $\ell_i$ is placed next to the central requester;
and otherwise it is placed on the bottom position next to the requester wire leaving the gadget, and similarly for $\nu'$ and the target configuration.
Because all considered valuations satisfy the formula, it is guaranteed that the central requester is connected.

\newcommand{\bridgeandswitch}[1]{
\!\!\!\!\begin{tabular}{c}
            $\texttt{BRIDGE}_#1$ \\
            + $\texttt{SWITCH}_#1$\end{tabular}\!\!\!\!
       }

\begin{figure}[h]
    \centering
    \pgfsetlayers{edgelayer,nodelayer}
    
    \begin{subfigure}[b]{0.3\linewidth}
        \centering
        \begin{tikzpicture}[
            scale=0.9, 
            every node/.style={scale=0.7}
        ]
        \begin{pgfonlayer}{nodelayer}
            \node [style={positiveclause}] (0) at (-0.5, 2) {$a \lor b \lor c$};
            \node [style={variable_node}] (3) at (0.25, 0.5) {$c$};
            \node [style={positiveclause}] (4) at (2.5, 2) {$e \lor f$};
            \node [style={variable_node}] (5) at (1.75, 0.5) {$e$};
            \node [style={variable_node}] (6) at (2.5, 0.5) {$f$};
            \node [style={variable_node}] (7) at (-0.5, 0.5) {$b$};
            \node [style=none] (8) at (3, 0.5) {};
            \node [style=none] (9) at (3, -1) {};
            \node [style=none] (10) at (-1.75, -1) {};
            \node [style=none] (11) at (-1.75, 0.5) {};
            \node [style={positiveclause}] (13) at (0, 2.5) {$a \lor d$};
            \node [style={variable_node}] (14) at (1, 0.5) {$d$};
            \node [style=none] (15) at (1, 2.5) {};
            \node [style=none] (16) at (-1.5, 2.5) {};
            \node [style=none] (17) at (-1.5, 0.75) {};
            \node [style={variable_node}] (18) at (-1.25, 0.5) {$a$};
            \node [style={negativeclause}] (19) at (0.75, -0.65) {$\neg a \lor \neg f$};
            \node [style={negativeclause}] (20) at (0.75, -0.2) {$\neg b \lor \neg e$};
        \end{pgfonlayer}
        \begin{pgfonlayer}{edgelayer}
            \draw (4) to (5);
            \draw (4) to (6);
            \draw (0) to (7);
            \draw (0) to (3);
            \draw [style={variablecycle}] (7) to (3);
            \draw [style={variablecycle}] (5) to (6);
            \draw [style={variablecycle}] (11.center) to (10.center);
            \draw [style={variablecycle}] (10.center) to (9.center);
            \draw [style={variablecycle}] (9.center) to (8.center);
            \draw [style={variablecycle}] (8.center) to (6);
            \draw (15.center) to (14);
            \draw (15.center) to (13);
            \draw [style={variablecycle}] (3) to (14);
            \draw [style={variablecycle}] (14) to (5);
            \draw (17.center) to (16.center);
            \draw (16.center) to (13);
            \draw [style={variablecycle}] (18) to (7);
            \draw [style={variablecycle}] (18) to (11.center);
            \draw (18) to (0);
            \draw (17.center) to (18);
            \draw (18) to (19);
            \draw (19) to (6);
            \draw (5) to (20);
            \draw (20) to (7);
        \end{pgfonlayer}
        \end{tikzpicture}

        \caption{A graph $\litClauGraphSign{\phi,\pi}$.}
    \end{subfigure}
    \hfill
    \begin{subfigure}[b]{0.6\linewidth}
        \centering
        \begin{tikzpicture}[
            scale=0.55, 
            every node/.style={scale=0.7}
        ]
        \begin{pgfonlayer}{nodelayer}
            \node [style=none] (2) at (-2.5, 1.75) {};
            \node [style=none] (3) at (0.25, 1.75) {};
            \node [style=none] (4) at (-1, 1.75) {};
            \node [style=gadget] (5) at (-1, 1.75) {$a \lor b \lor c$};
            \node [style=none] (9) at (-0.5, 1.5) {};
            \node [style=none] (12) at (-3.75, -0.5) {};
            \node [style=none] (16) at (-3.75, 3.25) {};
            \node [style=none] (17) at (-2.75, -3.75) {};
            \node [style=none] (21) at (-2.75, -1.25) {};
            \node [style=gadget] (22) at (-2.75, -1.25) {$\texttt{a}$};
            \node [style=none] (27) at (0.75, -1.25) {};
            \node [style=gadget] (28) at (0.75, -1.25) {$\texttt{c}$};
            \node [style=none] (31) at (-2.5, 3.25) {};
            \node [style=none] (32) at (1.5, 3.25) {};
            \node [style=gadget] (34) at (-1, 3.25) {$a \lor d$};
            \node [style=none] (41) at (0.75, 1.75) {};
            \node [style=none] (49) at (4.5, -1.25) {};
            \node [style=gadget] (50) at (4.5, -1.25) {$\texttt{e}$};
            \node [style=none] (52) at (4.5, -2.75) {};
            \node [style=none] (53) at (-1, -2.75) {};
            \node [style=none] (56) at (2.75, 3.25) {};
            \node [style=none] (58) at (6.5, -3.75) {};
            \node [style=gadget] (67) at (5.5, 3.25) {$e \lor f$};
            \node [style=none] (72) at (4.5, 3.25) {};
            \node [style=none] (73) at (6.5, 3.25) {};
            \node [style=gadget] (74) at (-1, -1.25) {$\texttt{b}$};
            \node [style=gadget] (75) at (2.75, -1.25) {$\texttt{d}$};
            \node [style=gadget] (76) at (6.5, -1.25) {$\texttt{f}$};
            \node [style=none] (78) at (5, 3) {};
            \node [style=none] (81) at (-0.5, -1.25) {};
            \node [style=none] (85) at (-2.25, -1.25) {};
            \node [style=none] (87) at (-0.5, -0.75) {};
            \node [style=none] (90) at (1.25, -1.25) {};
            \node [style=none] (92) at (1.25, 2) {};
            \node [style=none] (93) at (-0.25, 2) {};
            \node [style=none] (97) at (3.25, -1.25) {};
            \node [style=none] (98) at (3.25, 3.5) {};
            \node [style=none] (99) at (-0.5, 3.5) {};
            \node [style=none] (100) at (5, -1.25) {};
            \node [style=none] (104) at (7, 3.5) {};
            \node [style=none] (105) at (6, 3.5) {};
            \node [style=gadget] (116) at (1.75, -2.75) {$\neg b \lor \neg e$};
            \node [style=gadget] (117) at (1.75, -3.75) {$\neg a \lor \neg f$};
            \node [style=none] (129) at (1.5, -2.5) {};
            \node [style=none] (130) at (-0.5, -2.5) {};
            \node [style=none] (132) at (5, -3) {};
            \node [style=none] (133) at (2, -3) {};
            \node [style=none] (134) at (2, -4) {};
            \node [style=none] (135) at (7, -4) {};
            \node [style=none] (136) at (-2.25, -3.5) {};
            \node [style=none] (137) at (1.5, -3.5) {};
            \node [style=none] (141) at (-0.5, -0.25) {};
            \node [style=none] (142) at (1.25, -0.75) {};
            \node [style=none] (145) at (1.25, -0.25) {};
            \node [style=none] (146) at (3.25, -0.75) {};
            \node [style=none] (149) at (3.25, -0.25) {};
            \node [style=none] (150) at (5, -0.75) {};
            \node [style=none] (153) at (5, -0.25) {};
            \node [style=none] (154) at (7, -1) {};
            \node [style=none] (157) at (7, -0.25) {};
            \node [style=none] (158) at (-3, -0.5) {};
            \node [style=none] (159) at (-3, -1) {};
            \node [style=none] (160) at (-2.5, -1) {};
            \node [style=none] (161) at (-1.25, -1) {};
            \node [style=none] (162) at (-1.25, -0.5) {};
            \node [style=none] (163) at (-1, -0.5) {};
            \node [style=none] (164) at (0.5, -1) {};
            \node [style=none] (165) at (0.5, -0.5) {};
            \node [style=none] (166) at (0.75, -0.5) {};
            \node [style=none] (167) at (2.5, -1) {};
            \node [style=none] (168) at (2.5, -0.5) {};
            \node [style=none] (169) at (2.75, -0.5) {};
            \node [style=none] (170) at (4.25, -1) {};
            \node [style=none] (171) at (4.25, -0.5) {};
            \node [style=none] (172) at (4.5, -0.5) {};
            \node [style=none] (173) at (6.25, -1) {};
            \node [style=none] (174) at (6.25, -0.5) {};
            \node [style=none] (175) at (6.5, -0.5) {};
            \node [style=none] (176) at (7, -1.5) {};
            \node [style=none] (177) at (6.75, -1) {};
            \node [style=none] (178) at (6.75, -1.5) {};
        \end{pgfonlayer}
        \begin{pgfonlayer}{edgelayer}
            \draw [style={connectivity_requester_wire}] (5) to (3.center);
            \draw [style={connectivity_requester_wire}] (5) to (2.center);
            \draw [style={connectivity_requester_wire}] (16.center) to (12.center);
            \draw [style={connectivity_requester_wire}] (22) to (17.center);
            \draw [style={connectivity_requester_wire}] (34) to (32.center);
            \draw [style={connectivity_requester_wire}] (34) to (31.center);
            \draw [style={connectivity_requester_wire}] (41.center) to (3.center);
            \draw [style={connectivity_requester_wire}] (16.center) to (31.center);
            \draw [style={connectivity_requester_wire}] (56.center) to (32.center);
            \draw [style={movement_wire}] (75) to (28);
            \draw [style={movement_wire}] (28) to (74);
            \draw [style={movement_wire}] (74) to (22);
            \draw [style={movement_wire}] (75) to (50);
            \draw [style={movement_wire}] (50) to (76);
            \draw [style={connectivity_provider_wire}] (93.center) to (92.center);
            \draw [style={connectivity_provider_wire}] (99.center) to (98.center);
            \draw [style={connectivity_provider_wire}] (105.center) to (104.center);
            \draw [style={connectivity_requester_wire}] (74) to (53.center);
            \draw [style={connectivity_requester_wire}] (58.center) to (117);
            \draw [style={connectivity_requester_wire}] (117) to (17.center);
            \draw [style={connectivity_requester_wire}] (53.center) to (116);
            \draw [style={connectivity_requester_wire}] (116) to (52.center);
            \draw [style={connectivity_requester_wire}] (52.center) to (50);
            \draw [style={connectivity_requester_wire}] (73.center) to (67);
            \draw [style={connectivity_requester_wire}] (67) to (72.center);
            \draw [style={connectivity_requester_wire}] (76) to (58.center);
            \draw [style={connectivity_provider_wire}] (130.center) to (81.center);
            \draw [style={connectivity_provider_wire}] (130.center) to (129.center);
            \draw [style={connectivity_provider_wire}] (132.center) to (100.center);
            \draw [style={connectivity_provider_wire}] (133.center) to (132.center);
            \draw [style={connectivity_provider_wire}] (135.center) to (134.center);
            \draw [style={connectivity_provider_wire}] (85.center) to (136.center);
            \draw [style={connectivity_provider_wire}] (137.center) to (136.center);
            \draw [style={connectivity_provider_wire}] (81.center) to (87.center);
            \draw [style={connectivity_provider_wire}] (141.center) to (9.center);
            \draw [style={connectivity_provider_wire}] (100.center) to (150.center);
            \draw [style={connectivity_provider_wire}] (153.center) to (78.center);
            \draw [style={connectivity_provider_wire}] (104.center) to (157.center);
            \draw [style={connectivity_provider_wire}] (146.center) to (97.center);
            \draw [style={connectivity_provider_wire}] (149.center) to (98.center);
            \draw [style={connectivity_provider_wire}] (142.center) to (90.center);
            \draw [style={connectivity_provider_wire}] (145.center) to (92.center);
            \draw [style={connectivity_provider_wire}] (157.center) to (154.center);
            \draw [style={connectivity_provider_wire}] (153.center) to (150.center);
            \draw [style={connectivity_provider_wire}] (149.center) to (146.center);
            \draw [style={connectivity_provider_wire}] (145.center) to (142.center);
            \draw [style={connectivity_provider_wire}] (141.center) to (87.center);
            \draw [style={connectivity_requester_wire}] (73.center) to (76);
            \draw [style={connectivity_requester_wire}] (72.center) to (50);
            \draw [style={connectivity_requester_wire}] (41.center) to (28);
            \draw [style={connectivity_requester_wire}] (5) to (74);
            \draw [style={connectivity_requester_wire}] (56.center) to (75);
            \draw [style={connectivity_requester_wire}] (2.center) to (160.center);
            \draw [style={connectivity_requester_wire}] (159.center) to (158.center);
            \draw [style={connectivity_requester_wire}] (158.center) to (12.center);
            \draw [style={connectivity_requester_wire}] (161.center) to (162.center);
            \draw [style={connectivity_requester_wire}] (163.center) to (162.center);
            \draw [style={connectivity_requester_wire}] (164.center) to (165.center);
            \draw [style={connectivity_requester_wire}] (166.center) to (165.center);
            \draw [style={connectivity_requester_wire}] (167.center) to (168.center);
            \draw [style={connectivity_requester_wire}] (169.center) to (168.center);
            \draw [style={connectivity_requester_wire}] (170.center) to (171.center);
            \draw [style={connectivity_requester_wire}] (172.center) to (171.center);
            \draw [style={connectivity_requester_wire}] (173.center) to (174.center);
            \draw [style={connectivity_requester_wire}] (175.center) to (174.center);
            \draw [style={connectivity_provider_wire}] (154.center) to (177.center);
            \draw [style={connectivity_provider_wire}] (176.center) to (178.center);
            \draw [style={connectivity_provider_wire}] (176.center) to (135.center);
        \end{pgfonlayer}
        \end{tikzpicture}

        \caption{The corresponding grid that connects different gadgets.}
    \end{subfigure}
    
    \caption{Showing the PSPACE-hardness: overall transformation of $\litClauGraphSign{\phi,\pi}$ into a grid.}
    \label{fig:reduction_unbounded_example}
\end{figure}
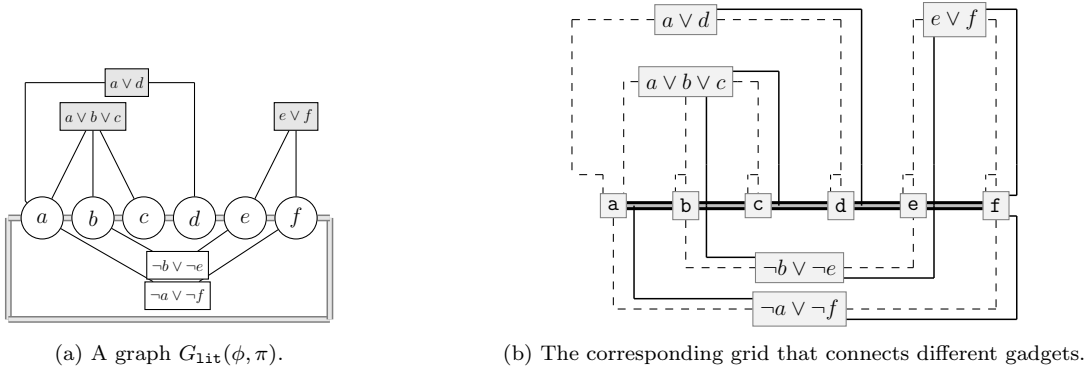

\begin{figure}[h]
	\centering
	\pgfsetlayers{edgelayer,nodelayer}

	\begin{tikzpicture}[
            scale=1, 
            every node/.style={scale=0.7}
        ]
	\begin{pgfonlayer}{nodelayer}
		\node [style=none] (3) at (-0.75, -0.5) {};
		\node [style=none] (4) at (-0.25, -0.5) {};
		\node [style=none] (5) at (-0.25, 1) {};
		\node [style=none] (6) at (-0.75, 1) {};
		\node [style=none] (8) at (-0.5, -0.25) {};
		\node [style=none] (17) at (-0.5, 0.75) {$x$};
		\node [style=none] (19) at (-1.25, -0.25) {$\dots$};
		\node [style=target] (55) at (0.75, -0.25) {$a_y$};
		\node [style=none] (69) at (0.5, -0.5) {};
		\node [style=none] (70) at (3.25, -0.5) {};
		\node [style=none] (71) at (3.25, 1) {};
		\node [style=none] (72) at (0.5, 1) {};
		\node [style=none] (73) at (2.75, 0.25) {};
		\node [style=none] (74) at (3, -0.25) {};
		\node [style=none] (75) at (0.75, 0.75) {$y$};
		\node [style=none] (76) at (3, 0.25) {};
		\node [style={green_agent}] (78) at (2.5, 0.25) {};
		\node [style=none] (79) at (1.75, 0.25) {};
		\node [style=none] (80) at (2.25, -0.25) {};
		\node [style=none] (81) at (1.75, -0.25) {};
		\node [style=target] (82) at (2, 0.25) {$b_y$};
		\node [style=none] (83) at (1.5, -0.25) {};
		\node [style={green_agent}] (84) at (1.25, -0.25) {};
		\node [style=none] (117) at (3.5, -0.25) {};
		\node [style=none] (118) at (4, -0.25) {$\dots$};
		\node [style=none] (121) at (-4.5, -0.5) {};
		\node [style=none] (122) at (-2.25, -0.5) {};
		\node [style=none] (123) at (-2.25, 1) {};
		\node [style=none] (124) at (-4.5, 1) {};
		\node [style=none] (125) at (-2.75, 0.25) {};
		\node [style=none] (126) at (-2.5, -0.25) {};
		\node [style=none] (127) at (-4.25, 0.75) {$\alpha$};
		\node [style=none] (128) at (-2.5, 0.25) {};
		\node [style={green_agent}] (129) at (-3, 0.25) {};
		\node [style=none] (130) at (-3.75, 0.25) {};
		\node [style=none] (131) at (-3.25, -0.25) {};
		\node [style=none] (132) at (-3.75, -0.25) {};
		\node [style=target] (133) at (-3.5, 0.25) {$b_{\alpha}$};
		\node [style=none] (134) at (-4, -0.25) {};
		\node [style=none] (136) at (-1.75, -0.25) {};
		\node [style=none] (137) at (4.5, -0.25) {};
		\node [style=target] (138) at (5, -0.25) {$a_{\omega}$};
		\node [style=none] (139) at (4.75, -0.5) {};
		\node [style=none] (140) at (7, -0.5) {};
		\node [style=none] (141) at (7, 1) {};
		\node [style=none] (142) at (4.75, 1) {};
		\node [style=none] (145) at (5, 0.75) {$\omega$};
		\node [style=none] (148) at (6, 0.25) {};
		\node [style=none] (149) at (6.5, -0.25) {};
		\node [style=none] (150) at (6, -0.25) {};
		\node [style=target] (151) at (6.25, 0.25) {$b_{\omega}$};
		\node [style=none] (152) at (5.75, -0.25) {};
		\node [style={green_agent}] (153) at (5.5, -0.25) {};
	\end{pgfonlayer}
	\begin{pgfonlayer}{edgelayer}
		\draw [style={paired_planar}] (5.center) to (6.center);
		\draw [style={paired_planar}] (3.center) to (4.center);
		\draw [style={paired_planar}] (5.center) to (4.center);
		\draw [style={paired_planar}] (71.center) to (72.center);
		\draw [style={paired_planar}] (69.center) to (70.center);
		\draw [style={movement_wire}] (76.center) to (74.center);
		\draw [style={movement_wire}] (73.center) to (76.center);
		\draw [style={movement_wire}] (79.center) to (82);
		\draw [style={paired_planar}] (72.center) to (69.center);
		\draw [style={movement_wire}] (81.center) to (83.center);
		\draw [style={movement_wire}] (74.center) to (117.center);
		\draw [style={paired_planar}] (71.center) to (70.center);
		\draw [style={paired_planar}] (123.center) to (124.center);
		\draw [style={paired_planar}] (121.center) to (122.center);
		\draw [style={movement_wire}] (128.center) to (126.center);
		\draw [style={movement_wire}] (125.center) to (128.center);
		\draw [style={movement_wire}] (131.center) to (132.center);
		\draw [style={movement_wire}] (132.center) to (130.center);
		\draw [style={movement_wire}] (130.center) to (133);
		\draw [style={paired_planar}] (124.center) to (121.center);
		\draw [style={movement_wire}] (132.center) to (134.center);
		\draw [style={paired_planar}] (123.center) to (122.center);
		\draw [style={movement_wire}] (136.center) to (126.center);
		\draw [style={paired_planar}] (141.center) to (142.center);
		\draw [style={paired_planar}] (139.center) to (140.center);
		\draw [style={movement_wire}] (149.center) to (150.center);
		\draw [style={movement_wire}] (150.center) to (148.center);
		\draw [style={movement_wire}] (148.center) to (151);
		\draw [style={paired_planar}] (142.center) to (139.center);
		\draw [style={movement_wire}] (150.center) to (152.center);
		\draw [style={paired_planar}] (141.center) to (140.center);
		\draw [style={movement_wire}] (138) to (137.center);
		\draw [style={movement_wire}] (8.center) to (55);
		\draw [style={movement_wire}] (81.center) to (80.center);
		\draw [style={movement_wire}] (81.center) to (79.center);
	\end{pgfonlayer}
	\end{tikzpicture}
	
	\caption{Illustration of the line of variable gadgets. All the agents are on their respective target positions. The leftmost variable is $\alpha$. The rightmost variable is $\omega$. The plain edges represent the movement lines.}
	\label{fig:variable_gadget_line}
\end{figure}
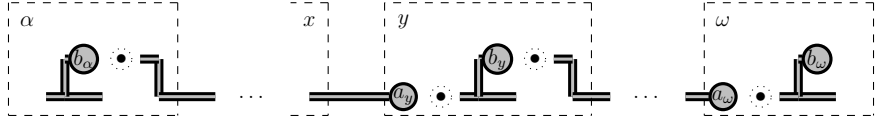

\bigskip
\noindent\textbf{CMAPF environment construction.}  The overall reduction is illustrated in \Cref{fig:reduction_unbounded_example}. 
The construction is similar to the reduction of the previous section (\Cref{claim:bounded2DCMAPFpolytime}) with the new gadgets we described.
One important difference is that instead of a simple provider wire, the line connecting variable gadgets is made of empty cells surrounded by connectivity providers (see bottom of \Cref{table:wires}).
All these connectivity providers are connected in the communication graph (see for example \Cref{fig:switch_and_bridge_pspace_hardness}): each variable gadget $x$, thanks to $\connectorgadget{x}$ transmits connectivity from left providers to right providers.
Moreover, the left and top isolated requesters are connected to either $a_x$ or $b_x$ such that providing connectivity to these requesters ensures connectivity of top and bottom providers of the line.
Here, isolated requesters correspond to requesters that are not directly part of a requester wire (see for instance the requester at the left of agent $a_y$ and the requester at the right of agent $a_x$ in \Cref{fig:switch_and_bridge_pspace_hardness}).
\Cref{fig:variable_gadget_line} illustrates the said line of variable gadgets.

\Cref{fig:variable_gadget_line} describes the source configuration of each agent that can move on the line: initially, all agents are on their target position.
Each variable gadget $x$ contains the agents $a_x, b_x$ as well as two isolated requesters (except the variable gadget for the first variable $\firstvariable$ which only contains agents $b_\firstvariable$).

\newcommand{\someagent}{i} %
\begin{claim}
Let $V$ be the set of variables in $\phi$. Consider the set of agents $A := \set{b_\alpha} \cup \bigcup_{x \in V \setminus \set{\alpha}} \{a_x,b_x\}$ (see \Cref{fig:variable_gadget_line}).
	\begin{enumerate}
	\item In each configuration where all variables are set\footnote{no agent $x$ in $\switch{x}$ is in the middle position}, either all agents in $A$ are placed next to a left or top isolated requester, or there is a unique agent free to move.
	\item From any configuration  where all variables are set, for all $\someagent \in A$, it is possible to move agents in $A\setminus\{i\}$ so that $\someagent$ is eventually free to move.
	\end{enumerate}
	\label{claim:unique_agent_move_on_line}
\end{claim}
\begin{proof}

	For proving Point 1, consider a configuration where all variables are set. Suppose that not all agents are next to a left or top requester. 
	On the one hand $|A| = 2|V| - 1$. 
	On the other hand, as shown in \Cref{fig:variable_gadget_line}, there are $1 + 2(|V| - 2) + 1 = 2|V|-2$ requesters in total that require connectivity from agents in $A$: 
	one requester in gadget for $\firstvariable$, 2 requesters for each inner variable gadgets, and one requester in the gadget for $\lastvariable$.
	So there is a single agent free to move
	(for instance \Cref{fig:variable_gadget_line} shows a configuration where only agent $b_{\lastvariable}$ is free to move while all other agents are 
	constrained by the requester it is next to).

	First we prove Point 2 from the initial configuration as shown in \Cref{fig:variable_gadget_line}.
	If $\someagent = b_{\omega}$ then $\someagent$ can already move freely in its connected component.
	Else, $b_{\omega}$ moves towards the left requester in $\bridge{\omega}$ to give the possibility to $a_{\omega}$ to move. If $\someagent = a_{\omega}$, we are done.
	If $\someagent \neq a_{\omega}$ then $a_{\omega}$ can move to the top requester in the gadget in the variable at the left of $\omega$.
	This process is repeated until $\someagent$ have been found and is free to move.
	 
	To finish the proof, consider any configuration $c$. We proved that $c$ can be reached from the initial configuration. As movements in CMAPF are reversible, the initial configuration is reachable from $c$. Thus, we can obtain a configuration where $\someagent$ is free to move from $c$.
\end{proof}

\begin{claim}
	The obtained 2D CMAPF instance $(G, s, t)$ is computable in poly-time in $(\phi, \nu, \nu')$. 
	\label{claim:2DCMAPFpolytime}
\end{claim}

\begin{proof}
	Similar to the proof of \Cref{claim:bounded2DCMAPFpolytime}.
\end{proof}

\begin{claim}
	$\nu \rightarrowflips{\phi} \nu'$ iff $s \rightarrowcmapf{G} t$.
	\label{claim:bounded2DCMAPFreductioncorrection}
\end{claim}

\begin{proof}
	\fbox{$\Rightarrow$} Suppose $\nu \rightarrowflips{\phi} \nu'$. The idea is to mimic the variable flips in order to reach $t$ from $s$. 
	We explain how a variable flip $\nu \rightarrowflip{x} \nu_1$ is simulated by movements of agents. Suppose we flip $x$ from true to false.
	\begin{itemize}
		\item First, for each clause containing literal $\ell_i=x$, we move agent $\ell_i$ in the clause gadget (see \Cref{fig:clause_gadget_PSPACE}) down to give extra connectivity to connectivity requesters. We can perform this move since the isolated requester in the clause gadget remains connected to the rest (by agent $\ell_j$ or $\ell_k$), because the clause remains true after flipping $x$ to false.
		\item In case of $\splitgadget{x}$ and $\connectorgadget{x}$ gadget (see \Cref{fig:connecter_split_gadget}), we move the agents~$s_x$ and $c_x$ down. This movement can be done since the requesters going to the clause are connected thanks to $\ell_i$.
		\item Now, thanks to Point 2 \Cref{claim:unique_agent_move_on_line}, we can reach a configuration in which the agent $b_x$ is free to move.
		Moreover Point 1 of \Cref{claim:unique_agent_move_on_line} implies that at that configuration, $b_x$ is the only agent on the variable gadgets line that can move. 
		Agent $b_x$ goes down and right in $\switch x$; so that $x$ is the only variable agent that can move during this step.
		Agent $x$ then goes down, setting the variable~$x$ to false.
		\item Finally, in each clause gadget containing $\lnot x$, agent $\lnot x$ goes next to the isolated requester.
	\end{itemize}

	Conversely if the variable $x$ flips from false to true the agents of components of variable gadget $\switch{x}$, $\connectorgadget{x}$ and $\splitgadget{x}$ goes from bottom to top position.
	By concatenating all the simulations of movements of agents for all variable flips, we get $s \rightarrowcmapf{G} t$.

	\fbox{$\Leftarrow$} Consider a sequence $s \rightarrowcmapf{G} t$. To each configuration of agents in the grid, we define the corresponding assignment in which $x$ is true when $x$ is up in $\switch x$, and false otherwise.
	By Point 1 of \Cref{claim:unique_agent_move_on_line}, there is at most one variable $x$ such that $b_x$ is at the rightmost position in $\switch x$. This means that two consecutive assignments are either equal, or obtained by a variable flip. 
	Notice that in the clause gadgets (see \Cref{fig:clause_gadget_PSPACE}), the isolated requester is connected. As either $\literal_1, \literal_2$ or $\literal_3$ is next to that isolated requester, the current assignment makes either $\literal_1, \literal_2$ or $\literal_3$ true. In other words, each assignment makes $\phi$ true.

	By removing consecutive equal assignments, we get a sequence of assignments $\nu \rightarrowflips{\phi} \nu'$.
\end{proof}

\section{Perspectives}
In this paper, we studied the linear and planar fragments of 3-SAT and 
used these to establish the computational complexity of the bounded and unbounded CMAPF problem in 2D.
Interestingly, our complexity results hold for anonymous (unlabeled agents), that is, when the targets are not assigned to specific agents but it is only required that \emph{some} agents should reach each target. 
Indeed, in our gadgets, each agent is trapped in its connected component and there is a single agent that can reach any target. So, the anonymous version of CMAPF is PSPACE-complete too, while the bounded version is also NP-complete.
To obtain our complexity results, we developed a theory based on cycle augmentation. Cycle augmentation is a theoretical problem on its own. We plan to exhibit other classes for which it is in P.
It would also be interesting to make bridges with other similar problems like Steiner trees \cite{brazil2014history} (which also asks to ``add'' an object to a graph) or two-page book embeddings \cite{hong2009two}.

Linear 3SAT is already applied for planning \cite{gupta2024collision} and geometrical problems \cite{arkin2018selecting}. In the future, a research track would be to be continue the study with linear literal-planar 3SAT and monotone linear planar 3SAT.
We plan to apply the complexity results for \LPTSAT Reconfiguration to other problems similar to CMAPF. For instance, other types of motion planning problems that could be proven to be PSPACE-complete, \emph{e.g.} tethered robots \cite{DBLP:journals/ral/PengSSS25} or snakes \cite{DBLP:journals/jair/GuptaSZ20}.

\section*{Author contributions}

The first author had the initiative of this work, and is responsible for most of the technical content.
The second and third authors have mostly contributed to improving the content and to the writing.

\bibliographystyle{elsarticle-num} 
\bibliography{ref}

\begin{appendices}
\section{Definitions and properties of planar graphs}
\label{section:appendix-planar-graphs}

In order to make our paper self-contained, we recall some properties of planar graphs that are used in \Cref{section:cycleaugmentation}.
Some of these results can be found in textbooks such as \cite{hararyGraph} while we also rely on
properties proved in \cite{klein2011flatworlds}.

Recall that a planar graph $G = (V,E)$ is an undirected \emph{multi-graph}\footnote{However for simplicity we continue to call them graphs.} that can be embedded in $\mathbb{R}^2$ in such a way that edges do not intersect except at their endpoints.

\subsection{Embedding}
Intuitively, an embedding $\Gamma_G$ of a planar graph $G = (V,E)$ is a drawing of $G$ in the plane in which edges do not cross. 
A \emph{face} is a connected region delimited by edges.

\begin{figure}[h]
	\centering
	\pgfsetlayers{edgelayer,nodelayer}
	
	\begin{subfigure}[b]{0.45\linewidth}
		\centering
		\begin{tikzpicture}[
			scale=1.0, 
			every node/.style={scale=0.8}
			]
			\begin{pgfonlayer}{nodelayer}
				\node [style={variable_node}] (0) at (-1, 1.25) {$u$};
				\node [style={variable_node}] (1) at (1.5, 1.5) {$v$};
				\node [style={variable_node}] (2) at (0.25, 0.75) {$w$};
				\node [style={variable_node}] (3) at (0.25, -0.75) {$x$};
				\node [style={variable_node}] (4) at (-1, -0.25) {$y$};
				\node [style={variable_node}] (5) at (2.75, 0) {$z$};
				\node [style=none] (6) at (-0.25, 0.25) {$F^1$};
				\node [style=none] (7) at (1.5, 0.25) {$F^2$};
				\node [style=none] (8) at (2.5, 1.5) {$F^{\infty}$};
			\end{pgfonlayer}
			\begin{pgfonlayer}{edgelayer}
				\draw (0) to (4);
				\draw (4) to (3);
				\draw (1) to (0);
				\draw (1) to (2);
				\draw (1) to (3);
				\draw (1) to (5);
				\draw (5) to (3);
			\end{pgfonlayer}
		\end{tikzpicture}
		
		\caption{A planar embedding of $G = (V,E)$. The faces are denoted by $F^1, F^2$ and $F^{\infty}$ such that $F^{\infty}$ represents the outer face.\label{figure:planarembedding}}
	\end{subfigure}
	\hfill
	\begin{subfigure}[b]{0.45\linewidth}
		\centering
		\begin{tikzpicture}[
			scale=0.9, 
			every node/.style={scale=0.8}
			]
			\begin{pgfonlayer}{nodelayer}
				\node [style={variable_node}] (0) at (-1, 1.25) {$u$};
				\node [style={variable_node}] (1) at (1.5, 1.5) {$v$};
				\node [style={variable_node}] (2) at (0.25, 0.75) {$w$};
				\node [style={variable_node}] (3) at (0.25, -0.75) {$x$};
				\node [style={variable_node}] (4) at (-1, -0.25) {$y$};
				\node [style={variable_node}] (5) at (2.75, 0) {$z$};
				\node [style={dual_g_face}] (9) at (1.5, 0.25) {$2$};
				\node [style={dual_g_face}] (10) at (2.75, 1.5) {$\infty$};
				\node [style={dual_g_face}] (11) at (-0.25, 0.25) {$1$};
				\node [style=none] (12) at (-1.5, 1.75) {};
				\node [style=none] (14) at (0.75, 1.2) {};
				\node [style=none] (15) at (-1, -0.75) {};
			\end{pgfonlayer}
			\begin{pgfonlayer}{edgelayer}
				\draw (0) to (4);
				\draw (4) to (3);
				\draw (1) to (0);
				\draw (1) to (2);
				\draw (1) to (3);
				\draw (1) to (5);
				\draw (5) to (3);
				\draw [style={dotted_edge}, bend left=45] (9) to (11);
				\draw [style={dotted_edge}] (9) to (10);
				\draw [style={dotted_edge}, bend right=90, looseness=3.00] (9) to (10);
				\draw [style={dotted_edge}, bend left=75] (11) to (10);
				\draw [style={dotted_edge}, bend left=45, looseness=1.50] (11) to (12.center);
				\draw [style={dotted_edge}, bend left=60, looseness=0.75] (12.center) to (10);
				\draw [style={dotted_edge}, bend left=90, looseness=0.75] (14.center) to (11);
				\draw [style={dotted_edge}, bend right=60] (14.center) to (11);
				\draw [style={dotted_edge}, bend right=75] (15.center) to (11);
				\draw [style={dotted_edge}, bend left=45] (15.center) to (12.center);
			\end{pgfonlayer}
		\end{tikzpicture}
		\caption{An embedding of $G$ and its dual $G^*$. $G^*$ is represented by dotted lines.\label{fig:dualgraph}}
		
	\end{subfigure}
	
	\caption{A planar embedding of the graph $G = (V,E)$.}
	\label{fig:planar_embedde_graph}
\end{figure}
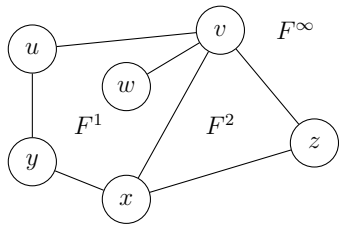
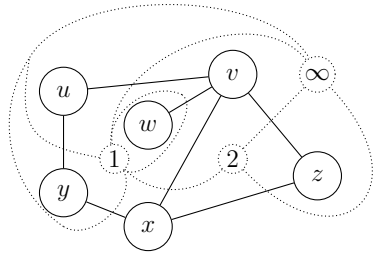

\begin{example}
	\Cref{figure:planarembedding} shows the embedding of a graph along with its three faces.
\end{example}

Here we recall the notion of combinatorial embedding \cite{klein2011flatworlds} that gives a more algebraic flavor to embeddings, completely focused on `abstract' edges.

\begin{definition}
    Given a graph $G = (V,E)$, an embedding $\Gamma_G$ of $G$ is a set of faces, where a face is a non-empty cyclic sequence of edges such that each edge of $E$ appears twice in~$\Gamma_G$.
\end{definition}

\begin{definition}\cite{klein2011flatworlds}
    An embedding $\Gamma_G$ of $G = (V,E)$ is \emph{planar} if $G$ can be drawn in the plane without crossing edges so that the faces in the drawing correspond to $\Gamma_G$.
    \label{theorem:planar_embedding}
\end{definition}

\begin{example}
	Figure \ref{figure:planarembedding} depicts the planar embedding $\Gamma_G = \{ F^1, F^2, F^{\infty} \}$ with
	\begin{align*}
		F^1 & = (uv, vw, vw, vx, xy, yu) \\
		F^2 & = (vz, zx, vx) \\
		F^{\infty} & = (uv, yu, yx, zx, vz) 
	\end{align*}
	Each edge indeed appears twice. For instance, $uv$ touches $F^1$ and $F^\infty$. The edge $vw$ is particular since it touches only face $F^1$, but appears twice in $F^1$.
\end{example}

As seen in the previous example, each face is a sequence representing a visit of the edges of the face $F$ in clockwise order
starting from an arbitrarily chosen vertex, unless $F$ is the outer face $F_{\infty}$ in which case the visit is in
counter-clockwise order.

For all $e \in E$, if $e$ appears in two distinct faces $F$ and $F'$, then we  define $\faces_{\Gamma_G}(e) = FF'$. For all $e \in E$, if $e$ appears in a single face $F$, we define $\faces_{\Gamma_G}(e) = FF$.

\begin{example}
	$\faces_{\Gamma_G}(uv) = F^1 F^\infty$ and $\faces_{\Gamma_G}(vw) = F^1 F^1$.
\end{example}

We will denote the edge sequence defining each face $F$ by $(F_1, \dots, F_k)$ for some~$k$.
We write $|F| = k$, and define for all $i\geq k+1$, $F_i = F_{(i \mod k+1)+1}$, that is, we see $F$ as an infinite sequence
repeating the finite sequence $(F_1, \dots, F_k)$.

\begin{definition}
	Recall that there are exactly two occurrences of each edge $e$ in $\Gamma_G$. We define the function 
	$\other: \Gamma_G \times \mathbb{N} \rightarrow \Gamma_G \times \mathbb{N}$ such that
	$$\other(F,i) = \begin{cases}
		(F',j) \text{ if $F_i$ appears in two distinct faces $F$ and $F'$ and $F'_j = F_i$,} \\
		(F, j) \text{ if $F_i$ appears only in $F$ with $i \neq j \text{ mod } |F|$.} 
	\end{cases}$$
	\label{def:other_function}
\end{definition}

\begin{example}
	We have $\other(F^2,2) = (F^\infty, 4)$,
	$F^2_2 = F^\infty_4 = zx$,
	and
	$\other(F^1,2) = (F^1, 3)$,
	$F^1_2 = F^1_3 = vw$.
\end{example}

\subsection{Dual graph}
\label{subsec:dual_graph}

Consider an embedding $\Gamma_G$, we define the dual graph $G^* = (V^*, E^*)$ whose vertices are the faces of $\Gamma_G$ and there is an $E^*$-edge between two faces when they touch i.e. they share a common edge.

\begin{definition}[Dual graph]
    Given an embedding $\Gamma_G$ of a connected planar graph $G=(V,E)$,
    $G^* = (V^*,E^*)$ denotes the \emph{dual graph} of $G$ and $\Gamma_G$
		with $V^* = \Gamma_G$ and $E^* = \multiset{ \faces_{\Gamma_G}(e) \mid e \in E}$
		where $\multiset{\cdot}$ denotes a multiset.
    \label{def:dual_graph}
\end{definition}

\begin{property}
	For a graph $G=(V,E)$ and an embedding $\Gamma_G$, $\faces_{\Gamma_G}$ is a bijection.
    \label{prop:bijections_associated_with_dual}
\end{property}

\begin{example}
	\Cref{fig:dualgraph} shows the dual $G^*$. Vertices are the faces $F^1, F^2, F^\infty$. For each edge in $G$ there is an edge in $G^*$ between the faces touching that edge, \emph{e.g.} the dual of $uv$ is $F^1F^\infty$. 
	If the edge touches a single face, there is a self loop, \emph{e.g.} $vw$ and $F^1F^1$.
\end{example}

Note that the multi-sets $E$ and $E^*$ have the same size and that $\faces_{\Gamma_G}$ is a bijection from $E$ into $E^*$.
From an embedding $\Gamma_G$, we can construct a corresponding ``dual" embedding  $\Gamma^*_G$ of $G^*$, as follows.

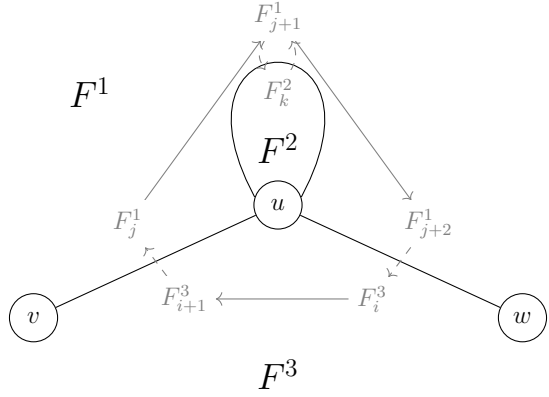
\begin{figure}[h]
	\centering
	\pgfsetlayers{edgelayer,nodelayer}
	\tikzstyle{cycleother} = [text=gray]
	\tikzstyle{cycleotheredge} = [draw=gray]
	\begin{tikzpicture}[
		scale=1.0, 
		every node/.style={scale=0.8, fill=white, draw=none}
		]
	\begin{pgfonlayer}{nodelayer}
		\node [style=none] (1) at (-0.25, -0.25) {};
		\node [style=none] (2) at (0.25, -0.25) {};
		\node [style=cycleother] (5) at (0, 1.25) {$F^2_k$};
		\node [style=cycleother] (7) at (-2, -0.5) {$F^1_j$};
		\node [style=none] (8) at (0, -2.5) {\Large $F^3$};
		\node [style=none] (9) at (-2.5, 1.25) {\Large $F^1$};
		\node [style={variable_node}] (10) at (-3.25, -1.75) {$v$};
		\node [style={variable_node}] (11) at (3.25, -1.75) {$w$};
		\node [style=cycleother] (13) at (-1.25, -1.5) {$F^3_{i+1}$};
		\node [style=cycleother] (14) at (1.25, -1.5) {$F^3_{i}$};
		\node [style=cycleother] (15) at (2, -0.5) {$F^1_{j+2}$};
		\node [style=none] (17) at (0, 0.5) {\Large $F^2$};
		\node [style={variable_node}] (0) at (0, -0.25) {$u$};
		\node [style=cycleother] (6) at (0, 2.25) {$F^1_{j+1}$};
	\end{pgfonlayer}
	\begin{pgfonlayer}{edgelayer}
		\draw [bend left=240, looseness=15.00] (2.center) to (1.center);
		\draw (0) to (10);
		\draw (0) to (11);
		\draw [style={cycleotheredge, directed_edge}, bend right] (5) to (6);
		\draw [style={cycleotheredge, directed_edge}, bend right, looseness=1.25] (6) to (5);
		\draw [style={cycleotheredge,directed_edge}] (15) to (14);
		\draw [style={cycleotheredge,directed_edge}] (13) to (7);
		\draw [->,cycleotheredge] (6) to (15);
		\draw [->,cycleotheredge] (14) to (13);
		\draw [->,cycleotheredge] (7) to (6);
	\end{pgfonlayer}
	\end{tikzpicture}

	\caption{Construction of the sequence of edges in clockwise order around $u$: $vu, uu, uw$. Consider $F^1 = \dots vu \cdot uu \cdot uw \dots$, $F^3 = \dots uw \cdot vu \dots$ and $F^2 = uu$. We arbitrarily select $F^3$. The edge $F^3_i = uw$ is the first occurrence of an edge incident to $u$ in $F^3$. Therefore $F^3_{i+1} = vu$ is selected to construct the sequence of edges for $u$. We apply $\other{}(F^3, i+1) = (F^1,j)$ (represented by dashed arrows) and we obtain the next edge $F^1_{j+1} = uu$ (represented by plain arrows). We continue the operations until retrieving the starting edge $vu$. }
	\label{fig:clockwise_computation}
\end{figure}

Recall that faces of $\Gamma^*_G$ are in correspondence with vertices of $G$. 
Let $v$ denote a vertex of $G$.
Let $e_0, \ldots,e_k$ denote a sequence of edges incident to $v$ in clockwise or counter-clockwise order.
This sequence is obtained by selecting a face $F \in \Gamma_G$ such that $F_i$ is the first edge in the sequence $F$ incident to $v$.
We define $e_0 := F_{i+1}$.
Then we define $e_n := F'_{j}$ and $e_{n+1} := F''_{k+1}$ such that $\other{}(F', j) = (F'', k)$.
The computation stops if $e_{n+1} = e_0$. Hence $e_k := e_n$.
 \Cref{fig:clockwise_computation} illustrates the principle for obtaining $e_0, \dots, e_k$.
We apply the bijection $\bijectionedgetodual$ to obtain a sequence of $E^*$-edges, which defines a face in~$\Gamma^*_G$:
\[\face^*(v) = (\bijectionedgetodual(e_0), \bijectionedgetodual(e_1), \dots).\]

\begin{definition}
	Given an embedding $\Gamma_G$, we define the embedding $\Gamma^*_G$ by $\Gamma^*_{G} = \set{ \face^*(v) \mid v \in V}$.
	\label{def:dual_construction}
\end{definition}

\begin{example}
The sequence of edges incident to $v$ in clockwise order is: 
	 $vx, vw, uv, vz$. Now, we take the image by $\bijectionedgetodual$ of each edge, and thus obtain
	 the $\Gamma^*_G$-face corresponding to $v$:
	$V := \face^*(v) = (F^2F^1, F^1F^1, F^1F^\infty, F^\infty F^2)$.
	In the same way, we define $\Gamma^*_G$-faces $U, Z, X, Y$ for respectively vertex $u, z, x, y$ and we get:
\begin{equation*}
    \begin{aligned}
        \Gamma_G^* & = \{ U, V, Z, X, Y \}, \\
        U & = (F^1 F^\infty, F^1 F^\infty), V = (F^2F^1, F^1 F^1, F^1 F^\infty, F^\infty F^2), \\ 
        Z & = (F^1 F^\infty, F^1 F^\infty),
        X = (F^2 F^1, F^1 F^\infty, F^\infty F^2), \\
        Y & = (F^\infty F^2, F^\infty F^2).
    \end{aligned}
\end{equation*}

\end{example}

Note that $\Gamma^*_G$ is an embedding of $G^*$. 
When $\Gamma_G$ is planar, note that the function $\face^*$ defines a bijection between the vertices $V$ of $G$ and the faces of $\Gamma^*_G$. In the sequel, when $\Gamma_G$ is planar, we suppose that $V$ and $\Gamma^*_G$ is the same set. We still use the notation $V$ for concreteness.

\subsection{Properties}

\begin{property}\cite{klein2011flatworlds}
    An embedding $\Gamma_G$ of $G = (V,E)$ is \emph{planar} if $|\Gamma_G^*| {-} |E| {+} |\Gamma_G| {=} 2$.
    \label{def:planar_embedding}
\end{property}

\begin{property}\cite{hararyGraph}
    If $G$ is planar, then $G^*$ is planar and connected.
    \label{prop:dual_connected}
\end{property}

\begin{property}\cite{klein2011flatworlds}
    If $\Gamma_G$ is planar and connected, then $\Gamma_G^*$ is planar and connected.
    \label{prop:dual_planar_connected}
\end{property}

\begin{property}\cite{klein2011flatworlds}
    If $\Gamma_G$ is planar and connected,	$\Gamma_G^{**} = \Gamma_G$ (up to a renaming of the edges)%
    \label{prop:if_g_conn_then_g_dual_dual_is_g}
\end{property}

\begin{property}\cite{klein2011flatworlds}
    Given a connected planar graph $G$ we can compute some embeddings $\Gamma_G$ and $\Gamma_G^*$ in polynomial time.
    \label{prop:dual_given_embedding_poly_time}
\end{property}

\subsection{Operations on Embeddings}
Now we present some simple operations on embeddings $\Gamma_G$ that can be computed in poly-time \cite{klein2011flatworlds}.
Such operations are standard and described in literature dedicated to data structures  for embeddings of planar graphs \cite{tamassia1988dynamic}.
However, here, we present some variants that preserve planarity and connectivity.
We present both the modification of the operations on the graph itself and the embedding, although the graph can be infered from the embedding (vertices are elements of $\Gamma_G^*$).

\begin{figure}[h]
    \centering
    \begin{tikzpicture}[xscale=0.5, yscale=0.5, baseline=0mm]
        \draw[face, line width=2mm,draw=white] (2, 2) to[in=0, out=0] (2, -1) -- (-2,-1) to[in=180, out=180] (-2, 2) -- cycle;
        \node (v0) at (0, 2) {$u$};
    \draw (-2, 2) edge node[above] {$e_1$} (v0);
    \node (v1) at (0, -1) {};
    \node (v2) at (0, -1) {};
    \node (v3) at (0, -1) {$v$};
    \draw (v3) edge node[below] {$e_3$} (2, -1);
    \draw (-2, -1) edge node[below] {$e_4$} (v3);
    \draw (v0) edge node[above] {$e_2$} (2, 2);
        \draw (2, 2) edge[dotted, in=0, out=0] (2, -1);
        \draw (-2, 2) edge[dotted, in=180, out=180] (-2, -1);
        \node[face] at (0, 0.5) {$F$};
    \end{tikzpicture}
    \superleftrightarrow{Edge addition}{Edge deletion}
    \begin{tikzpicture}[xscale=0.5, yscale=0.5, baseline=0mm]
        \draw[face, line width=2mm,draw=white] (2, 2) to[in=0, out=0] (2, -1) -- (0,-1) -- (0, 2) -- cycle;
        \draw[face, line width=2mm,draw=white] (-2, 2) to[in=180, out=180] (-2, -1) -- (0,-1) -- (0, 2) -- cycle;   \node (v0) at (0, 2) {$u$};
        \node (v0) at (0, 2) {$u$};
        \draw (-2, 2) edge (v0);
        \node (v1) at (0, -1) {};
        \node (v2) at (0, -1) {};
        \node (v3) at (0, -1) {$v$};
        \draw (v3) edge (2, -1);
        \draw (-2, -1) edge (v3);
        \draw (v0) edge (2, 2); 
        \draw (2, 2) edge[dotted, in=0, out=0] (2, -1);
        \draw (-2, 2) edge[dotted, in=180, out=180] (-2, -1);
        \node[face] at (-1, 0.5) {$F^1$};
        \node[face] at (1, 0.5) {$F^2$};
        \draw (v0) edge (v3);
    \end{tikzpicture}

    \caption{Edge addition and edge deletion. We recall here that $e_1$ may be equal to $e_2$ and $e_3$ may be equal to $e_4$. If $e_1 = e_2 = e_3 = e_4$ the edge $e$ is a self-loop.}
    \label{fig:edge-addition-deletion}
\end{figure}
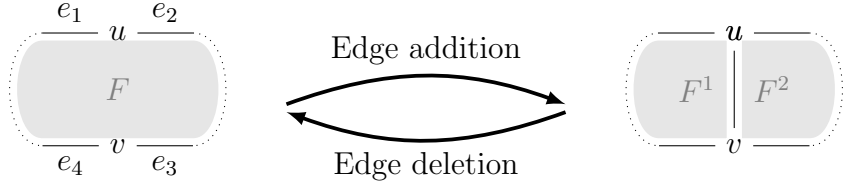

\begin{definition}[Edge addition]
    Consider a planar embedding $\Gamma_G$ of a connected planar graph $G = (V,E)$ that contains at least one edge. 
    
    \begin{itemize}
    	\item Given $u, v \in V$ and a face $F$, we define the addition of edge $e = uv$ in face $F$ containing\footnote{A face contains a vertex $u$ if there is some edge in the sequence that is incident to $u$.} $u$ and $v$ by
    \begin{align*}
    	G{+}e _{\in F} & := (V,E \cup \{e\}) \\
    	\Gamma_{G }+ e_{\in F} & := (\Gamma_G \setminus \{F\}) \cup \{ F^1, F^2\}
    \end{align*}
   
   where $F = e_4 \dots e_1e_2 \dots e_3$ where $e_1, e_2$ are incident to $u$ and $e_3, e_4$ are incident to $v$ (in some cases, $e_1, e_2, e_3, e_4$ may be equal) and $F^1 = e_1 \cdot e \cdot e_4 \dots$ and $F^2 = e_3 \cdot e \cdot e_2 \dots$.

   \item Given $u \in V$, and $v \not \in V$ and a face $F$, we define the addition of vertex $v$ and edge $e = uv$ in face $F$ by
   \begin{align*}
   	G{+}e_{\in F} & := (V{\cup}\{v\},E \cup \{e\}) \\
   	\Gamma_{G }+ e_{\in F} & :=  (\Gamma_G \setminus \{F\}) \cup \{ e \cdot e \cdot F\}
   \end{align*}
   \end{itemize}
    \label{def:edge_addition}
\end{definition}

\begin{definition}[Edge deletion]
    Consider a planar embedding $\Gamma_G$ of a connected planar graph $G = (V,E)$.
    Given an edge $e=uv$ that appears in two distinct\footnote{We could define also the deletion of an edge that appears in a single face but it is not useful in our work.} faces $F^1$ and $F^2$, the deletion of edge $e$ is defined by:
    \begin{align*}
    	G{-}e & = (V,E \setminus \{e\}) \\
        \Gamma_{G}{-}e & = (\Gamma_G \setminus \{F^1,F^2\}) \cup F   
    \end{align*}

	where $F = \dots e_1e_4 \dots e_3e_2 \dots$ where $F^1 = \dots e_1ee_2 \dots$ and $F^2 = \dots e_3ee_4 \dots$, where $e_1, e_2$ are incident to $u$ and $e_3, e_4$ are incident to $v$
	(possibly $e_1 = e_2$ and $e_3 = e_4$).
	\label{def:edge_deletion}
\end{definition}

In the previous definition, $F$ is the resulted face after deleting $e$, and merging $F^1$ with $F^2$.

 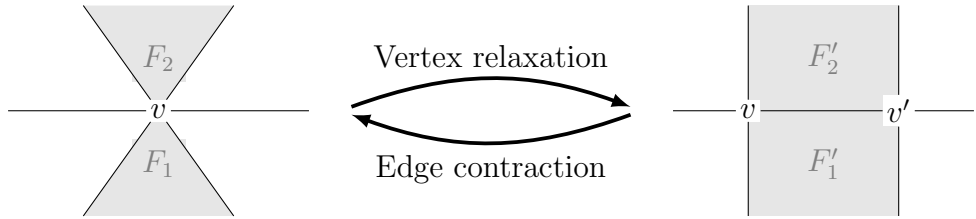
\begin{figure}[t]
    \centering
    
\begin{tikzpicture}[yscale=0.7, baseline=0mm]
	\draw[face] (-1, 2) -- (0, 0) -- (1, 2) -- cycle;
	\draw[face] (-1, -2) -- (0, 0) -- (1, -2) -- cycle;
	\node[primal] (v0) at (0, 0) {$v$};
	\node[face] (v2) at (0, -1) {$F_1$};
	\node[face] (v1) at (0, 1) {$F_2$};
	\draw (v0) edge (-1, -2);
	\draw (v0) edge (1, -2);
	\draw (v0) edge (1, 2);
	\draw (v0) edge (-1, 2);
	\draw (v0) edge (-2, 0);
	\draw (v0) edge (2, 0);
\end{tikzpicture}
\superleftrightarrow{Vertex relaxation}{Edge contraction}
\begin{tikzpicture}[yscale=0.7, baseline=0mm]	
	\draw[face] (-1, 2) -- (-1, 0) -- (1,0) -- (1, 2) -- cycle;
	\draw[face] (-1, -2) -- (-1, 0) -- (1,0) -- (1, -2) -- cycle;
	\node[primal] (v0) at (-1, 0) {$v$};
	\node[face] (v2) at (0, -1) {$F'_1$};
	\node[face] (v1) at (0, 1) {$F'_2$};
	\node[primal] (v3) at (1, 0) {$v'$};
	\draw (v0) edge (-1, -2);
	\draw (v3) edge (1, -2);
	\draw (v3) edge (1, 2);
	\draw (v0) edge (-1, 2);
	\draw (v0) edge (-2, 0);
	\draw (v3) edge (2, 0);
	\draw (v0) edge (v3);
\end{tikzpicture}

    \caption{Vertex relaxation and edge contraction. Vertex relaxation of vertex $v$ wrt to $F_1$ and $F_2$ creates an edge from $v$ to a copy $v'$ while still separating $F_1$ and $F_2$ (becoming $F_1'$ and $F_2'$). An edge contraction of $vv'$ merges vertices $v$ and $v'$.}
    \label{fig:vertexrelaxationedgecontraction}
\end{figure}

\begin{figure}[t]
	\centering
	\begin{tikzpicture}[yscale=0.5, xscale=0.5, baseline=0mm]
		\draw[face,line width=2mm, draw=white] (2, 2) to[in=0, out=0] (2, -1) -- (-2,-1) to[in=180, out=180] (-2, 2) -- cycle;
		\node (v0) at (0, 2) {$F_2$};
		\draw (-2, 2) edge node[above] {} (v0);
		\node (v1) at (0, -1) {};
		\node (v2) at (0, -1) {};
		\node (v3) at (0, -1) {$F_1$};
		\draw (v3) edge node[below] {} (2, -1);
		\draw (-2, -1) edge node[below] {} (v3);
		\draw (v0) edge node[above] {} (2, 2);
		\draw (2, 2) edge[dotted, in=0, out=0] (2, -1);
		\draw (-2, 2) edge[dotted, in=180, out=180] (-2, -1);
		\node[face] at (0, 0.5) {$v$};
	\end{tikzpicture}
	\superleftrightarrow{Edge addition}{Edge deletion}
	\begin{tikzpicture}[yscale=0.5, xscale=0.5, baseline=0mm]
		\draw[face,line width=2mm, draw=white] (2, 2) to[in=0, out=0] (2, -1) -- (0,-1) -- (0, 2) -- cycle;
		\draw[face,line width=2mm, draw=white] (-2, 2) to[in=180, out=180] (-2, -1) -- (0,-1) -- (0, 2) -- cycle;  
		\node (v0) at (0, 2) {$F_2'$};
		\draw (-2, 2) edge (v0);
		\node (v1) at (0, -1) {};
		\node (v2) at (0, -1) {};
		\node (v3) at (0, -1) {$F_1'$};
		\draw (v3) edge (2, -1);
		\draw (-2, -1) edge (v3);
		\draw (v0) edge (2, 2); 
		\draw (2, 2) edge[dotted, in=0, out=0] (2, -1);
		\draw (-2, 2) edge[dotted, in=180, out=180] (-2, -1);
		\node[face] at (-1, 0.5) {$v$};
		\node[face] at (1, 0.55) {$v'$};
		\draw (v0) edge (v3);
	\end{tikzpicture}
	
	\caption{The dual operations of vertex relaxation and edge contraction are respectively edge addition and edge deletion in the dual graph.}
	\label{fig:dualofvertexrelationandedgecontraction}
\end{figure}
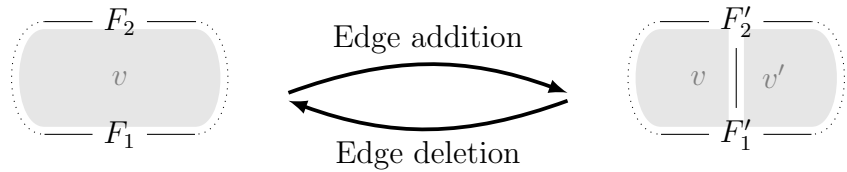

\begin{definition}[Vertex Relaxation]
    Consider a planar embedding $\Gamma_G$ of a connected planar graph $G = (V,E)$.
    Given $v \in V$, two distinct faces $F^1, F^2 \in \Gamma_G$ %
    touching
     $v$, the vertex relaxation of vertex $v$ into the edge $e = vv'$ %
    separating
     faces $F^1, F^2$ is defined by:
    \begin{align*}
    	G{\times}v _{\in F^1, F^2} & := (V \cup \{v'\},E \cup \{e\}) \\
    	\Gamma_{G}{\times}v_{\in F^1, F^2} & = (\Gamma_G \setminus \{F^1, F^2\}) \cup \{F'^1,  F'^2\}
    \end{align*}

    where we write $F^1 = \dots e_1e_2 \dots$ and $F^2 = \dots e_3e_4 \dots$ with $e_1, e_2, e_3, e_4$ incident to $v$, and then $F'^1 = \dots e_1ee_2 \dots$ and $F'^2 = \dots e_3ee_4 \dots$.
    \label{def:vertex_relaxation}
 \end{definition}

The vertex relaxation operation consists in expanding a vertex $v$ into an edge $e = vv'$.
Vertex relaxation can be performed by adding an edge in the dual embedding (\Cref{prop:vertex_relaxation_duality_edge_addition}).

 \begin{definition}[Edge contraction]
    Consider a planar embedding $\Gamma_G$ of a connected planar graph $G = (V,E)$.
\newcommand{\renamingfunction}{\rho}
	Given an edge $e = vv'$ appearing in cycles of length $\geq 3$, the \emph{edge contraction} of $e$ is defined by:
    \begin{align*}
		{G}{/}e & = (V \setminus \set{v'}, \multiset{\renamingfunction(x)\renamingfunction(y) \mid xy \in E \setminus \set{vv'}}) \\
        \Gamma_{G}{/}e & = (\Gamma_G \setminus \{F'^1,  F'^2\}) \cup \{F^1,  F^2\}
    \end{align*}

	where $
			\renamingfunction(x) = \begin{cases}
				x ~\text{if}~ x \in V \setminus \{v'\} \\
				v ~\text{otherwise}
			\end{cases}
		$
	and we write $F'^1 = \dots e_1ee_2 \dots$ and $F'^2 = \dots e_3ee_4 \dots$ with $F^1 = \dots e_1e_2 \dots$ and $F^2 = \dots e_3e_4 \dots$.

    \label{def:edge_contraction}
 \end{definition}

 Note that edge contraction is the inverse operation of the vertex relaxation (see \Cref{fig:vertexrelaxationedgecontraction}).%

 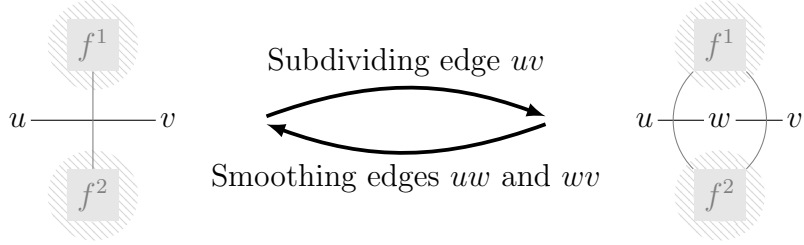
\begin{figure}[t]
 	\centering
 	
 	\begin{tikzpicture}[baseline=0mm]
 		\node[primal] (v1) at (-1, 0) {$u$};
 		\node[primal] (v2) at (1,0) {$v$};
 		\facenode{0}{1}
 		\facenode{0}{-1}
 		\node[face] (f1) at (0, 1) {$f^1$};
 		\node[face] (f2) at (0, -1) {$f^2$};
 		\draw (v1) edge (v2);
 		\draw[dual] (f1) edge (f2);
 	\end{tikzpicture}
 	\superleftrightarrow{Subdividing edge $uv$}{Smoothing edges $uw$ and $wv$}
 	\begin{tikzpicture}[baseline=0mm]
 		\node[primal] (v1) at (-1, 0) {$u$};
 		\node[primal] (v3) at (0, 0) {$w$};
 		\node[primal] (v2) at (1,0) {$v$};
 		\facenode{0}{1}
 		\facenode{0}{-1}
 		\node[face] (f1) at (0, 1) {$f^1$};
 		\node[face] (f2) at (0, -1) {$f^2$};
 		\draw (v1) edge (v3);
 		\draw (v3) edge (v2);
 		\draw[dual] (f1) edge[bend right=45] (f2);
 		\draw[dual] (f1) edge[bend left=45] (f2);
 	\end{tikzpicture}

 	\caption{Edge subdivision and edge smoothing. Edge subdivision consists in adding a middle vertex $w$ while edge smoothing consists in removing a middle vertex $w$.}
 	\label{fig:smoothing_subdivision}
 \end{figure}

\begin{definition}[Edge smoothing]
	Edge smoothing of edges $uw$ and $wv$ with $\degree{G}(w) = 2$ is the edge contraction of $uw$ (or $wv$).
	\label{def:edge_smoothing}
\end{definition}

\begin{definition}[Edge subdivision]
    Edge subdivision of edge $uv$ consists in a vertex relaxation of $v$ into $uw$,$wv$ with $\degree{G}(w) = 2$.
    \label{def:edge_subdivsion}
\end{definition}

Note that edge smoothing is the inverse operation of edge subdivision which consists in adding an intermediate vertex $w$ on the edge $uv$ (see Figure~\ref{fig:smoothing_subdivision}).

\subsection{Duality Among Operations}

We now state the duality between vertex relaxation and edge addition duality (see \Cref{fig:vertexrelaxationedgecontraction} and \ref{fig:dualofvertexrelationandedgecontraction}). We give two equivalent statements (one in the primal/dual, the second in the dual/primal).

\begin{property}[Vertex Relaxation and Edge Addition Duality]
	Consider an embedding $\Gamma_G$ and its dual $\Gamma^*_G$.
	Let $e = F^1F^2$ be a future edge to be added in $\Gamma^*_G$ inside the face $v$ with $F^1 \neq F^2$.
	The face $v$ of $\Gamma^*_G$ is a vertex of $G$ adjacent to the faces $F^1$ and $F^2$ of $\Gamma_G$.
	Then the edge addition of $e$ in the face $v$ of $\Gamma^*_G$ is a vertex relaxation of $v$ in $\Gamma_G$:
	$$(\Gamma^*_G{+}e_{\in v})^* = \Gamma_G{\times}v_{\in F^1,F^2}$$
    \label{prop:vertex_relaxation_duality_edge_addition}
\end{property}

\newcommand{\edgeAdditionEmbed}{\Gamma^*_G{+}e_{\in v}}

\begin{proof}
	Consider $F^1, F^2 \in \Gamma_G$ such that $e = F^1F^2$ is not in $\Gamma^*_G$.
	We perform an edge addition inside the face $v \in \Gamma^*_G$ with $v$ a vertex of $G$.
    
	First we describe the faces $F^1$ and $F^2$ in $\Gamma_G$.
	Since $G$ is connected planar, we have $\Gamma_G^{**} = \Gamma_G$ by \Cref{prop:if_g_conn_then_g_dual_dual_is_g}.
	Therefore $\Gamma_G$ is the dual of $\Gamma^*_G$.
	Recall that $v = e_4 \dots e_1 e_2 \dots e_3$ in $\Gamma^*_G$ (\Cref{def:edge_addition}).
	We define $i$ and $j$ such that $v_i = e_1, ~ v_{i+1} = e_2$ and $v_j = e_3, ~ v_{j+1} = e_4$.
    Let $\Psi$ and $\Omega$ be two faces of $\Gamma^*_G$ such that $\other(\Psi,k) = (v,i)$ and $\other(\Omega,k') = (v,j)$ (\Cref{def:other_function})
    By construction of the dual embedding (see \Cref{def:dual_construction}) we have: 
    \begin{equation*}
        \begin{split}
            F^1 & = \dots \faces_{\Gamma^*_G}(\Psi_k) \cdot \faces_{\Gamma^*_G}(v_{i+1}) \dots \\ 
			& = \dots \faces_{\Gamma^*_G}(e_1) \cdot \faces_{\Gamma^*_G}(e_2) \dots \\
            F^2 & = \dots \faces_{\Gamma^*_G}(\Omega_{k'}) \cdot \faces_{\Gamma^*_G}(v_{j+1}) \dots \\ 
			& = \dots \faces_{\Gamma^*_G}(e_3) \cdot \faces_{\Gamma^*_G}(e_4) \dots
        \end{split}
    \end{equation*}

	Now consider the dual embedding of $\edgeAdditionEmbed$ that we name $(\Gamma_G)'$.
	Recall that since $e$ is an edge between two existing vertices of $G^*$, after the addition of $e$ in $\Gamma^*_G$, we have $\edgeAdditionEmbed = (\Gamma_G \setminus \{v\}) \cup \{ v^1, v^2\}$ such that $v^1 = \dots e_1 \cdot e \cdot e_4 \dots $ and $v^2 = \dots e_3 \cdot e \cdot e_2 \dots$ (see \Cref{def:edge_addition}).

	We now define $i$ and $j$ such that $v^1_i = e_1$ and $v^2_j = e_3$.
	Since the only faces that have been modified in $\edgeAdditionEmbed$ is $v$ (replaced by $v^1$, $v^2$) we know that $\Psi_k = e_1$ and $\Omega_{k'} = e_3$.
	Hence we have $\other(\Psi,k) = (v^1,i)$ and $\other(\Omega,k') = (v^2,j)$.
	Therefore we obtain the following equalities:
	$v^1_{i+1} = v^2_{j+1} = e$ and $\other(v^1,i+1) = (v^2,j+1)$.
	Then we obtain the two following faces in the dual:

	\begin{equation*}
        \begin{split}
            F'^1 & = \dots \faces_{\Gamma^*_G}(\Psi_k) \cdot \faces_{\Gamma^*_G}(v^1_{i+1}) \cdot \faces_{\Gamma^*_G}(v^2_{j+2}) \dots\\ 
			& = \dots \faces_{\Gamma^*_G}(e_1) \cdot \faces_{\Gamma^*_G}(e) \cdot \faces_{\Gamma^*_G}(e_2) \dots \\
			F'^2 & = \dots \faces_{\Gamma^*_G}(\Omega_{k'}) \cdot \faces_{\Gamma^*_G}(v^2_{j+1}) \cdot \faces_{\Gamma^*_G}(v^1_{i+2}) \dots \\ 
			& = \dots \faces_{\Gamma^*_G}(e_3) \cdot \faces_{\Gamma^*_G}(e) \cdot \faces_{\Gamma^*_G}(e_4) \dots 
        \end{split}
    \end{equation*}

	Since $e$ is the only edge that have been added to $\Gamma^*_G$, the only edge that have been added to $\Gamma_G$ is $\faces_{\Gamma^*_G}(e)$ and we described both of the faces where $\faces_{\Gamma^*_G}(e)$ occurs ($F'^1$ and $F'^2$).
	Therefore we have $(\Gamma_G)' = (\Gamma_G \setminus \{F^1, F^2\}) \cup \{F'^1,  F'^2\}$ with $F^1, F^2$ and $F'^1, F'^2$ corresponding to the definition of \Cref{def:vertex_relaxation}.
	In conclusion we have $(\Gamma_G)' = \Gamma_G{\times}v_{\in F^1,F^2}$ implying:
	$$(\Gamma^*_G{+}e_{\in v})^* = \Gamma_G{\times}v_{\in F^1,F^2}$$
\end{proof}

\begin{property}[Edge Contraction and Edge Deletion Duality\cite{klein2011flatworlds}]	
	Consider an embedding $\Gamma_G$ and an edge $e$ in $G$.
	We have:
	$$(\Gamma^*_G{-}\bijectionedgetodual(e))^* = \Gamma_G / e$$
	\label{prop:edge_contraction}
\end{property}

\begin{property}[Self Loop Contraction\cite{klein2011flatworlds}]
	Consider an embedding $\Gamma_G$ and an edge $uv$ in $G$ such that $\deg_{G}(v) = 1$.
	In $\Gamma^*_G$, the edge $\bijectionedgetodual(uv)$ is a self-loop forming a single edge face.
	We have $|\Gamma^*_G| - 1 = |(\Gamma_G{-}uv)^*|$.
	\label{prop:self_loop_deletion}
\end{property}

\subsection{Planarity and Connectivity Preservation}

\begin{property}
    Edge deletion and addition preserve planarity and connectivity.
    \label{prop:operations_preserve_planar_and_conn}
\end{property}

\begin{proof}
    Suppose that the embedding $\Gamma_G$ is connected and planar i.e. by \Cref{def:planar_embedding} $|V| - |E| + |\Gamma_G| = 2$ with $|V| = |\Gamma^*_G|$.
    
    \begin{itemize}
        \item Consider the case where an edge $e=uw$ is added between $u,w \in V$. By \Cref{def:edge_addition}, the number of vertices does not change,
					but there is one additional edge, and we have $|\Gamma_G{+}e_{\in F}| = |\Gamma_G|+1$. 
					We get $|V| - (|E|+1) + (|\Gamma_G|+1) = 2$, thus planarity is preserved.        
        \item Consider the case where an edge $e=uw$ is added between $u \in V$ and $w \not \in V$. 
		Since $w$ is of degree $1$ in $G$ by \Cref{prop:self_loop_deletion} we have $|(\Gamma_G{+}e\in F)^*| - 1 = |\Gamma^*_G| = |V|$.
					In this case, the number of vertices is incremented by 1, and so is the number of edges, but the number of faces does not change.
					We get $|V|+1 - (|E|+1) + |\Gamma_G| = 2$, and planarity is preserved.
        \item Now consider the case where the edge $e=uw$ is deleted with $u,w \in V$. By \Cref{def:edge_deletion}, the number of edges and the number of faces are decremented each by 1,
					while the number of vertices is unchanged.
					We get $|V| - (|E|-1) + |\Gamma_G|-1 = 2$, and planarity is preserved.
	\end{itemize}\end{proof}

\begin{property}
    Vertex relaxation and edge contraction preserve planarity and connectivity.
	\label{prop:relax_contrac_planar_preserve}
\end{property}

\begin{proof}
	Edge relaxation adds a vertex connected to an existing vertex, and increments the number of edges by 1, without changing the number of faces; thus it preserves planarity and connectivity.
	Edge contraction is the reverse operation so it also preserves planarity and connectivity.
\end{proof}

\begin{property}\cite{tamassia1988dynamic}
    Edge deletion, edge addition, vertex relaxation and edge contraction are computable in polynomial time.
    \label{prop:poly_time_embedding_modif}
\end{property}

As described in \Cref{def:edge_addition} and \Cref{def:edge_deletion} the edge addition and deletion consist in adding or deleting an element in at most two cyclic sequences which can be performed in poly-time.
Moreover the computation of a dual embedding as described in \ref{subsec:dual_graph} can be performed in poly-time since the function \other{} can be implemented in the following way: search the other occurence of a given edge $e$ in all cyclic sequences.
The combinatorial embeddings can be represented by data structures such as doubly-connected edge lists \cite{muller1978finding} in which each operation described above is computable in poly-time.

\subsection{Subgraph Embedded Inside a Specific Face}

In this section we consider the following connected and planar graphs $G$, $H$ and $G \cup H$.
We obtain $\Gamma_G$ (resp. $\Gamma_H$) by removing all occurrences of each edge of $H$ (resp. $G$) in an embedding $\Gamma_{G \cup H}$.
We can perform this operation since $\Gamma_G$ and $\Gamma_H$ are connected and planar.

\begin{definition}[Graph Embedded Inside a Face]
	The subgraph $H$ is embedded in a face $F \in \Gamma_G$ in $\Gamma_{G\cup H}$ if $\Gamma_{G\cup H}$ contains all the faces of $\Gamma_G$ but $F$.
	Formally $H \sqsubset_{\Gamma_{G\cup H}} F$ if $((\Gamma_G \setminus \{F\}) \subset \Gamma_{G \cup H} ~\text{and}~ F \not \in \Gamma_{G \cup H})$.
    \label{def:graph_embedded_inside_face}
\end{definition}

Intuitively, this definition means that $H$ is embedded in the face $F$ of $\Gamma_G$ if the only face of $\Gamma_G$ that is no more in $\Gamma_{G \cup H}$ is $F$.
Indeed, $F$ can be divided by adding successively the edges of $H$ and disappear in $\Gamma_{G \cup H}$: $F$ is the only face that has been modified from $\Gamma_G$ to $\Gamma_{G \cup H}$.

\begin{example}
	Consider two graphs $G = (\{1,g,2,b\}, \{1g,g2,2b,b1\})$ and $H = (\{g,a,b\}, \{ga, ab\})$ in \Cref{figure:algorithmmergingfacesthensmoothing}.
	Consider now the embeddings $\Gamma_G$ and $\Gamma_{G \cup H}$. All the faces of $\Gamma_G$ are inside $\Gamma_{G \cup H}$ except a face $F$ that contains the subgraph $H$ since $H$ cut $F$ in two:
	by \Cref{def:graph_embedded_inside_face} $H$ is embedded in the face $F$ of $\Gamma_G$.
\end{example}

\begin{property}[Subgraph Location After Vertex Relaxation]
	Consider $H \sqsubset_{\Gamma_{G \cup H}} F^1$.
	Let $\Gamma_{R} = \Gamma_{G \cup H} {\times} v _{\in A,B}$ be the embedding of $\Gamma_{G \cup H}$ after the relaxation of $v$.
	If $v \in V(G) \setminus V(H)$, then $H \sqsubset_{\Gamma_{R}} F'^1$ with $F'^1$ being the face obtained after the relaxation of $v$ in $\Gamma_G$ by \Cref{def:vertex_relaxation}. 
	\label{prop:location_after_vertex_relaxation}
\end{property}

\begin{proof}
	By \Cref{def:vertex_relaxation}, a vertex relaxation implies the addition of an edge $e = vv'$ in faces $F^1$ and $F^2$: we obtain from $F^1 = \dots e_1 \cdot e_2 \dots$ the face $F'^1 = \dots e_1 \cdot vv' \cdot e_2 \dots$.
	By \Cref{prop:relax_contrac_planar_preserve} since $\Gamma_{G \cup H}$ is planar and connected $\Gamma_R$ is planar and connected.
	Moreover, $v$ and $v'$ are not vertices of $H$ then $H$ is left unchanged from $\Gamma_{G \cup H}$ to $\Gamma_R$.
	We can conclude that $H \sqsubset_{\Gamma_{R}} F'^1$ since all the edges of $F^1$ are in $F'^1$.
\end{proof}

\begin{property}[Subgraph Location After Edge Deletion]
	Consider $H \sqsubset_{\Gamma_{G \cup H}} F^1$ and $e = uv$, an edge that separates $F^1$ and $F^2$.
	Let $\Gamma_{D} = \Gamma_{G \cup H}{-}e$ be the embedding of $\Gamma_{G \cup H}$ after the deletion of $e$.
	If $u,v \in V(G) \setminus V(H)$, then $H \sqsubset_{\Gamma_{D}} F$ with $F$ being the face obtained after the deletion of $uv$ in $\Gamma_G$ in \Cref{def:edge_deletion}.
	\label{prop:location_after_edge_deletion}
\end{property}

\begin{proof}
	By \Cref{def:edge_deletion}, an edge deletion implies that $F^1$ and $F^2$ are merged into a single face $F$: from $F^1 = \dots e_1 \cdot uv \cdot e_2 \dots$ we obtain $F = \dots e_1 \cdot e_4 \dots e_3 \cdot e_2 \dots$.
	Thanks to \Cref{prop:operations_preserve_planar_and_conn} and by the same arguments used in the proof of \Cref{prop:location_after_vertex_relaxation} we have $H \sqsubset_{\Gamma_D} F$ since all the edges of $F^1$ are in $F$.
\end{proof}

\begin{example}
	\Cref{figure:algorithmmergingfacesthensmoothing} shows the subgraph $H_1: g - a - b$ embedded inside face $F_1 = 1b2g$ and $H_2: h - i -j$ embedded inside face $F_2: 3h2j$. After vertex relaxation of $2$ into $22'$, $H_1$ is embedded inside face $1b2'2g$ while $H_2$ is embedded inside face $3h22'j$. After edge deletion of $22'$, both $H_1$ and $H_2$ are embedded inside face $1b2'j3h2g$.
\end{example}

\subsection{Planar Eulerian Graph}
Given a \emph{connected} planar graph $G = (V, E)$, an \emph{Eulerian cycle} in $G$ is a closed walk that traverses each edge of $G$ exactly once. 

\begin{definition}[Eulerian Graph]
    Graph $G$ is Eulerian iff  $G$ is connected and every vertex has an even degree.
    \label{prop:eulerian_graph_prop}
\end{definition}

\begin{property}\cite{van2001course}
    A connected planar graph $G$ is Eulerian iff its dual $G^*$ is bipartite.
    \label{prop:bipartite_dual_eulerian}
\end{property}

In the following lemma, $G$ is planar and Eulerian. Therefore, by Property~\ref{prop:bipartite_dual_eulerian}, the dual graph $G^*$ is bipartite. So the faces of $G$ are either blue or red.

\begin{property}
	Let $G$ be a planar Eulerian graph. $G$ has two faces iff $G$ is a cycle.
    \label{prop:planar_eulerian_g_two_faces_is_cycle}
\end{property}

\begin{proof}
    As $G$ is planar, the Euler formula applies and we get $|V| - |E| + 2  = 2$, hence $|V| = |E|$.
    By the handshake lemma, we have the following equation: 
    \[
    \sum_{v \in V}{\degree{G}(v)} = 2|E| = 2|V|
    \]
    Since $G$ is Eulerian we have for all $v \in V$, $\degree{G}(v) = 2k$ with $k > 0$.
    This implies that for all $v \in V$, $\degree{G}(v) = 2$.
    As $G$ is connected, it implies that $G$ is a cycle.
    Reciprocally, if $G$ is a cycle, then given any embedding, there is two faces: the inside and the outside ones.
\end{proof}

\end{appendices}

\end{document}